\documentclass[12pt,letterpaper]{JHEP3}

\usepackage{amsmath,amssymb,amsfonts,amsxtra,graphics,graphicx,amsthm,verbatim}

\newcommand{\be}{\begin{equation}}
\newcommand{\ee}{\end{equation}}
\newcommand{\beq}{\begin{equation}}
\newcommand{\eeq}{\end{equation}}
\newcommand{\ba}{\begin{array}}
\newcommand{\ea}{\end{array}}
\newcommand{\bea}{\begin{eqnarray}}
\newcommand{\eea}{\end{eqnarray}}
\newcommand{\ben}{\begin{enumerate}}
\newcommand{\een}{\end{enumerate}}
\newcommand{\bean}{\begin{eqnarray*}}
\newcommand{\eean}{\end{eqnarray*}}
\newcommand{\eref}[1]{(\ref{#1})}

\newcommand{\tref}[1]{Table~\ref{#1}}

\newcommand{\BC}{\mathbb{C}}

\newcommand{\BZ}{\mathbb{Z}}
\newcommand{\BF}{\mathbb{F}}
\newcommand{\CC}{{\cal C}}

\newcommand{\ra}{\rightarrow}

\title{On the Classification of Brane Tilings}

\author{John Davey, Amihay Hanany, Jurgis Pasukonis \\
Theoretical Physics Group, The Blackett Laboratory \\
Imperial College London, Prince Consort Road\\
London,  SW7 2AZ,  UK \\
{\tt jpdavey, a.hanany, jurgis.pasukonis08@imperial.ac.uk}}

\abstract{We present a computationally efficient algorithm that can be used to generate all possible brane tilings. Brane tilings represent the largest class of superconformal theories with known AdS duals in 3+1 and also 2+1 dimensions and have proved useful for describing the physics of both D3 branes and also M2 branes probing Calabi-Yau singularities. This algorithm has been implemented and is used to generate all possible brane tilings with at most 6 superpotential terms, including consistent and inconsistent brane tilings. The collection of inconsistent tilings found in this work form the most comprehensive study of such objects to date.}

\preprint{Imperial/TP/09/AH/04}

\begin{document}

\section{Introduction}
Brane tilings were originally developed to help understand the superconformal gauge theories that live on D3 branes probing toric Calabi-Yau singularities \cite{HananyKennaway} \cite{FrancoHanany}. These ideas are reviewed in \cite{Kennaway07} and \cite{YamazakiRev}. In some sense these theories are dual to Type IIB string theory in a background of $AdS_5 \times X_5$, where $X_5$ is a Sasaki-Einstein Manifold \cite{Kehagias} \cite{KlebanovWitten} \cite{HullEtAl} \cite{Plesser}. One use of brane tilings is in determining exactly which string theory is dual to a given superconformal theory \cite{FrancoHanany}.\\

Recently, it has been shown that brane tilings can also be used to describe supersymmetric quiver Chern-Simons (CS) theories \cite{TilingCS} \cite{TilingM2}. This discovery has not only increased our knowledge of these interesting theories but has also furthered our understanding of the physics of M2 branes probing toric Calabi-Yau 4-fold singularities.\\

Not every tiling corresponding to a 3+1 dimensional theory is a consistent brane tiling \cite{Zig-Zag}. However it appears that brane tilings used to describe 2+1 dimensional Chern-Simons theories have a more relaxed set of consistency conditions. There are therefore a large number of tilings that were once ignored are now thought to be important in the study of M2 branes. For this reason in this paper we do not see the 3+1 dimensional consistency condition as a selection criteria for a tiling, but rather a feature of a tiling.\\

In this paper we present a computationally efficient algorithm that can in principle be used to generate a complete list of brane tilings. This list includes all known 3+1 dimensional tilings with at most 6 superpotential terms and forms the largest collection of inconsistent tilings computed to date. This list also includes a set of models that do not have (3+1)d parents \cite{Orphans}. At the heart of the algorithm is the enumeration of all possible quivers \cite{Tax} \cite{YangsTeam}. For each quiver found, all superpotentials satisfying the toric condition \cite{ToricD} are generated and each of these theories are then tested as to whether they can admit a tiling description. As each tiling corresponds to both a superpotential and a quiver, we can be sure that this search is truly exhaustive. We then use an implementation of the algorithm to compute all possible brane tilings with 2, 4 and 6 superpotential terms.\\

Although several families of tilings (for instance $Y_{p,q}$ \cite{FrancoHanany}, $X_{p,q}$ \cite{Infinite}, $Z_{p,q}$ \cite{Oota}, $L^{a,b,c}$ \cite{Labc}, the del Pezzo surfaces \cite{FrancoHanany}, \cite{delPezz} and the pseudo del Pezzo surfaces \cite{PdP}) are known, to this date no truly complete list of brane tilings has been constructed. Such work would help classify both a large class of string theory backgrounds and also the largest class of SCFTs with known AdS duals. The algorithm in this paper makes it possible to generate such a list. Admittedly there are computational limitations to the algorithm, although an ordinary desktop computer is capable of finding all of the tilings listed in this paper in a short period of time.\\

\section{Some Background}
Before we discuss the classification algorithm, let us review some of the basics of brane tilings.\\

\begin{figure}
\begin{center}
\begin{tabular}{ccc}
\includegraphics[height=5cm]{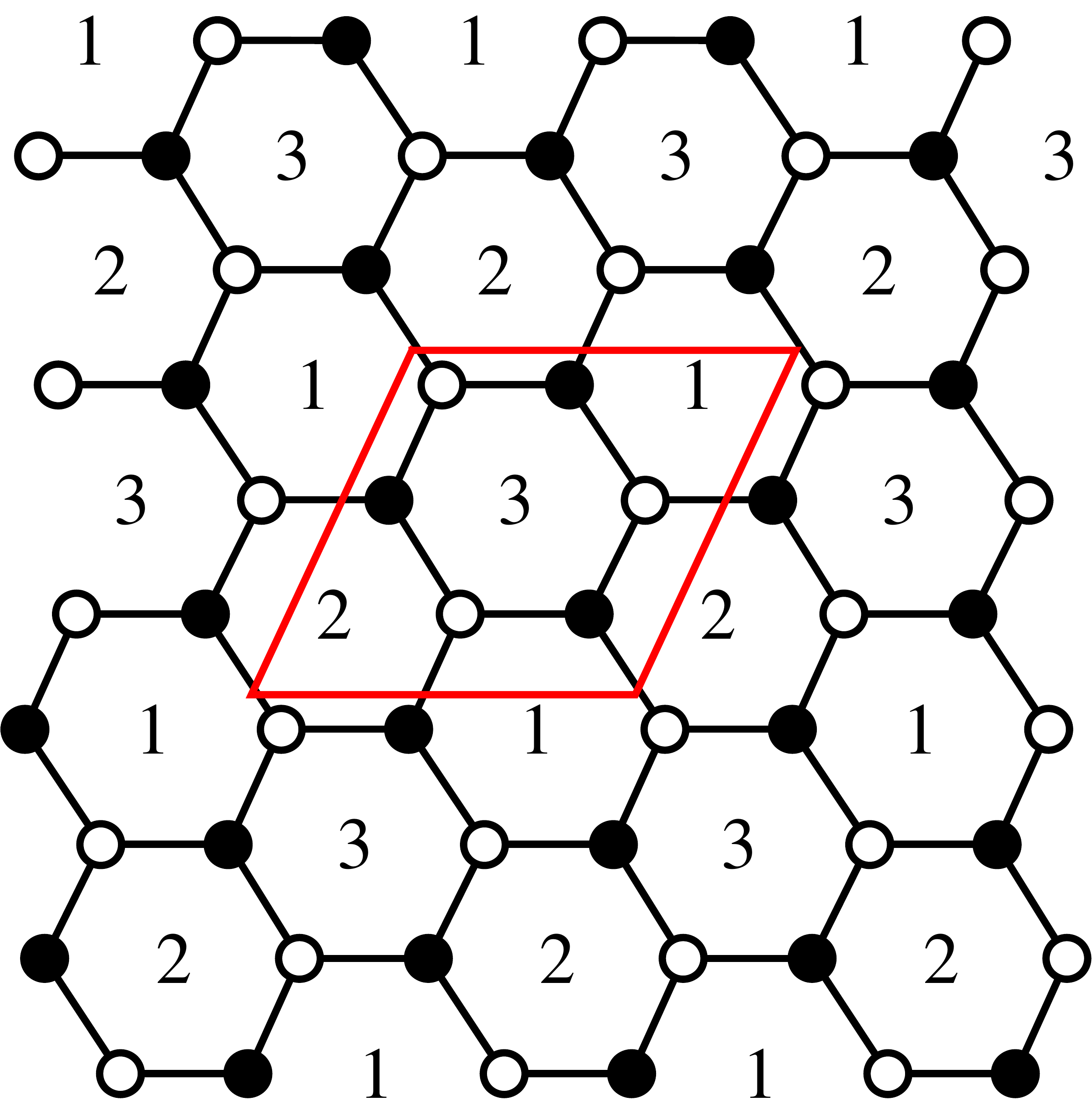}
\end{tabular}
\end{center}
\caption{A Typical Brane Tiling. The fundamental domain is drawn in red. A 2-torus can be formed by identifying opposite edges of this red parallelogram.}
\label{fig:btilingintro}
\end{figure}

A brane tiling (or dimer model) is a periodic bipartite graph on the plane. Alternatively, we may draw it on the surface of a 2-torus by taking the smallest repeating structure (known as the fundamental domain) and identifying opposite edges \cite{HananyKennaway}. The bipartite nature of the graph allows us to colour the nodes either white or black such that white nodes only connect to black nodes and vice versa. A typical brane tiling is given (Figure \ref{fig:btilingintro}). For this tiling, the smallest repeating unit consists of 6 nodes (3 black and 3 white) and 9 edges. Brane tilings can be used to describe the world volume physics of both D3 and M2 branes.\\

\subsection{Brane Tilings for D3 Brane Theories}
Brane tilings were originally developed to describe certain 3+1 dimensional superconformal field theories (SCFTs) that arise in Type IIB string theory.\\

Specifically, let us consider Type IIB string theory on $AdS_5 \times X_5$, where $X_5$ is a Sasaki-Einstein manifold. This string theory can be thought of as the gravity dual of a gauge theory living in a stack of D3 branes placed at the conical singularity of $Y_6$ \label{Y6}, the cone over $X_5$ \cite{KlebanovWitten}. In this work we take $Y_6$ to be a (non-compact) toric Calabi-Yau 3-fold. The world-volume theory of the D3-branes placed at the singularity is defined by a Brane Tiling in the UV.\\

There is a simple dictionary between a tiling and the 3+1 dimensional gauge theory that it represents. Every face in the tiling corresponds to a $U(N)$ gauge group. Each edge in the tiling corresponds to a chiral field that transforms under a bi-fundamental representation of the two gauge groups that the edge sits next to in the tiling, with an orientation defined by the bipartite nature of the tiling. White (black) nodes in the tiling correspond to positive (negative) superpotential terms. Each term is a gauge invariant quantity formed by tracing over the fields that the node connects to. The relationship between a tiling, its graph dual - the periodic quiver and the gauge theory it represents is given (Table \ref{t:bt}). One can fully reconstruct a quiver gauge theory's Lagrangian with knowledge of the tiling.\\

\begin{table}[h]
\begin{center}
\begin{tabular}{c|c|c}
Tiling & Periodic Quiver & Gauge Theory\\
\hline
Face & Node & $U(N)$ Gauge Group\\
Edge & Edge & Bi-fundamental Chiral Field \\
Node & Face & Superpotential Term
\end{tabular}
\end{center}
\caption{The relationship between brane tilings, periodic quivers and the field theories they represent}
\label{t:bt}
\end{table}

A quantum field theory can be defined by specifying an ultraviolet fixed point together with an infrared fixed point connected by a renomalisation group flow \cite{Amax1} \cite{Amax2}. Every quantum field theory is thought to flow to some conformal field theory at low energies. Quantum field theories corresponding to brane tilings are not exceptions to this rule and every such theory flows to a supersymmetric conformal field theory (SCFT) at low energies. The IR limit of a large class of quantum field theories corresponding to brane tilings is known, although some `inconsistent' brane tilings exist which correspond to theories that have unknown IR properties. These inconsistent tilings can correspond to gauge theories that are tachionic \cite{Tachion}, while others are fractional Seiberg duals \cite{FractionalDuals} or mutations \cite{Mutations}. Luckily there is a simple and elegant consistency check we can perform on a tiling.\\

It is thought that a tiling representing a 3+1 dimensional gauge theory is consistent if and only if it has the same number of gauge groups as there are cycles for D-branes to wrap in the dual gravity theory \cite{Zig-Zag} \cite{Gulotta}. A glance at the tiling is sufficient to find the number of gauge groups of the quiver theory however the method we employ to count the number of gauge groups from the string theory side is a little more involved. One way of counting the relevant cycles is by computing the area enclosed by toric diagram produced by applying the fast forward algorithm to the tiling \cite{FrancoHanany}. Many of the tilings later shown are labeled consistent or inconsistent based on this check.\\

\subsection{Brane Tilings for M2 brane Theories}
$AdS_4 \times X_7$ M-theory backgrounds, where $X_7$ is a seven dimensional Sasaki-Einstein manifold are known to preserve $\cal{N}$=2 supersymmetry and have been the subject of much recent investigation  \cite{HullEtAl} \cite{Plesser} \cite{Ueda}. These backgrounds arise as near horizon geometries of M2 branes probing the singular tip of $C(X_7)$ (the cone over $X_7$), which is known to be a Calabi-Yau 4-fold singularity. Brane tilings are proving to be useful tools in making the correspondence between these Calabi-Yau 4-fold singularities and their dual 2+1 dimensional Chern-Simons theories precise \cite{TilingCS} \cite{TilingM2}. Tilings can also be used to examine the structure of the master space, the mesonic moduli space, and the baryonic moduli space of these Chern-Simons theories \cite{Davey:2009sr}.\\

Conveniently, and perhaps not so surprisingly \cite{TilingCS}, the tilings used to describe M2 brane theories are very similar to the original D3 brane tilings.  They are both periodic, bipartite tilings of the torus $T^2$ with every face corresponding to a $U(N)$ gauge group, every edge depicting a chiral superfield, transforming under a bi-fundamental representation of the gauge groups it borders, and each white (black) node representing a positive (negative) superpotential term. The main difference between the two types of tiling is that Chern-Simons theories have a set of levels that must be chosen in order to completely specify the theory \cite{TilingM2}. These levels are integer valued, and a different level is associated to each gauge group.\\

One important feature of brane tilings that are used to describe 2+1 dimensional Chern-Simons theories is that they are thought to have a more relaxed set of consistency conditions.  This means there are many tilings that are interesting in the study of M2 branes that were not studied in any detail previously. These inconsistent tilings may be useful in the study of smooth toric Fano threefolds \cite{Fano}. In the remainder of this work we therefore completely relax the usual tiling consistency conditions \cite{Zig-Zag} \cite{Gulotta}; that is to say we focus on attempting to generate all periodic bipartite tilings of the plane.\\

\section{Classification Algorithm}
Let us now face the challenge of classifying the periodic, bipartite, two-dimensional tilings of the plane. 
Given the combinatoric complexity of the problem, we choose a purely computational approach and present a clearly defined algorithm that could, in principle, list all the tilings in increasing complexity.\\

The total number of these tilings is, of course, infinite, so the first thing we must do is choose some parameters to help organize the classification. The natural parameters of a tiling are the number of nodes in the fundamental domain of the tiling $N_T$ and the number of tiles $G$. The number of edges in the fundamental domain $E$ is then fixed by the Euler condition:
\be
E=G+N_T.
\ee
From the quiver gauge theory perspective these numbers are the number of nodes in the quiver $G$, the number of fields $E$ and the number of terms in the superpotential $N_T$.\\

Unfortunately working directly with tilings is computationally quite difficult. As the main objects of a tiling are geometrical, it is not obvious how to set up a systematic calculation of the possible periodic tilings with some parameters $(N_T, G)$, especially without making any a priori assumptions about the shapes of the tiles. For that reason we choose \emph{quiver gauge theories} as our main working objects \cite{Orphans} \cite{Tax} \cite{YangsTeam}. The method we choose is to enumerate all possible quivers and superpotentials, and then check which ones admit a tiling description. As each brane tiling corresponds to a quiver gauge theory, we can be sure that every tiling will be generated.\\

In summary, we propose the following algorithm for the classification of tilings:
\begin{enumerate}
\item 
Fix the order parameters $(N_T, G)$. 
\item 
Enumerate all distinct \emph{irreducible} quivers with $G$ nodes and $E=G+N_T$ fields.
\item 
For each quiver enumerate all possible superpotential terms satisfying the toric condition. This gives the full list of possible quiver gauge theories for $(N_T, G)$.
\item 
Try to reconstruct the tiling for each quiver gauge theory. If we succeed, we add it to the classification, otherwise we conclude that the gauge theory doesn't have a tiling description.
\end{enumerate}

Each step here requires further explanation. But let us postpone this and introduce the concept of doubling, which will provide some important insights and an understanding of the term irreducible quiver.\\

\subsection{The Doubling Process and Quadratic-Node Tilings}

\begin{figure}[h]
\begin{center}
\includegraphics{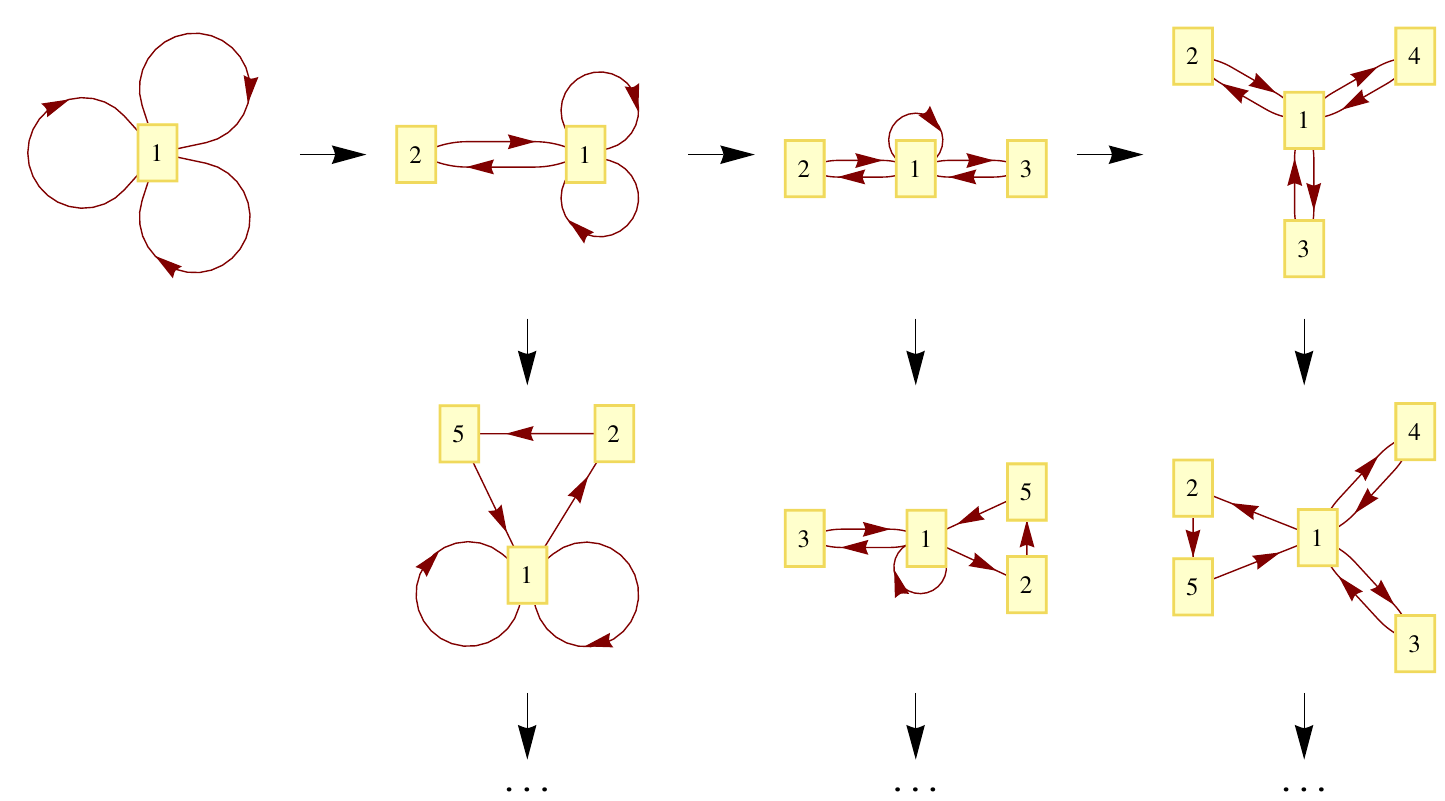}
\end{center}
\caption{Quivers generated by applying doubling to $\BC^3$.}
\label{fig:doubling-C3}
\end{figure}

Let us consider an operation on a quiver where we replace an edge with two edges, both connected to a node of valence 2. We shall call this process doubling. This process defines a new theory when applied to any of the fields in a quiver. For example, starting with a simple $\BC^3$ model we can construct an infinite number of models by repeatedly applying the doubling procedure (see Figure~\ref{fig:doubling-C3}). In some sense the process has a dual action on the brane tiling. An edge in the tiling is replaced by two edges and a face surrounded by only these 2 edges. This is known as a double bond \cite{TilingM2} \cite{Davey:2009sr}.\\

\begin{figure}[h]
\begin{center}
\includegraphics{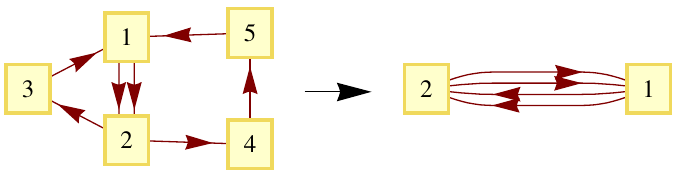}
\end{center}
\caption{Reduction of a quiver by removal of single-in, single-out nodes.}
\label{fig:reduction1}
\end{figure}

This doubling process is always reversible. If we are given a brane tiling with double-(or multi-)bonds, we can always remove them by the process of ``higgsing". By higgsing the right fields we can remove all nodes of valence 2 from the quiver (Figure~\ref{fig:reduction1}). Let us call quivers with at lease one node of valence two ``reducible". If a quiver isn't reducible it is said to be ``irreducible'\\

Now we understand the action of the doubling process on both the tiling and the quiver, we may consider from now on only irreducible quivers (or tilings with no double-(or multi-)bonds. All reducible quivers can easily be generated by applying the doubling process to the set of irreducible quivers. This is a crucial observation, because it lets us effectively ignore an infinite ``direction" in the space of tilings, thus allowing us to concentrate on the much smaller class of brane tilings, which are not related by this simple transformation.\\

\begin{figure}[h]
\begin{center}
\begin{tabular}{ccc}
\begin{minipage}{5cm}
\includegraphics[width=5cm]{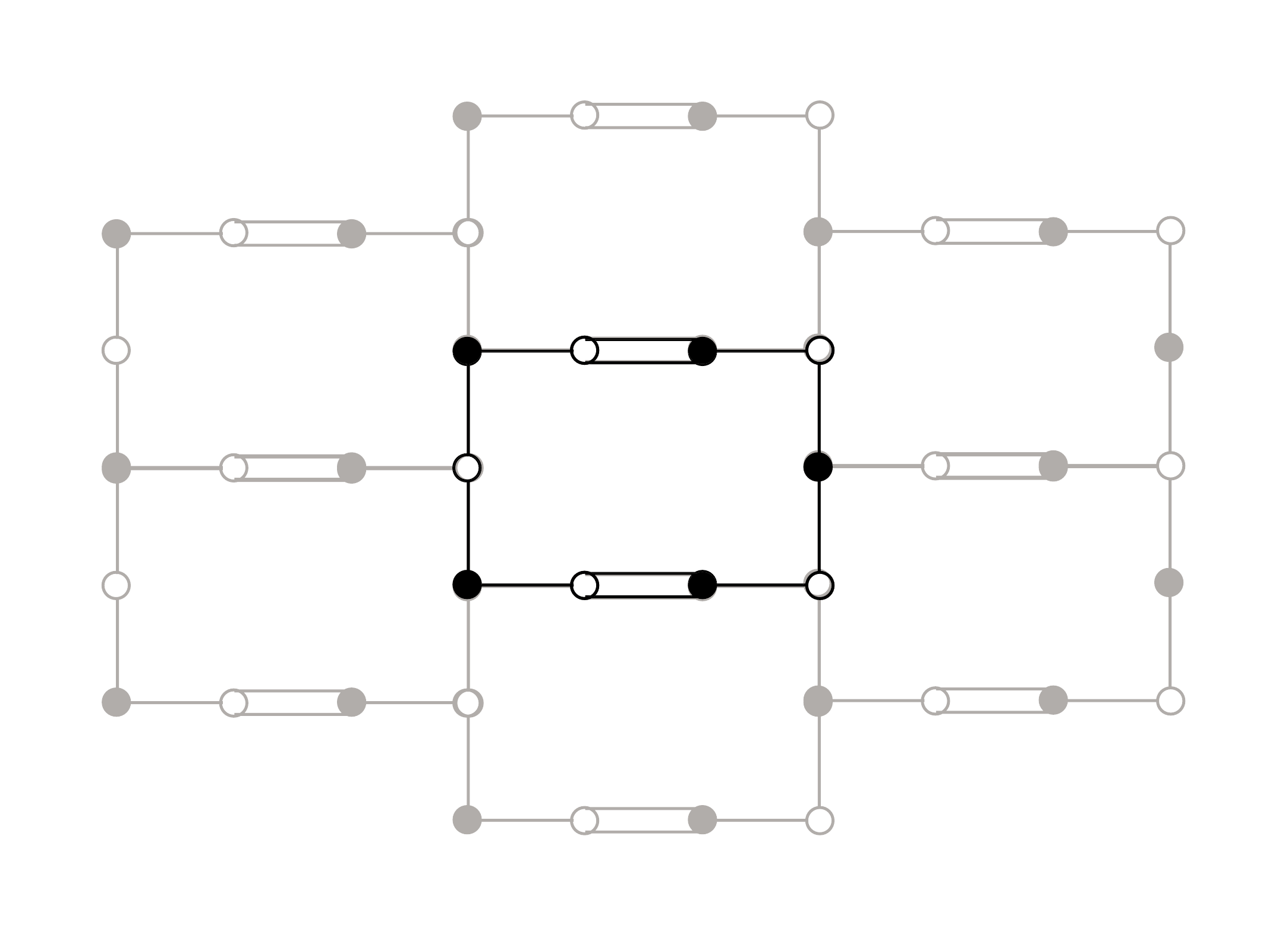}
\end{minipage} 
&  $\rightarrow$ &
\begin{minipage}{5cm}
\includegraphics[width=5cm]{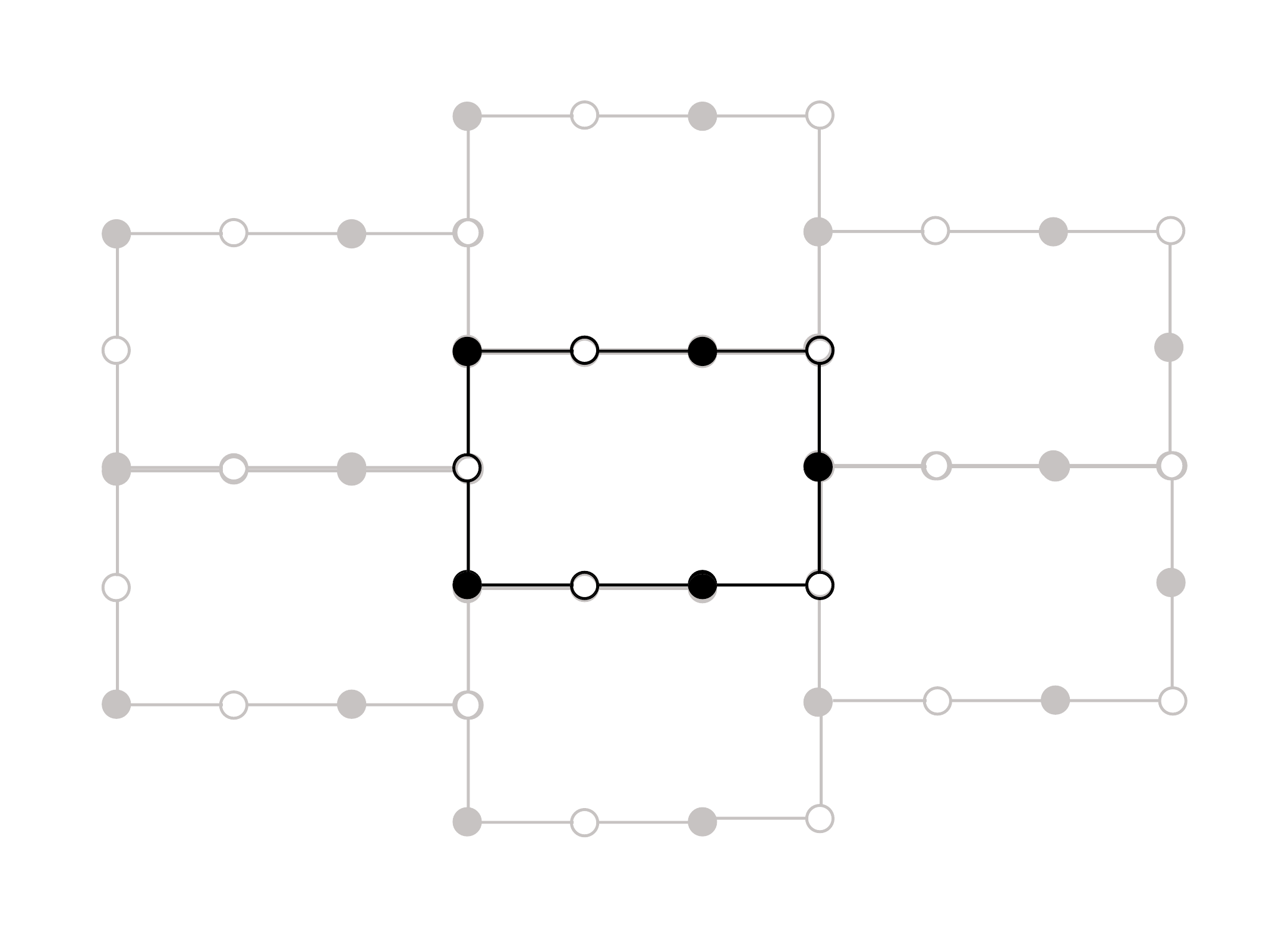}
\end{minipage} 
 \\
\begin{minipage}{5cm}
\begin{center} \includegraphics{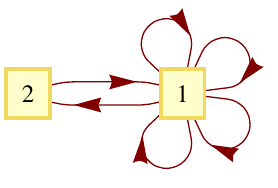} \end{center}
\end{minipage} 
& $\rightarrow$ &
\begin{minipage}{5cm}
\begin{center} \includegraphics{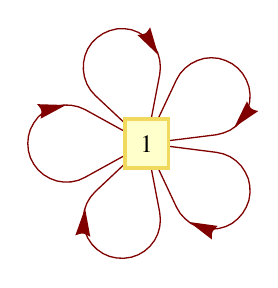} \end{center}
\end{minipage} 
\end{tabular}
\end{center}
\caption{Reduction of a quiver resulting in quadratic-node tiling.}
\label{fig:reduction2}
\end{figure}

We have to note, however, that there is one caveat in the argument above. For some reducible quivers the higgsing procedure results in a brane tiling, which has nodes connected only by two edges, as seen in Figure~\ref{fig:reduction2}. This means that the corresponding quiver gauge theory will have a superpotential with quadratic terms in it. We call such models quadratic-node tilings. \\

\begin{figure}[h]
\begin{center}
\begin{tabular}{ccc}
\begin{minipage}{5cm}
\includegraphics[width=5cm]{N4-base.pdf}
\end{minipage} 
&  $\rightarrow$ &
\begin{minipage}{5cm}
\includegraphics[width=5cm]{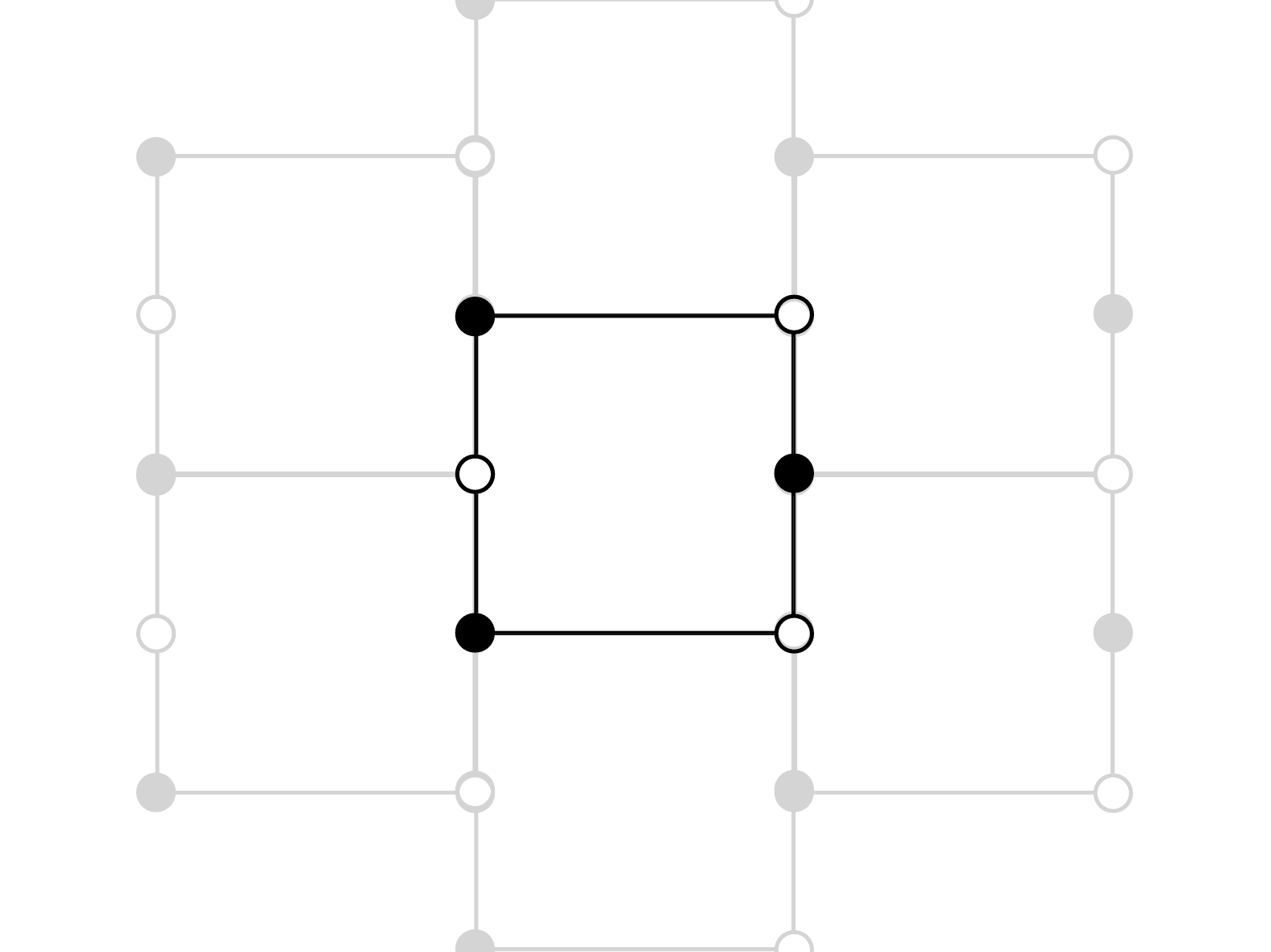}
\end{minipage} 
\end{tabular}
\end{center}
\caption{Reduction of a quadratic-node tiling.}
\label{fig:reduction3}
\end{figure}

The quadratic-node tilings are perfectly valid as bipartite tilings of a plane, however, they are not normally considered in the context of quiver gauge theories on D3 or M2 branes. This is because the quadratic superpotential terms indicate massive fields, which become non-dynamical in the infrared limit \cite{FrancoHanany}. Since we are interested in analyzing the IR limit of these gauge theories, the massive fields should be integrated out using their equations of motion. The corresponding effect on the tiling is that the quadratic node can be removed, gluing the two adjacent nodes together (see Figure~\ref{fig:reduction3}).\\

For this reason we exclude the tilings with quadratic nodes from our classification. However, this means that the models where quadratic nodes are only absent because of multi-edges (such as the one in Figure~\ref{fig:reduction2}) can not be recovered from the irreducible quivers simply by the doubling procedure. To get back such tilings from the classification in this paper we would have to combine the doubling procedure together with an insertion of two extra nodes.\\

With the procedure of doubling in mind, we restrict our attention to the classification of brane tilings without multi-edges and without quadratic nodes. Let us now go over the steps in the classification algorithm in more detail.\\

\subsection{Order parameters}

Let us recall the two parameters we are going to use to order our classification - $(N_T, G)$. Constraints placed on the tilings discussed above allow us to put limits on the possible values of $G$ for each value of $N_T$.\\

Firstly, let us consider the requirement that the quiver is irreducible. This is equivalent to saying that there should be no nodes in the quiver of valency 2. As the nodes must have the same number of incoming and outgoing edges\footnote{This is a consequence of the bipartite nature of the tiling.}, each node should be of valency 4 or higher. We also have the following relationship for any quiver:
\be
E = \frac{1}{2}\sum_{i=1}^{G}n_i,
\label{eq:adjacencies}
\ee
where $n_i$ is the order of node $i$. We can therefore conclude that we need a minimum of
\be
E_{min} = 2 G
\ee
fields for irreducible quivers. Using $E=G+N_T$ we have
\be
(N_T)_{min} = G
\ee
for fixed $G$ or, for fixed order parameter $N_T$, we have a maximum value of
\be
G_{max} = N_T
\ee
for order parameter $G$, if we only consider irreducible quivers. Most importantly, this tells us that for given $N_T$ there are a finite number of models with irreducible quivers. This statement is clear from the brane tiling perspective. If the number of nodes in a tiling is fixed, there are only a finite number of edges that can be added to the tiling that keep it irreducible.\\

A lower bound for $G$ for fixed $N_T$ can also be found. As the tilings are irreducible, this means the minimum order of all nodes is 3. Let us use \eref{eq:adjacencies} on the tiling, counting only edges and nodes in the fundamental domain. Now the edges are again fields and the nodes are the superpotential terms, giving us the bound:
\be
E_{min} = \frac{3}{2} N_T.
\ee
Using $E=G+N_T$ we get
\be
G_{min} = \frac{1}{2} N_T,
\ee
which is our lower bound on the parameter $G$ for given $N_T$, and so we have for fixed $N_T$\\
\be
\frac{1}{2} N_T \leq G \leq N_T
\ee
It is now clear how to organize the classification. We will consider each $N_T$ in an increasing order, exploring the possible $G$ values at each step. The number of possible superpotential terms $N_T$ is, of course, still unbounded, and we will be limited only by our technical abilities and curiosity. In this paper we explore the models up to \emph{six} terms in the superpotential. The summary of order parameters considered is given in \tref{tab:order}.\\

\begin{table}
\centering
\begin{tabular}{c|c|c|c|c}
$N_T$ & $G_{min}$ & $G_{max}$ & $E_{min}$ & $E_{max}$ \\
\hline
2 & 1 & 2 & 3 & 4 \\
4 & 2 & 4 & 6 & 8 \\
6 & 3 & 6 & 9 & 12
\end{tabular}
\caption{Values of order parameters explored.}
\label{tab:order}
\end{table}

\subsection{Finding Quivers}

Once we fix the parameters $(N_T, G)$, the next step is to enumerate all of the possible quiver graphs with a given number of nodes $G$ and edges $E$. The task is quite straightforward, but it has to be handled with a little care, to avoid the algorithm becoming too computationally expensive as $G$ and $E$ grow larger.\\

A naive approach would be to consider all possible ways of connecting $G$ nodes with $E$ edges. With $G(G-1)$ ways of drawing a directed edge, we would have the order of
\be
(G(G-1))^E
\ee
possible graphs to consider, which is clearly too large for, say, $G=6, E=12$. However, we are only interested in a very small fraction of these graphs. Nodes of quivers that correspond to brane tilings must have the same number of incoming as they do outgoing edges. This is known as the Calabi-Yau condition on the quiver theory and corresponds to the anomaly cancellation condition in 3+1 dimensions \cite{FrancoHanany}.\\

\begin{figure}[h]
\centering
\includegraphics{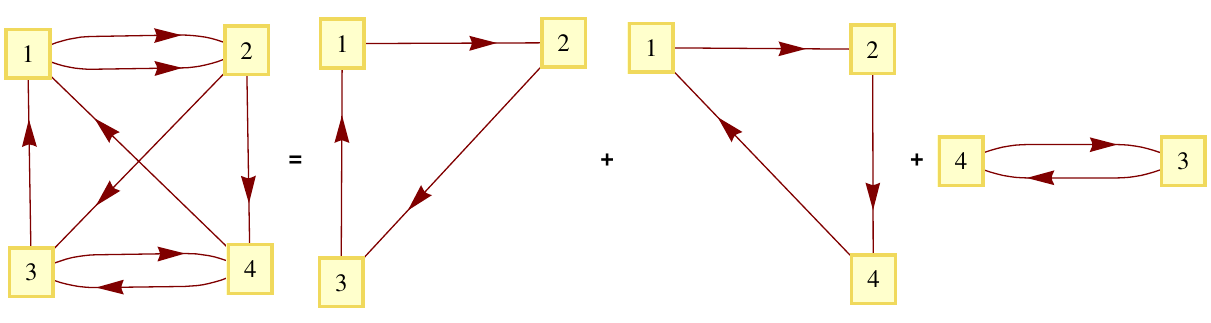}
\caption{Decomposition of a graph into cycles.}
\label{fig:decomposition}
\end{figure}

The key idea of this efficient algorithm for finding all possible quivers is to incorporate this Calabi-Yau condition into the construction of the quiver. We achieve this by making the following observation: a graph has the same number of incoming and outgoing edges at each node if and only if it can be decomposed into a sum of cycles. By ``sum" we mean that we take the union of nodes and the union of edges from the constituent cycles, while keeping the labels of the nodes intact (so that $1\ra 2\ra 3\ra 1$ is different from $1\ra 2\ra 4\ra 1$). An example of such a decomposition is shown in Figure~\ref{fig:decomposition}.\\

In order to build a complete list of quivers for a given $G$ and $E$, we must first consider all of the possible cycles over $G$ nodes. Then we take combinations of those cycles such that the total number of fields adds up to $E$. This way we have all of the quivers that satisfy the Calabi-Yau condition. This approach is significantly faster than the naive method, and is efficient enough to easily produce all of the quivers found in this paper using a modern computer.\\

\subsection{Finding Superpotentials}
\label{sec:superpotentials}
After finding the quivers, we must construct all possible quiver gauge theories. This is done by finding all of the superpotentials $W$ that could be associated with each quiver. By considering two important features these special quiver gauge theories must have, we can efficiently find all possible consistent superpotentials.\\

There are two useful constraints on the form of a quiver gauge theory's superpotential that we should consider. The first is that each term in $W$ has to be gauge-invariant. With the bi-fundamental (or adjoint) nature of the fields, this means that a field ``ending" on a group factor $g$ has to be contracted with a field ``starting" on $g$. Adopting the usual notation that a field $X_{ij}$ is fundamental under the group factor $i$ and anti-fundamental under $j$, a typical term will look like
\be
Tr(X_{12}X_{23}X_{31}),
\ee
where the trace gives the contraction between the last and the first fields. This condition has a nice interpretation in the quiver picture: gauge-invariant  terms are just \emph{cycles} in the quiver. From this observation, we can see that the cycles generated in the previous step of the algorithm will allow us to quickly generate all possible superpotentials.\\

The second constraint on the superpotential that we find useful is the toric condition \cite{ToricD}. It states that each field in the quiver gauge theory should appear in the superpotential exactly twice: once in a positive term and once in a negative term. The bipartite nature of the tiling is a manifestation of this toric condition. For every quiver, we take all of the cycles that make up the quiver and find all ways of combining them into superpotentials that satisfy this toric condition. However only a small fraction of these models can actually admit a tiling description, and for that we need the final step in the algorithm.\\

\subsection{Reconstructing Tilings}

The final step in the algorithm is to check for whether a given quiver gauge theory can correspond to a brane tiling and then to find this tiling.\\

The way we proceed is by using an object called a periodic quiver \cite{FrancoHanany}. It is the graph dual of the tiling, where nodes are groups, fields are edges and faces are superpotential terms. Since the data generated so far comprises of a list of quivers and superpotentials, the task of finding the tilings reduces to whether we can unfold the quivers into bi-periodic graphs of the plane. If we can find a periodic quiver from an ordinary quiver and a superpotential, then we know that the model admits a tiling description, and we can easily find its dual graph, the brane tiling.\\

The algorithm used to produce the tilings is as follows. We are given the quiver $Q$ and superpotential $W$. The idea is that we try to build up the fundamental domain of the periodic quiver. To do that, firstly, we represent each term in $W$ by a polygon with edges around its perimeter representing fields. We choose the fields have a clockwise orientation for positive terms and a counter-clockwise orientation for negative terms. These polygons will be the faces of the periodic quiver. \\

 Next, we fit these polygons together into one shape by gluing edges that represent the same field together. The process is always possible due to the toric condition on the superpotential. The shape generated is our candidate for the fundamental domain. The test this shape must pass is whether we can identify opposite edges in a way such that the shape forms a 2-torus. If we can do this we have found a periodic quiver and so a brane tiling.\\

\begin{figure}[h]
\centering
\includegraphics[width=3.5cm]{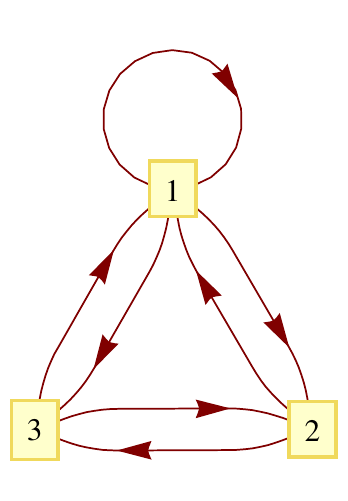} \\
\caption{$SPP$ quiver.}
\label{fig:spp-quiver}
\end{figure}

Let us illustrate this procedure with an example known as the suspended pinch point \cite{FrancoHanany}. The quiver is shown in Figure~\ref{fig:spp-quiver} and the superpotential is the following:
\be
W = \phi _1^{}.X_{12}^{}.X_{21}^{} -\phi _1^{}.X_{13}^{}.X_{31}^{} 
-X_{12}^{}.X_{23}^{}.X_{32}^{}.X_{21}^{} +X_{13}^{}.X_{32}^{}.X_{23}^{}.X_{31}^{}
\ee

\begin{figure}[h]
\centering
\begin{tabular}{c}
\includegraphics{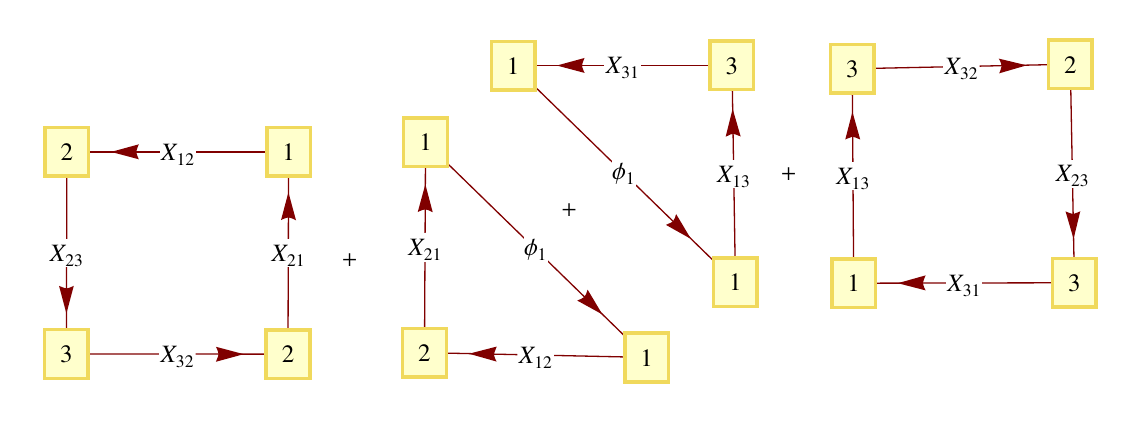} \\
$\downarrow$ \\
\includegraphics{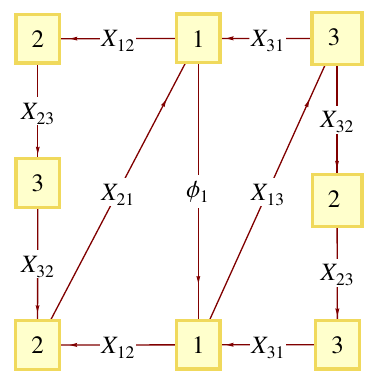}
\end{tabular}
\caption{Combining the superpotential terms into a fundamental domain of the periodic quiver.}
\label{fig:tiling1}
\end{figure}

There are four terms in the superpotential, which we represent by four polygons - two ``triangles" corresponding to the cubic terms and two ``squares" corresponding to the quartic terms (Figure~\ref{fig:tiling1}). Recall that the arrows around the faces go clockwise for positive and counter-clockwise for negative terms. We can now treat the problem just like a jigsaw puzzle: we have to put these pieces together allowing only edges corresponding to the same field to touch. If it is possible to fit these pieces together to form a 2-torus, we will have generated a graph that can be flattened out to form a periodic quiver.\\

Let us consider the SPP model and glue the four terms together into one shape, by identifying the three fields $X_{21}$, $\phi_1$ and $X_{13}$. This shape is our candidate for the fundamental domain. It is unimportant as to which three fields we pick to glue together; a different choice will just result in generating a different fundamental domain of the periodic quiver. Next we attempt to deform the shape into a rectangle that can be used to tile the plane. If this is possible we have found the model's periodic quiver\footnote{
In some more complicated cases, it is possible to generate a shape that has a pair of identical fields adjacent to each other. We simply glue together all of these repeating edges, until we have a shape with no such repeated edges. We then test whether this shape can be used to tile the plane.}.\\

We can see in Figure~\ref{fig:tiling1} that it is possible to find a periodic quiver for the SPP. By glancing at the rectangle, we can see that it is possible to use it to tile the plane with only edges corresponding to identical fields touching. We can equivalently see that the shape generated is really a 2-torus. The top and bottom sides of the rectangle can be identified directly along $(X_{12}, X_{31})$, effectively turning the rectangle into a cylinder. Then the ends of the cylinder each consist of $(X_{32}, X_{23})$, and even though they are not exactly the same on the rectangle, the cylinder can be ``twisted" so that the ends are correctly identified.\\

A key part of the algorithm is this important check for whether the resulting fundamental domain can be wrapped to make a torus. A given quiver gauge theory admits a tiling description if and only if that is possible. A simple shape that fails this check is one that has fields $(\phi_1, \phi_1, \phi_2, \phi_2)$ forming the perimeter of a rectangle.\\

\begin{figure}[h]
\begin{center}
\begin{tabular}{ccc}
\begin{minipage}{5cm}
\includegraphics[width=5cm]{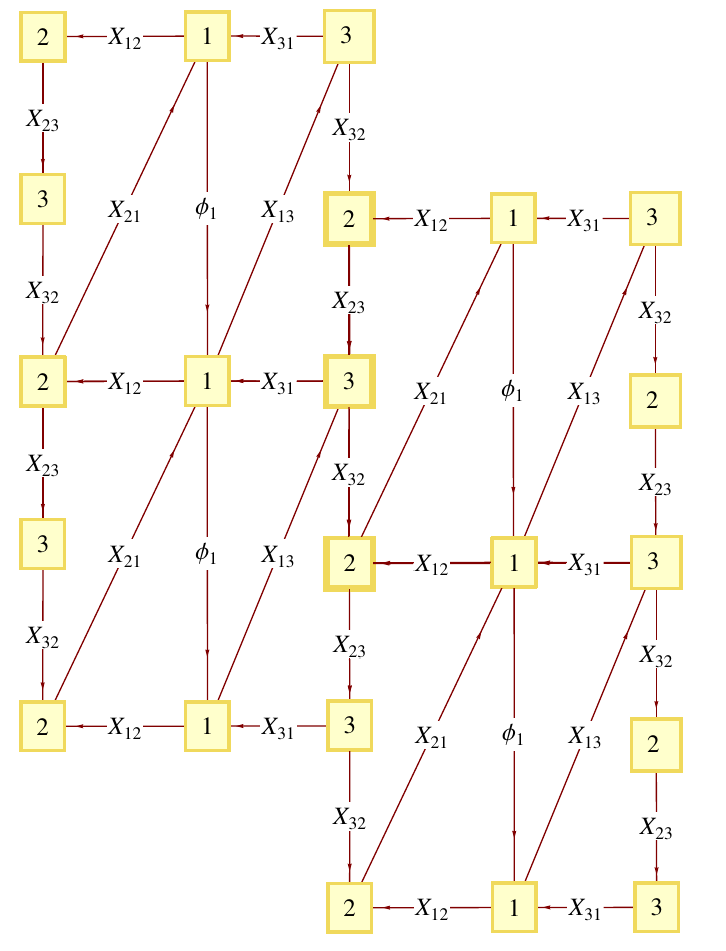}
\end{minipage} 
&  $\rightarrow$ &
\begin{minipage}{5cm}
\includegraphics[width=5cm]{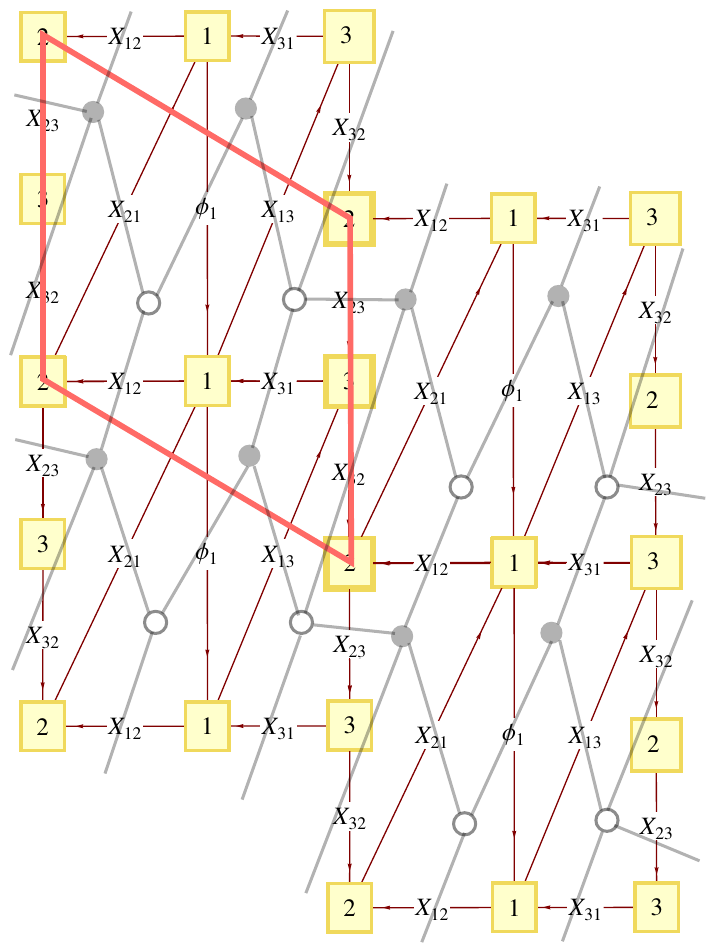}
\end{minipage} 
\end{tabular}
\end{center}
\caption{From periodic quiver to brane tiling for $SPP$.}
\label{fig:tiling2}
\end{figure}

If the construction of a periodic quiver works, we can easily extract the brane tiling from it by finding the dual graph. Firstly, we draw the periodic quiver with our ``fundamental rectangle". Then we insert a white or black node at the center of each face according to whether the arrows go clockwise or counter-clockwise around the perimeter of the face. By replacing edges as in Figure~\ref{fig:tiling2} we build the dual graph (the brane tiling). In the case of the SPP, we see that the tiling consists of two hexagons with one of them divided by a diagonal.\\

The reader should note that while the algorithm generates a complete list of tilings, it fails to produce aesthetically pleasing pictures. In order to display the tiling in terms of regular geometrical shapes, such as hexagons, squares or octagons we must either find the pattern by inspection, or to rely on existing algorithms that display large planar graphs neatly. The results shown later in this paper were generated by a combination of these methods.\\


\section{A Model Overview}

We have used an implementation of the algorithm described in this paper to generate all irreducible tilings that have at most 6 superpotential terms. Despite the inefficiency of our implementation, the computation only took a couple of hours on an ordinary desktop computer. In this section we will briefly discuss the models found. A full list of all of the tilings generated is given in Appendix~\ref{sec:catalog}.\\

\begin{figure}[h]
\begin{center}
\begin{tabular}{cc}
\begin{tabular}[b]{c}
\includegraphics[height=3.5cm]{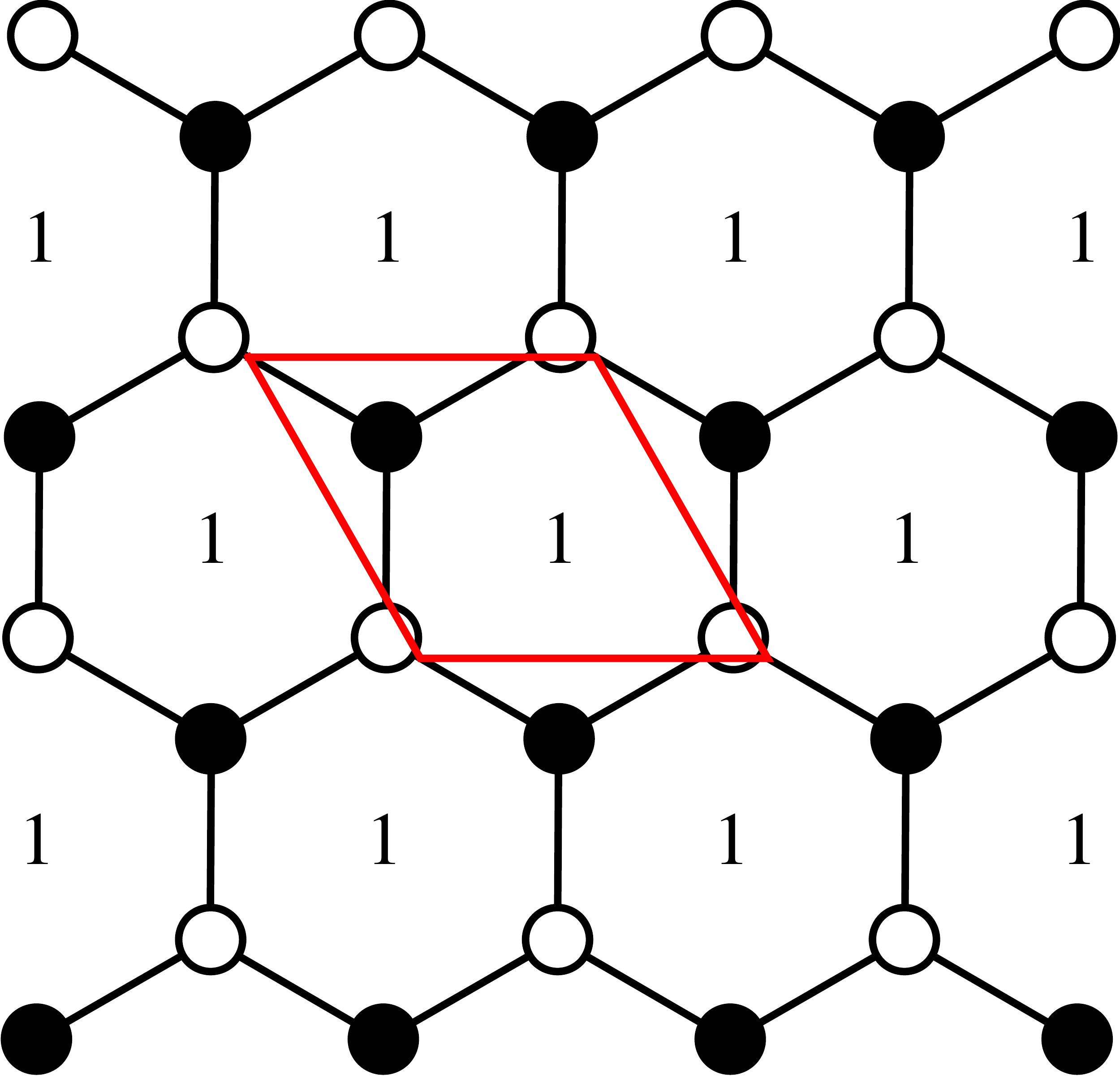} \\
(1.1) $\BC^3$
\end{tabular}
&
\begin{tabular}[b]{c}
\includegraphics[height=3.5cm]{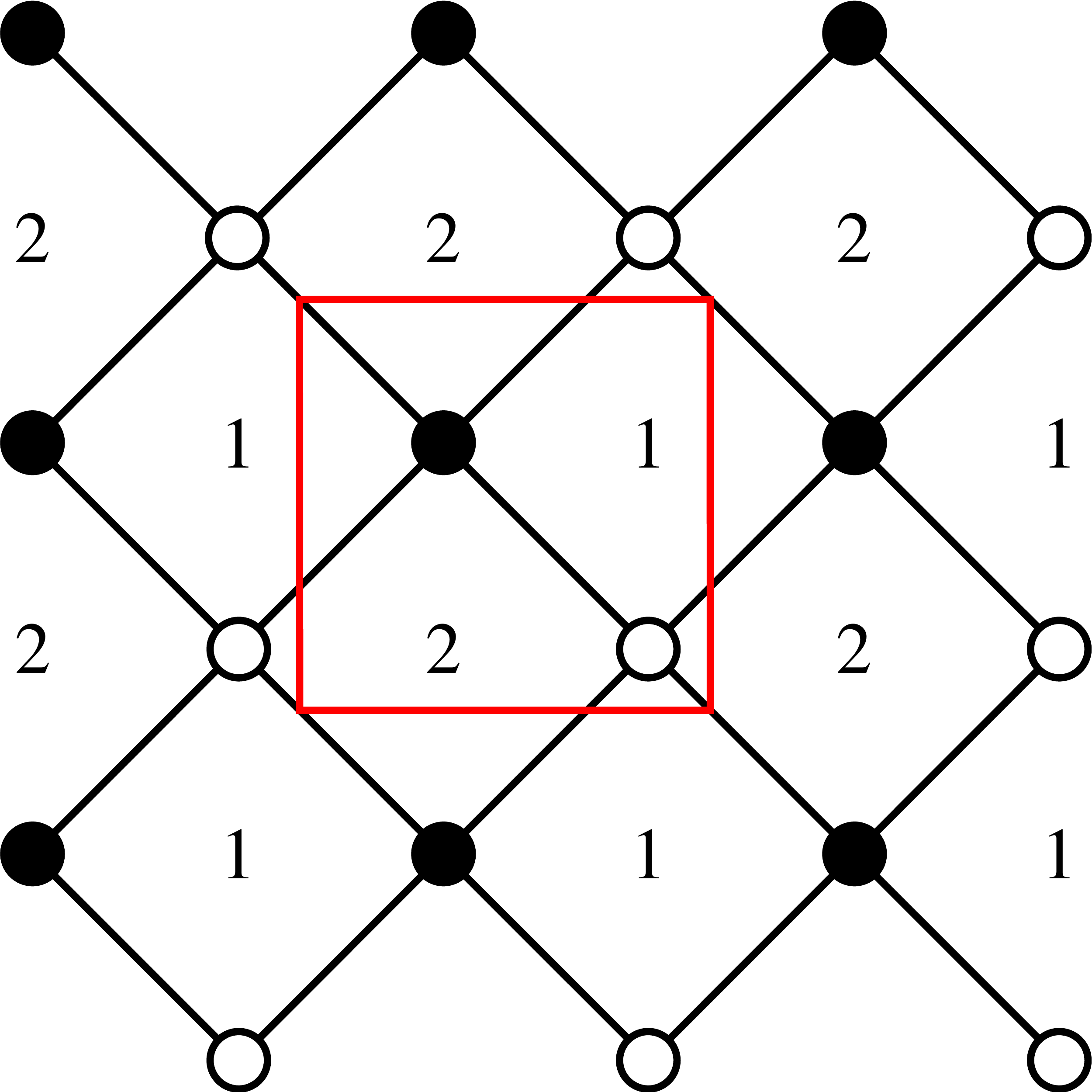} \\
(1.2) $\cal{C}$
\end{tabular}
\end{tabular}
\end{center}
\caption{Consistent tilings with two superpotential terms.}
\label{fig:tilings2}
\end{figure}

We start by considering the case of just two terms in the superpotential. Here we only have a choice of one or two gauge groups, and we find one possible tiling for each case. These are the most familiar models: the $\BC^3$ with the one-hexagon tiling and the conifold $\CC$ with the two-square tiling (see Figure~\ref{fig:tilings2}). Both of them are consistent tilings \cite{HananyKennaway}. Also the conifold tiling has proved useful in studying the ABJM theory on M2 branes \cite{TilingM2}.\\

\begin{figure}[h]
\begin{center}
\begin{tabular}{ccc}
\begin{tabular}[b]{c}
\includegraphics[width=3.5cm]{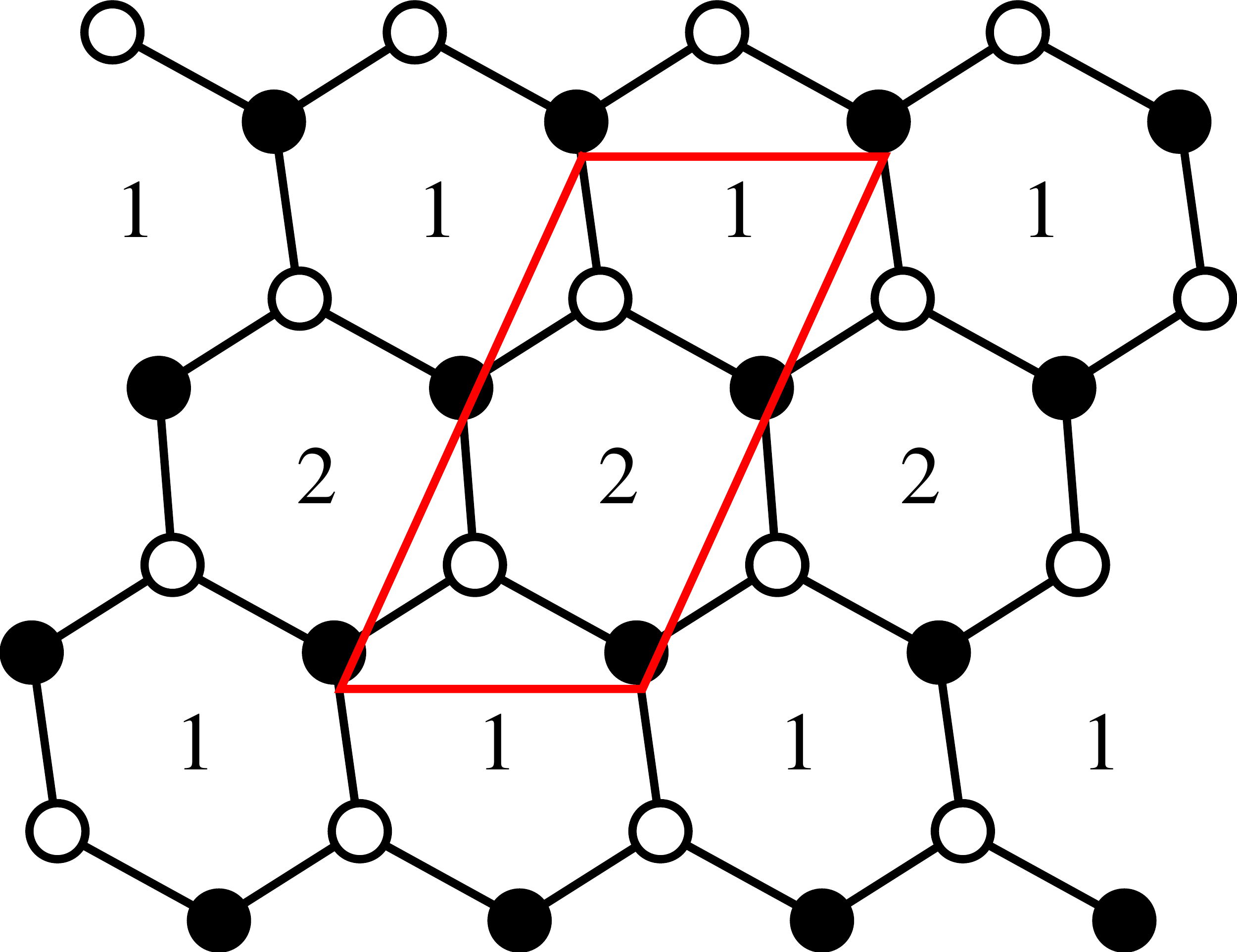} \\
(2.1) $\BC^2/\BZ_2 \times \BC$
\end{tabular}
&
\begin{tabular}[b]{c}
\includegraphics[width=3.5cm]{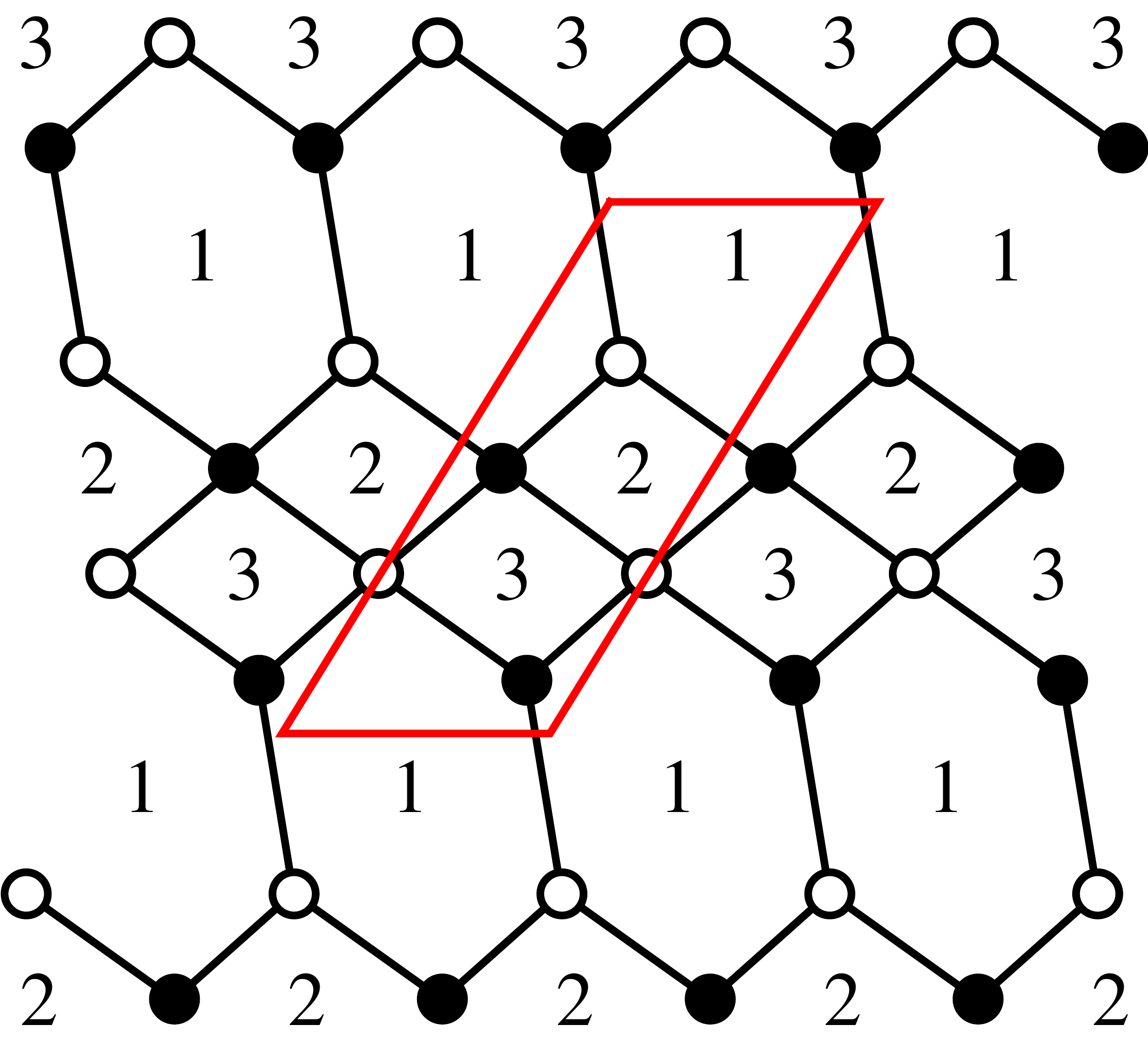} \\
(2.2) $SPP$
\end{tabular}
&
\begin{tabular}[b]{c}
\includegraphics[width=3.5cm]{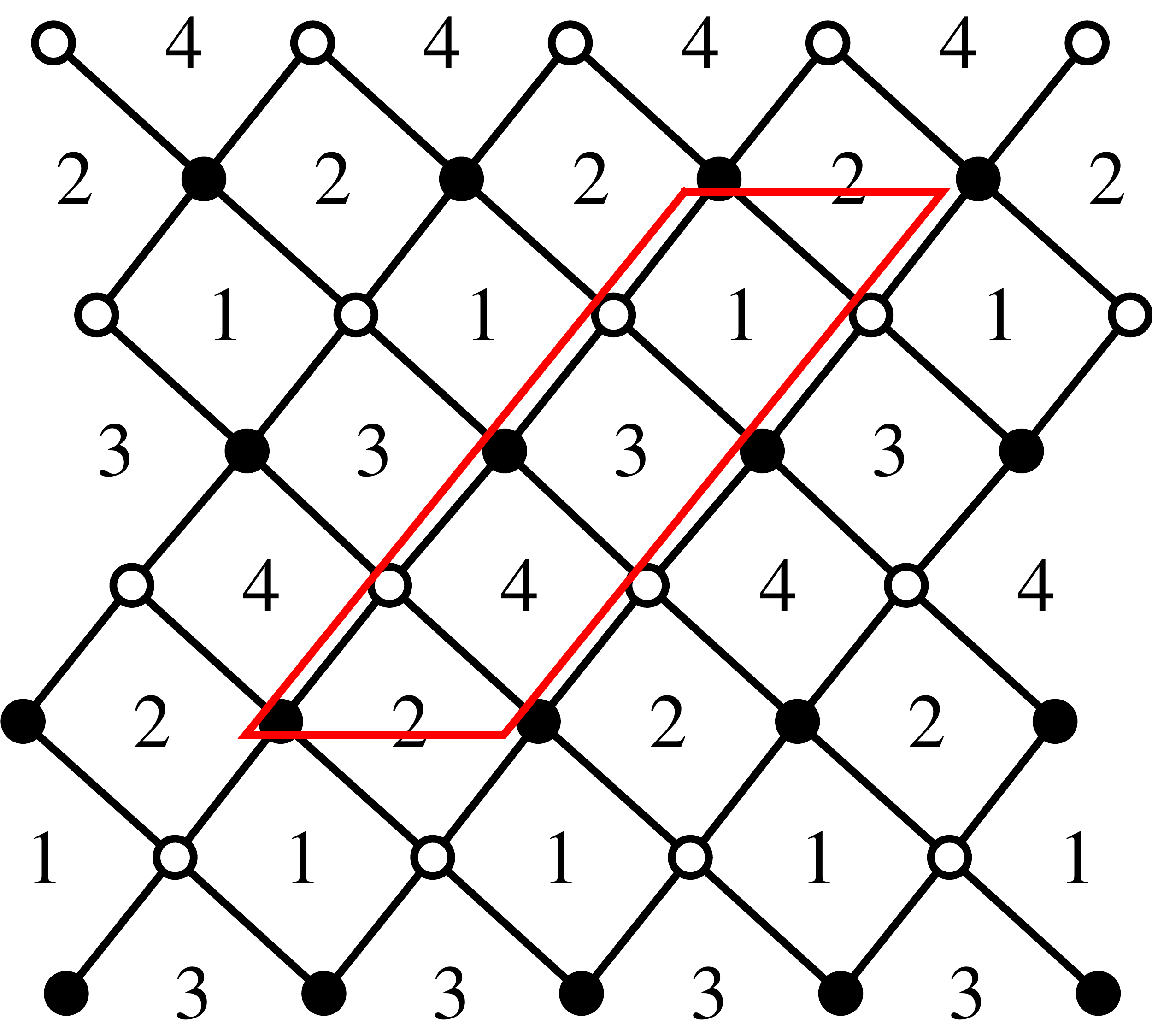} \\
(2.4) $L^{222}$ (I)
\end{tabular}
\\
&
\begin{tabular}[b]{c}
\includegraphics[width=3.5cm]{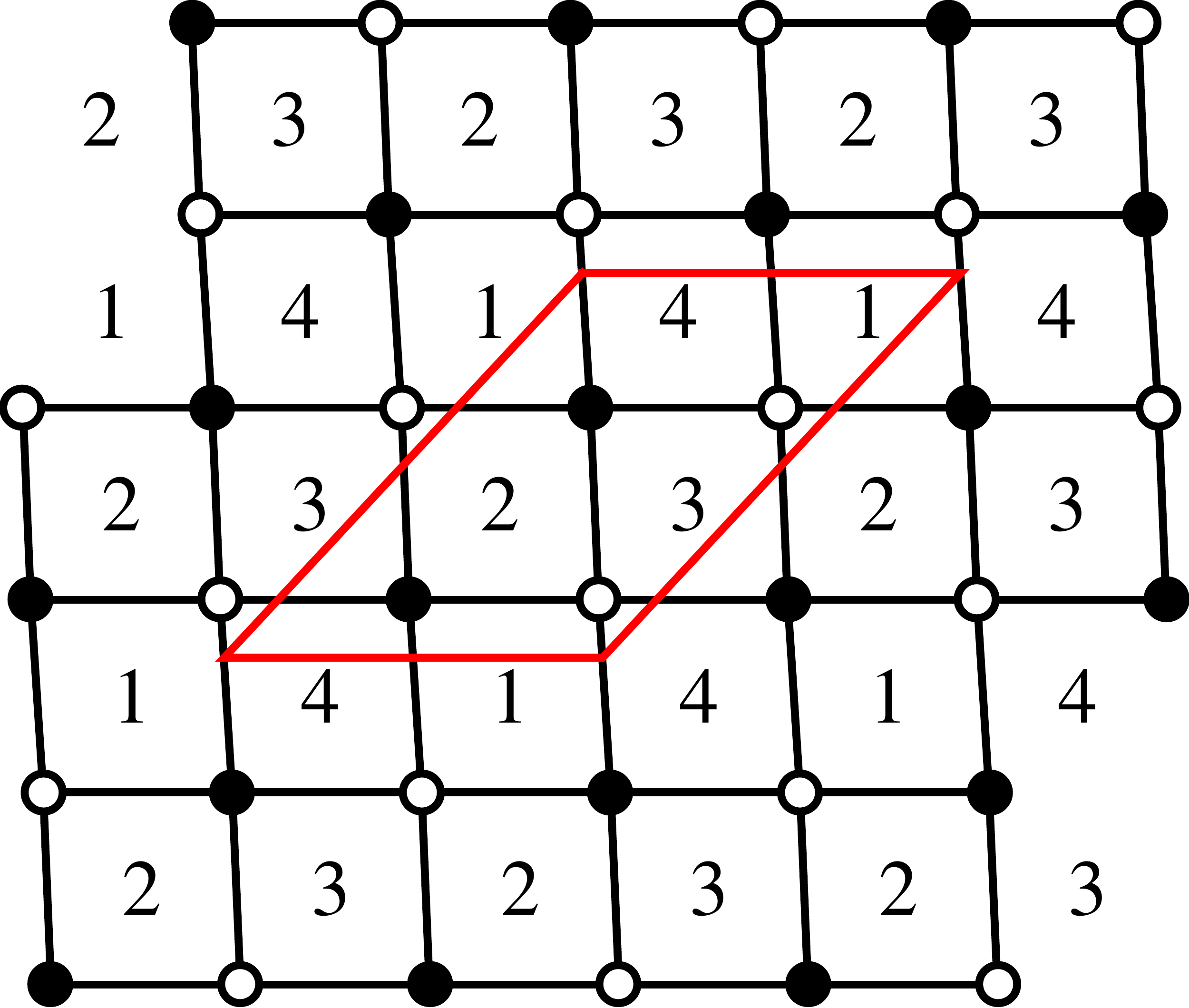} \\
(2.5) $\BF_0$ (I)
\end{tabular}
\end{tabular}
\end{center}
\caption{Consistent tilings with four superpotential terms.}
\label{fig:tilings4}
\end{figure}

Let us now consider the 6 tilings generated with with four superpotential terms. With the minimal possibility of two gauge groups and six fields we find the two-hexagon model, or $\BC^2 / \BZ_2 \times \BC$ \cite{HananyKennaway}. Among the models with three and four groups we have the $SPP$, Phase I of $\BF_0$ and Phase I of $L^{222}$ (Figure~\ref{fig:tilings4}). We also find the two models labeled as (2.3) and (2.6), which are inconsistent in 3+1 dimensions (Appendix~\ref{sec:catalog}).\\

Another elegant way of generating all of the models with four superpotential terms comes from considering the hexagon as the fundamental unit of a tiling. Let us start with the two-hexagon tiling. Adding new edges to a tiling keeps the number of superpotential terms the same but increases the number of gauge groups. We can find all tilings with 4 superpotential terms by adding edges across faces of the two-hexagon model. We find that there are two inequivalent ways of adding one diagonal to one of the hexagons, which give the models with three gauge groups (2.2) and (2.3) (see Appendix~\ref{sec:catalog}). If we add a 2nd diagonal to the tilings we find the remaining three tilings with four gauge groups. This procedure of finding the tilings by adding diagonals also works for the case with two superpotential terms. We start with the basic one-hexagon tiling and find the conifold model by adding one diagonal.\\

\begin{figure}[h]

\begin{center}
\begin{tabular}{ccc}

\begin{tabular}[b]{c}
\includegraphics*[height=3.5cm]{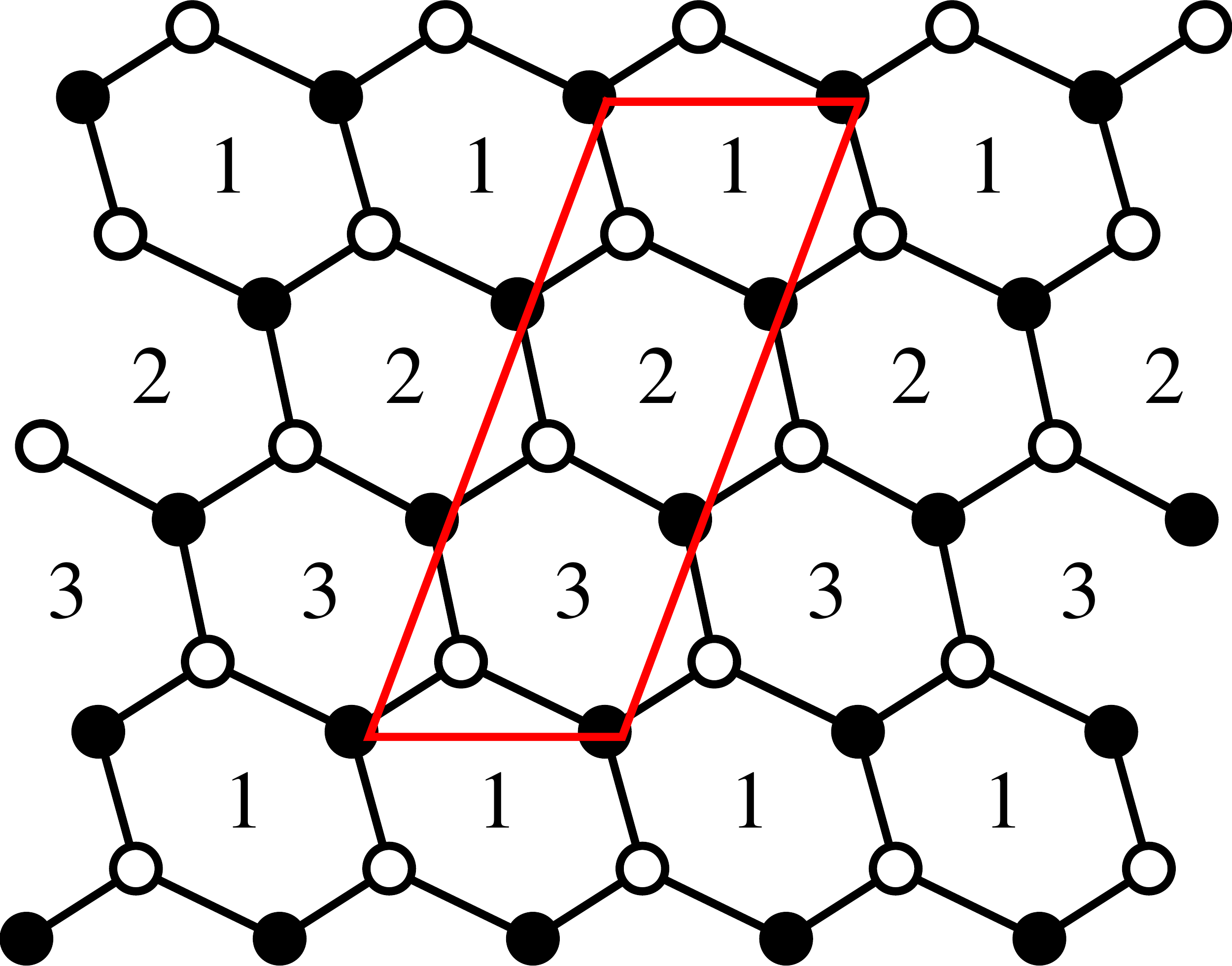} \\
(3.1) $\BC^2/\BZ_3 \times \BC$
\end{tabular}
&
\begin{tabular}[b]{c}
\includegraphics*[height=3.5cm]{N6-G3-1-tiling.pdf} \\
(3.2) $\BC^3 / \BZ_3 $
\end{tabular}
&

\\

\begin{tabular}[b]{c}
\includegraphics*[height=3.5cm]{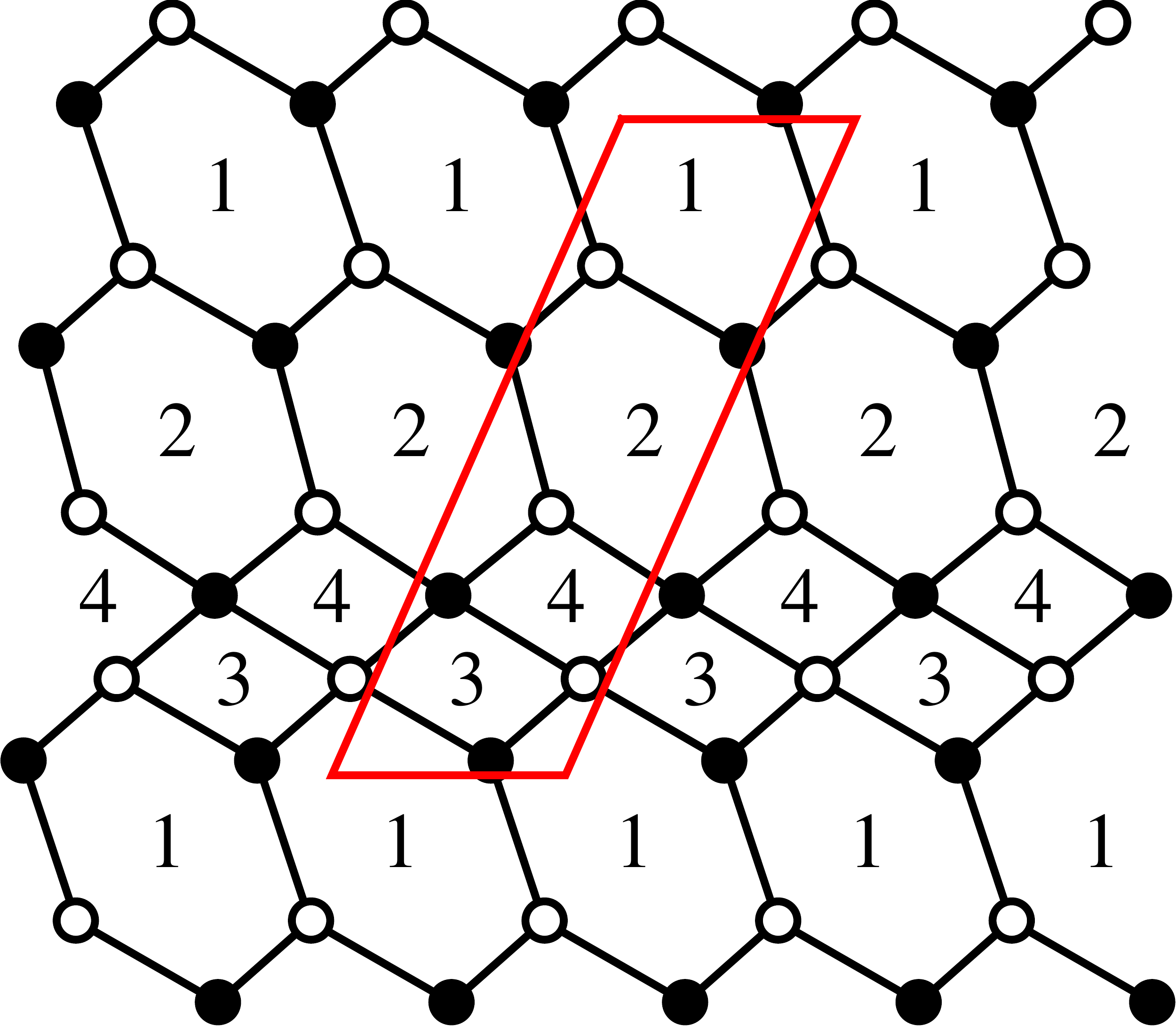} \\
(3.4) $ L^{131} $
\end{tabular}
&
\begin{tabular}[b]{c}
\includegraphics*[height=3.5cm]{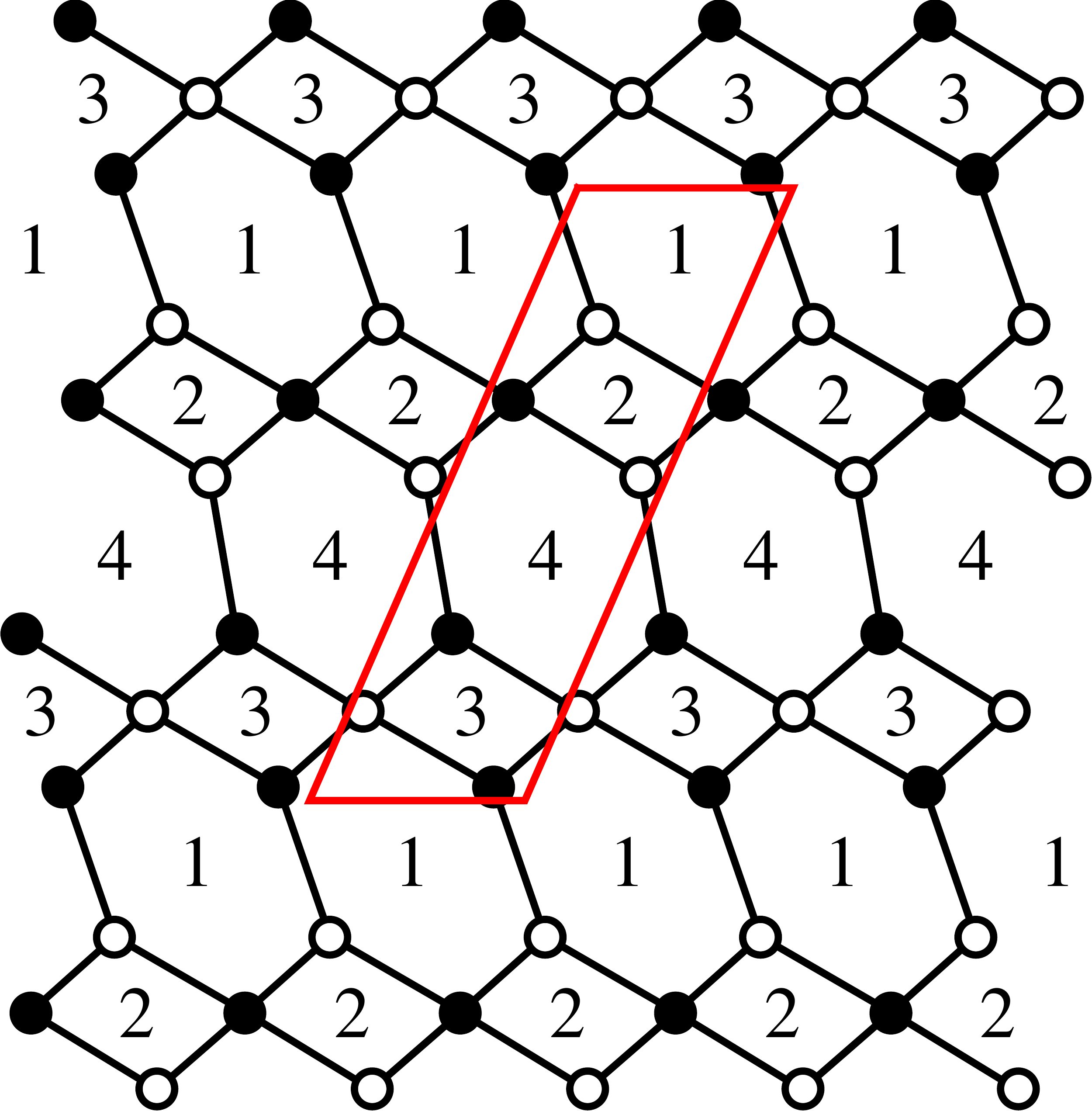} \\
(3.5) $ L^{222} $ (II)
\end{tabular}
&
\begin{tabular}[b]{c}
\includegraphics*[height=3.5cm]{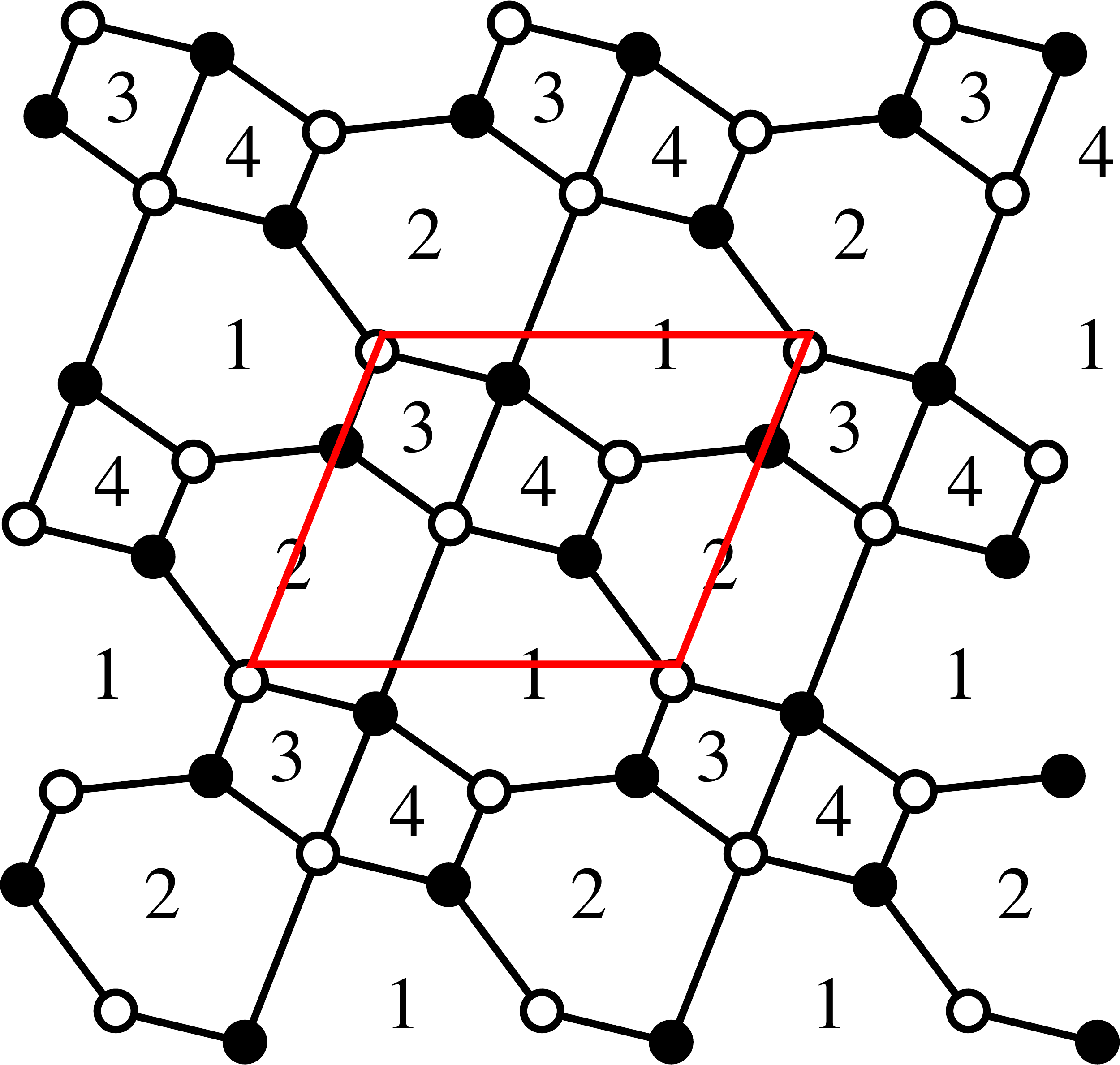} \\
(3.6) $ dP_1 $
\end{tabular}

\\

\begin{tabular}[b]{c}
\includegraphics*[height=3.5cm]{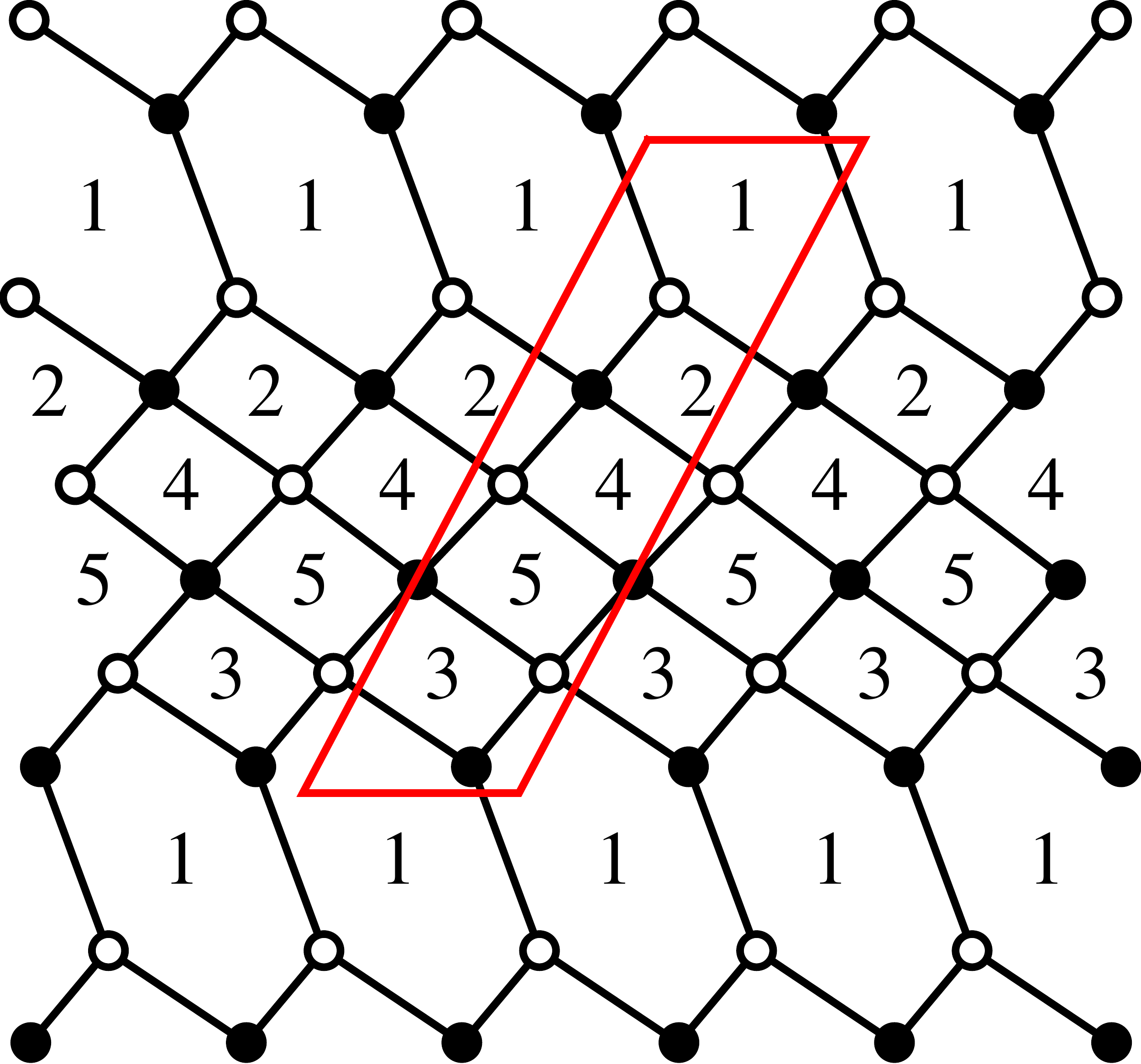} \\
(3.13) $ L^{232} $ (I)
\end{tabular}
&
\begin{tabular}[b]{c}
\includegraphics*[height=3.5cm]{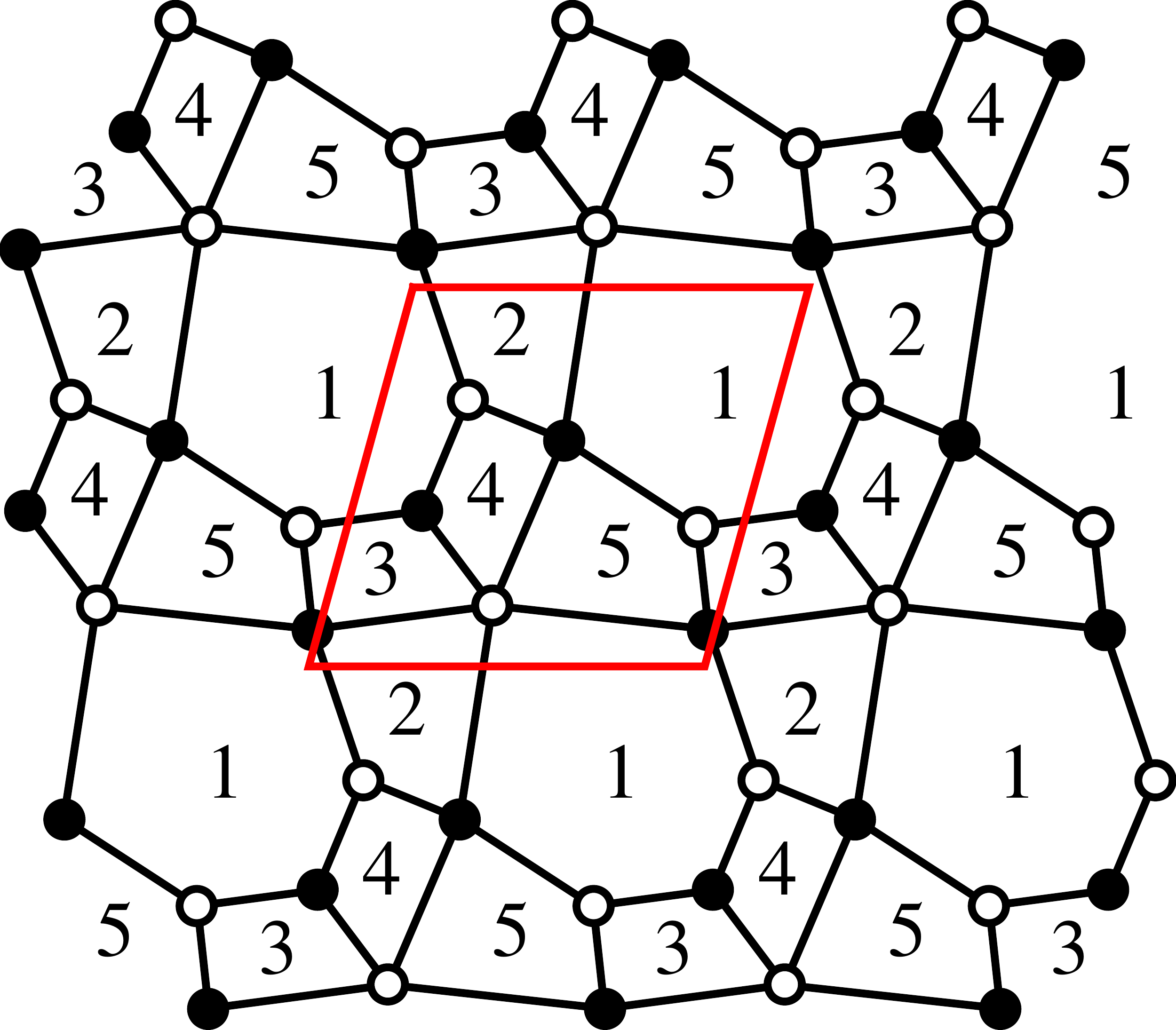} \\
(3.14) $ dP_2 $ (I)
\end{tabular}
&

\\

\begin{tabular}[b]{c}
\includegraphics*[height=3.5cm]{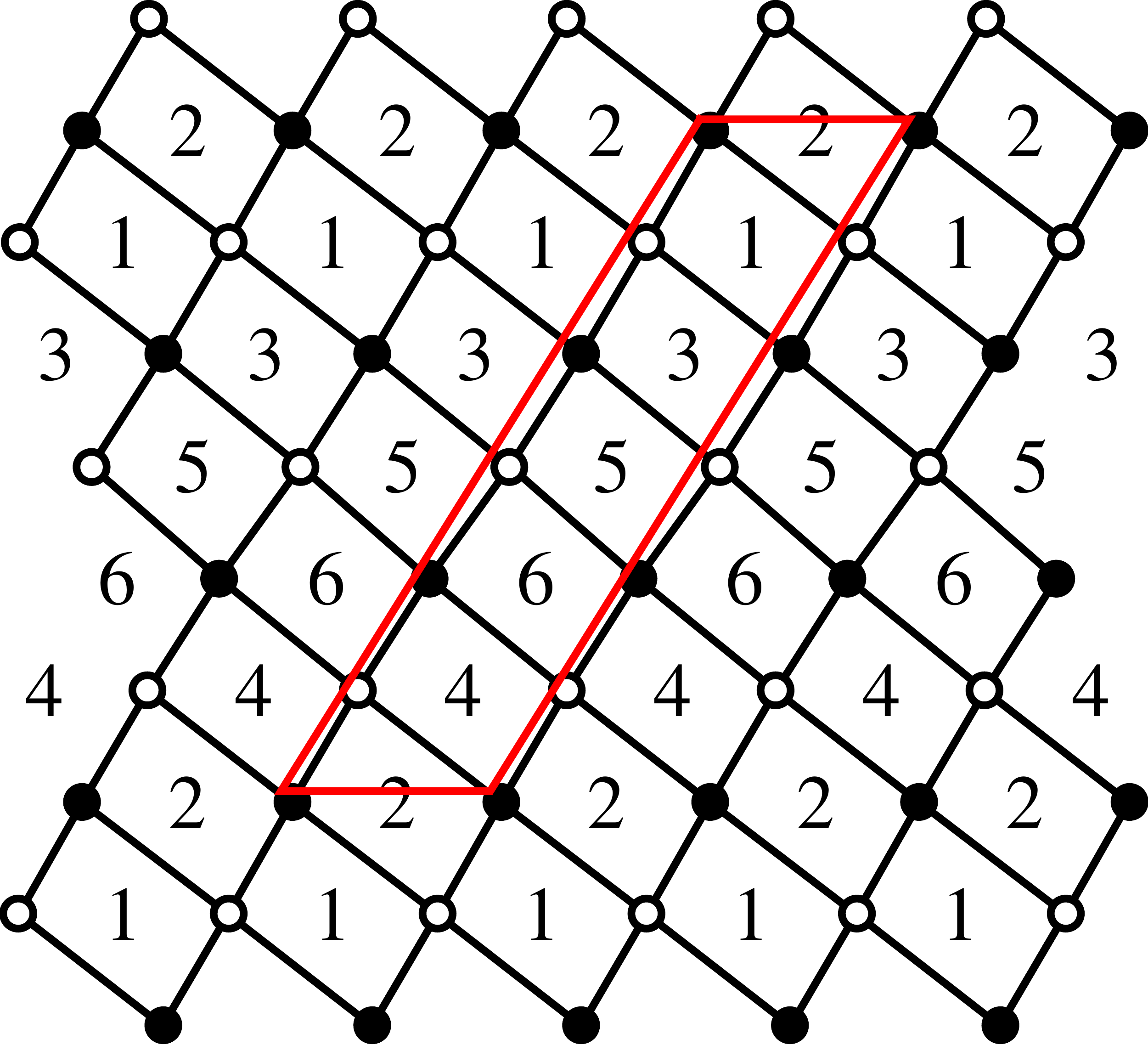} \\
(3.26)$ L^{333} $ (I)
\end{tabular}
&
\begin{tabular}[b]{c}
\includegraphics*[height=3.5cm]{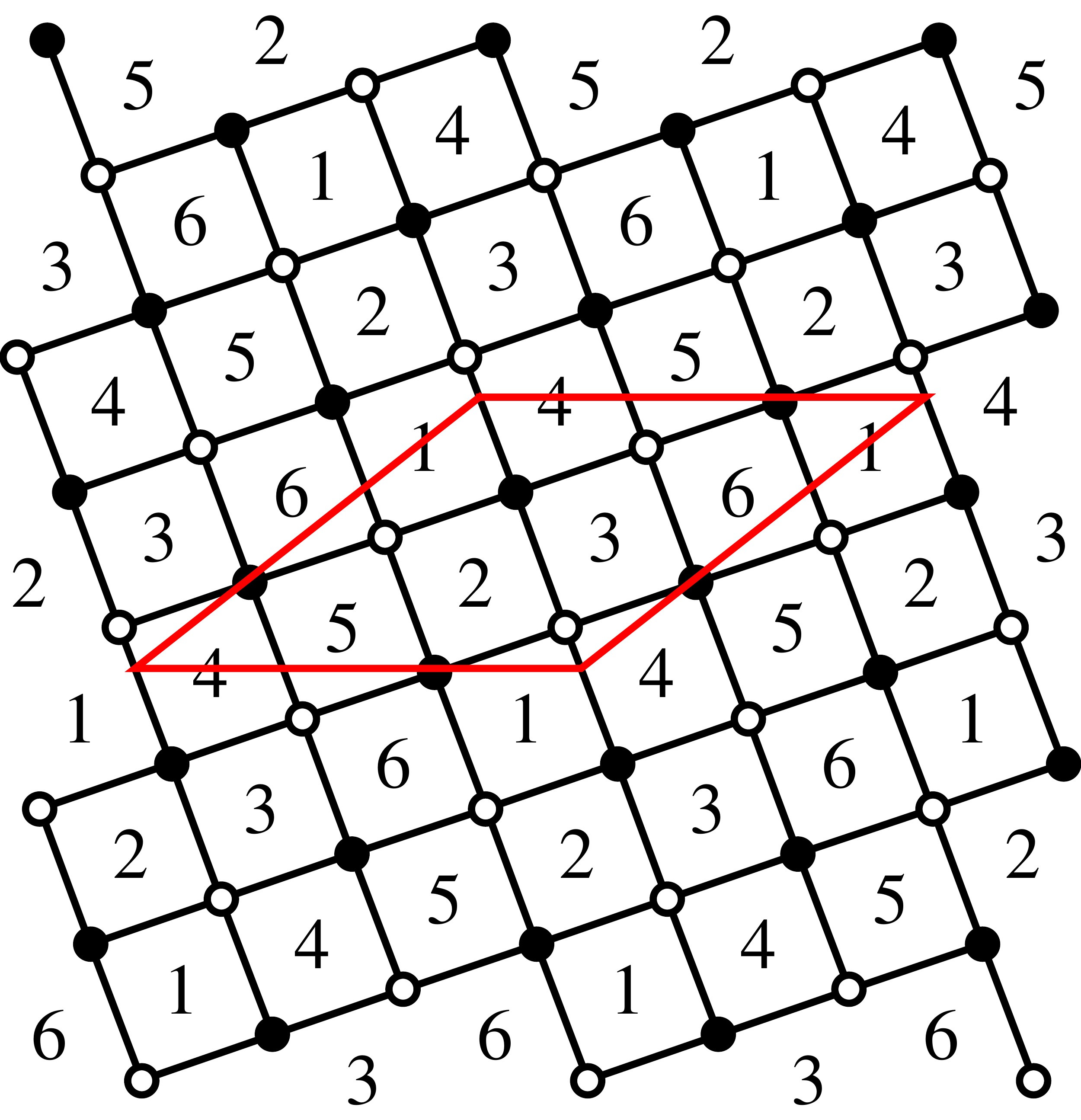} \\
(3.27) $ Y^{3,0} $ (I)
\end{tabular}
&
\begin{tabular}[b]{c}
\includegraphics*[height=3.5cm]{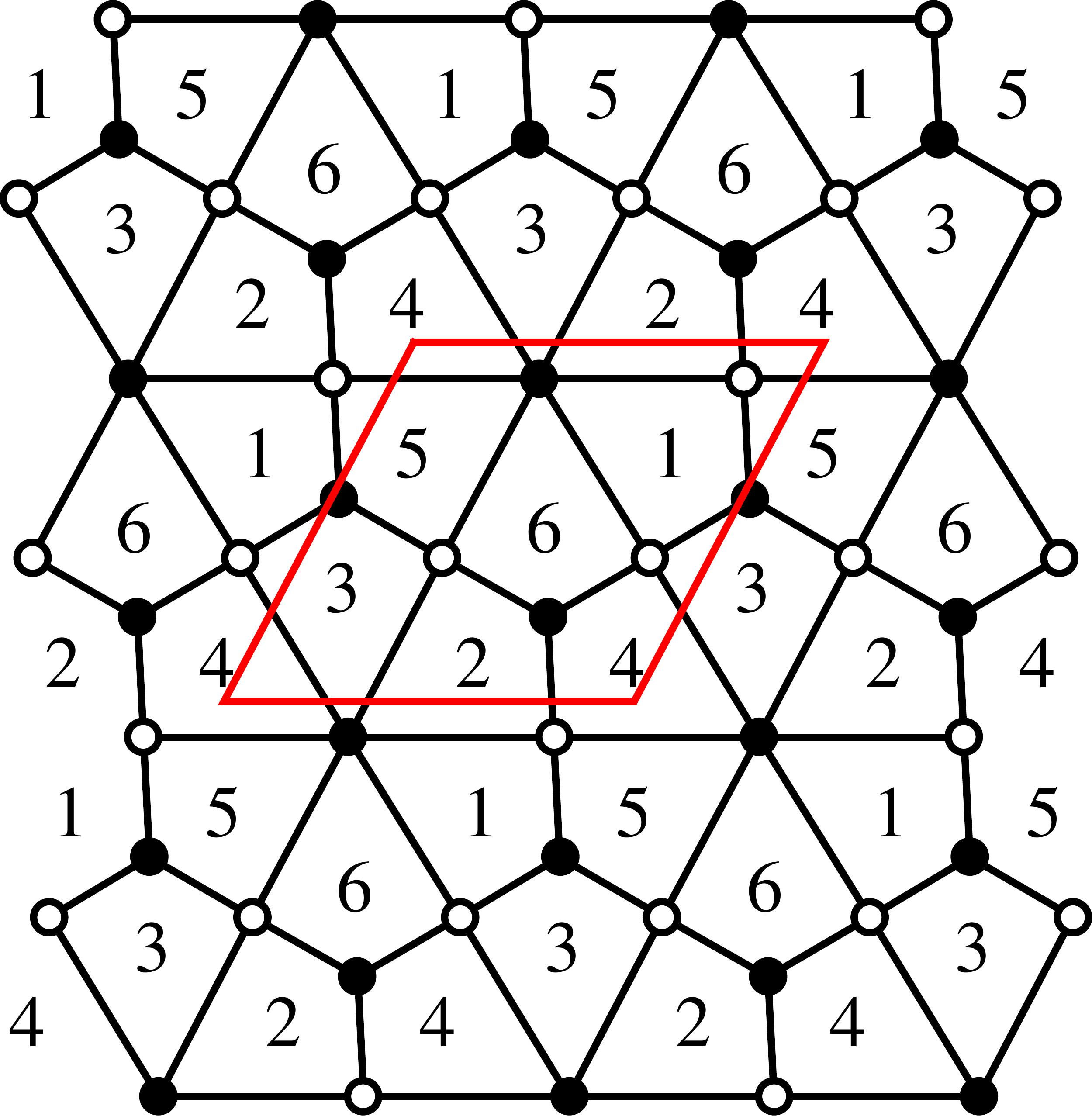} \\
(3.28) $ dP_3 $ (I)
\end{tabular}

\end{tabular}
\end{center}

\caption{Consistent tilings with six superpotential terms.}
\label{fig:tilings6}
\end{figure}

Let us now consider the models with six terms in the superpotential. Our algorithm generates a total of 37 different tilings, each having three to six gauge groups. Of these tilings, only 10 models give rise to consistent tilings. We find that all of the consistent tilings are either phases of $L^{aba}$ or $Y^{p,q}$ families, or one of the del-Pezzo surfaces. Specifically, we find the models $dP_0$ (or $\BC^3 / \BZ_3$), $dP_1$, $dP_2$, $dP_3$, $L^{030}$ (or $\BC^2 / \BZ_3 \times \BC$), $L^{131}$, another phase of $L^{222}$, $L^{232}$, $L^{333}$ and $Y^{3,0}$ (see Figure~\ref{fig:tilings6}). The other models are not as familiar, because they fail the usual 3+1 dimensional consistency condition.\\

We may wish to gain further insight by considering the procedure of adding diagonals to a 3 hexagon tiling, as we did for the 2 hexagon tiling. Unfortunately we are unable to generate all tilings with 6 superpotential terms using this method as there is a model that has an octagon. However, we may achieve the result if we start with some larger ``base figure" than a hexagon. We can consider a model with six superpotential terms, but only \emph{one} gauge group in the quiver and seven adjoint fields. Such a model would inevitably have quadratic terms in the superpotential, and so is not included in this classification.\\

By considering all possible base figures and by adding diagonals to these base figures in all possible ways, it should be possible to generate all of the tilings in this paper in a matter of minutes rather than hours as this is a computationally a lot easier. This idea is a possible direction for further work as it could help us to find more complex tilings without the need for greater computational power.\\

\section{Acknowledgements}

J.D. would like to thank Yang-Hui He, Noppadol Mekareeya and Giuseppe Torri for many enlightening discussions, the STFC for his studentship and also Jenny Forrester, Alexander Summers, Thomas Sutherland, David Weir, the Ettore Majorana Foundation and Centre for Scientific Culture and the Swiss Federal Institute of Technology for hospitality during the Summer of 2009.
A.H. would like to thank the kind hospitality of the KITP in Santa Barbara, the Galileo Galilei Institute for Theoretical Physics, the INFN, the Benasque Center for Theoretical Physics, the Institute for Advanced Study in Princeton, and the Simons Center for Geometry and Physics during the various stages of this work. This research was supported in part by the National Science Foundation under Grant No. PHY05-51164.\\

\ \clearpage

\appendix
\section{Appendix: Tiling catalog}
\label{sec:catalog}

In this section we give the details of all of the tilings found from an implementation of the algorithm discussed in this paper. For each tiling we provide the quiver with the superpotential, the brane tiling and the corresponding toric diagram of the D3 brane theory. The conventional name of the model is included next to the toric diagram, and the ``(inc.)" suffix indicates that the associated D3 brane theory is inconsistent. For models with a toric diagram which has more than one known phase we also include the phase number in parenthesis.\\

In each table for given order parameters we give the consistent models first, and then list the remaining tilings according to the area of their associated toric diagram.\\


\subsection{Two superpotential terms}


\begin{table}[ht]

\begin{center}
\begin{tabular}{c|c|c|c|c}
\# & Quiver & Tiling & Toric Diagram & Superpotential\\
\hline \hline

(1.1) &
\includegraphics[width=2.5cm]{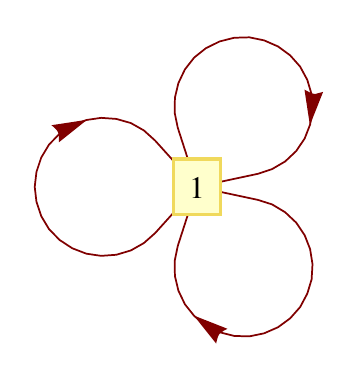} &
\includegraphics[height=3.5cm]{N2-G1-1-tiling.pdf} 
&
\begin{tabular}[b]{c}
\includegraphics[width=2.0cm]{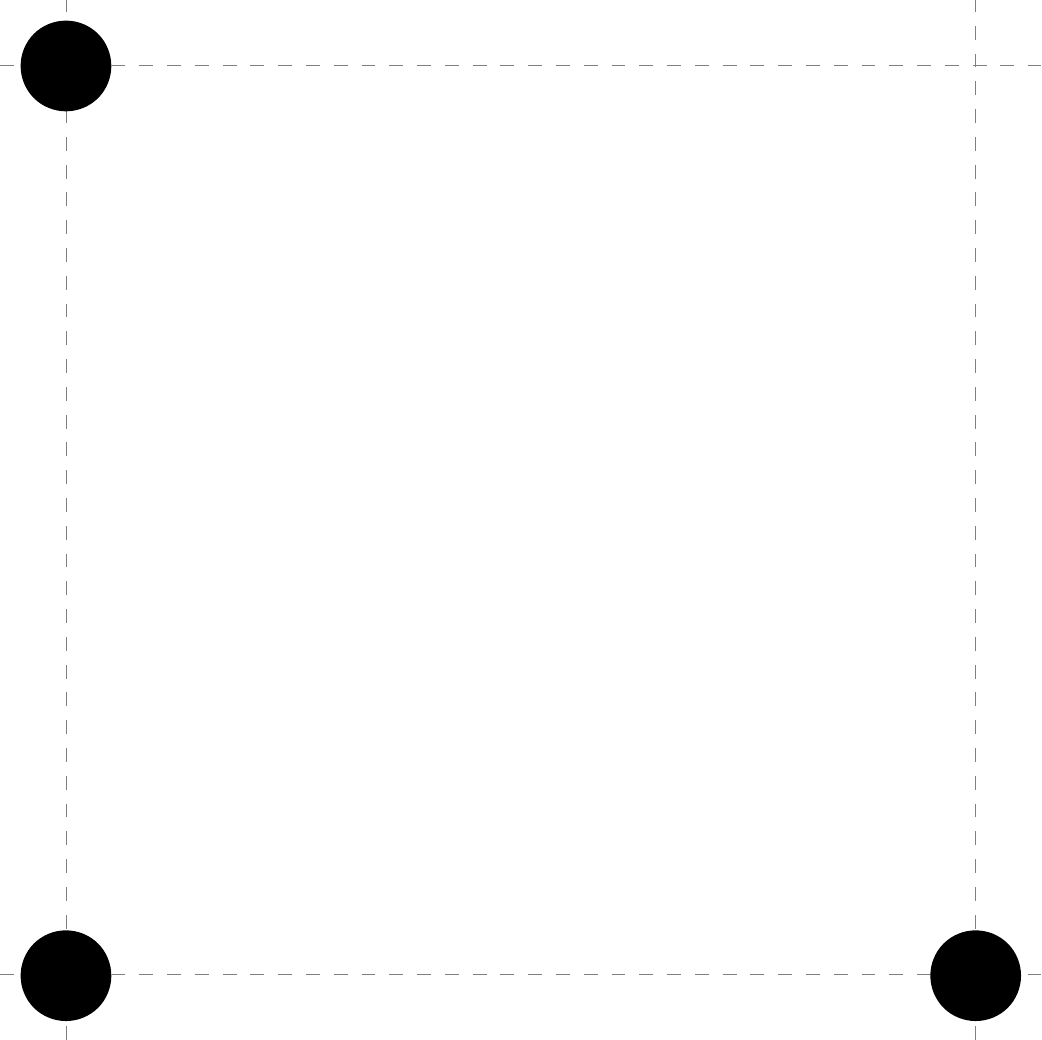} \\
$\BC^3$
\end{tabular}
&
\begin{tabular}[b]{c} 
$\phi _1^1.\phi _1^2.\phi _1^3$ \\
$-\phi _1^1.\phi _1^3.\phi _1^2$ 
\end{tabular}
\\ \hline

(1.2) &
\includegraphics[width=2.5cm]{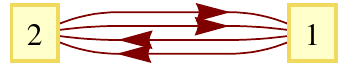} &
\includegraphics[height=3.5cm]{N2-G2-2-tiling.pdf} 
&
\begin{tabular}[b]{c}
\includegraphics[width=2.0cm]{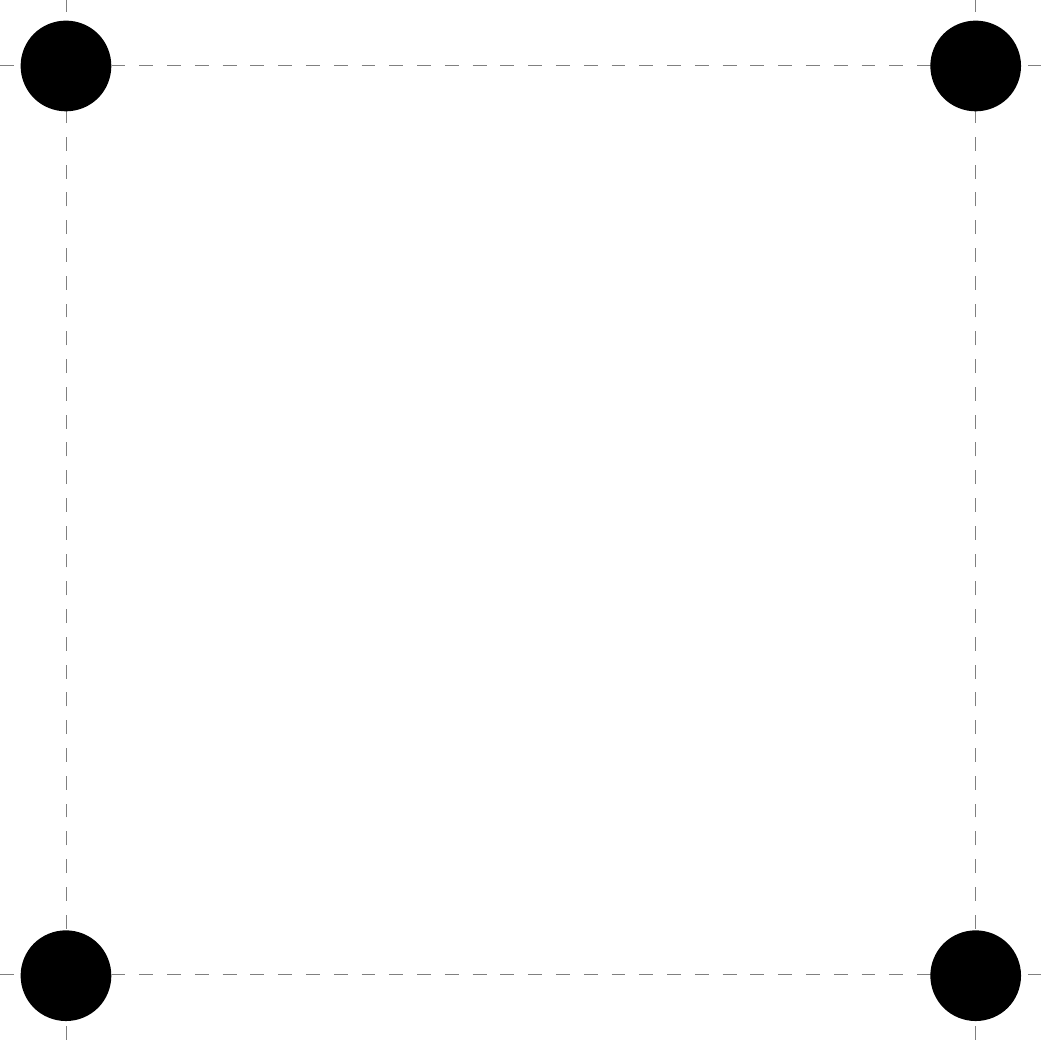}  \\
$\cal{C}$
\end{tabular}
&
\begin{tabular}[b]{c} 
$X_{12}^1.X_{21}^1.X_{12}^2.X_{21}^2$\\
$-X_{12}^1.X_{21}^2.X_{12}^2.X_{21}^1$
\end{tabular}

\end{tabular}
\end{center}

\caption{Tilings with 2 superpotential terms}
\label{t:tilings2}
\end{table}

\ \clearpage


\subsection{Four superpotential terms}


\begin{table}[h]

\begin{center}

\begin{tabular}{c|c|c|c|c}
\# & Quiver & Tiling & Toric Diagram & Superpotential\\
\hline \hline

(2.1) &
\includegraphics[width=2.5cm]{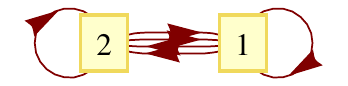} &
\includegraphics[height=3.5cm]{N4-G2-1-tiling.pdf} &
\begin{tabular}[b]{c}
\includegraphics[height=2.4cm,angle=90]{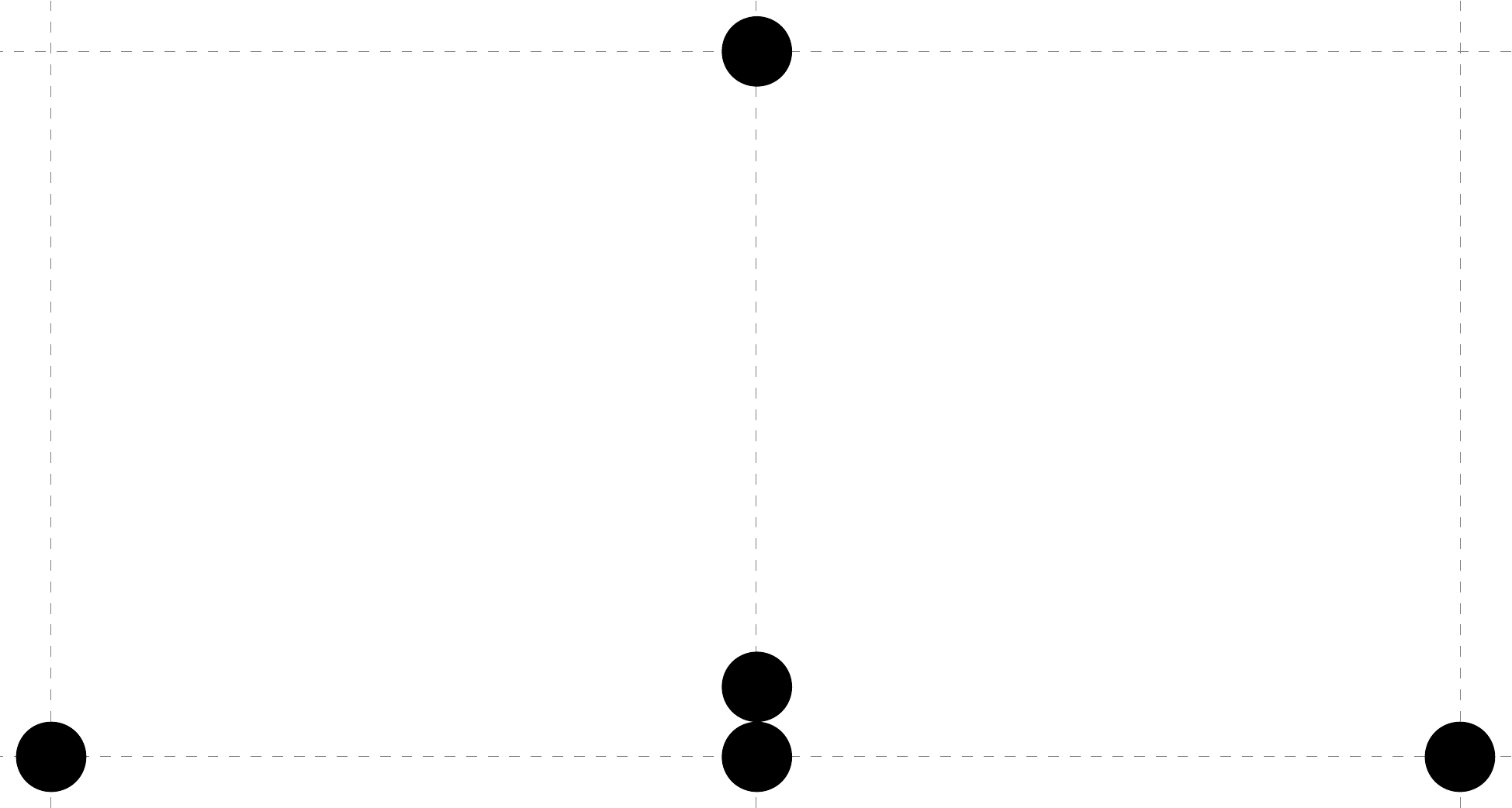}  \\
$\BC^2/\BZ_2 \times \BC$
\end{tabular}
&
\begin{tabular}[b]{c} 
$-X_{12}^1.\phi _2^{}.X_{21}^1$ \\
$+X_{12}^2.\phi _2^{}.X_{21}^2$ \\
$+\phi _1^{}.X_{12}^1.X_{21}^1$ \\
$-\phi _1^{}.X_{12}^2.X_{21}^2$ 
\end{tabular}

\label{Tilings with 4 superpotential terms and 2 gauge groups}
\end{tabular}

\end{center}

\caption{Tilings with 4 superpotential terms and 2 gauge groups}
\label{t:tilings4-2}
\end{table}


\begin{table}[h]

\begin{center}
\begin{tabular}{c|c|c|c|c}
\# & Quiver & Tiling & Toric Diagram & Superpotential\\
\hline \hline

(2.2) &
\includegraphics[width=2.5cm]{N4-G3-3-quiver.pdf} &
\includegraphics[height=3.5cm]{N4-G3-3-tiling.pdf} &
\begin{tabular}[b]{c}
\includegraphics[height=2.4cm,angle=90]{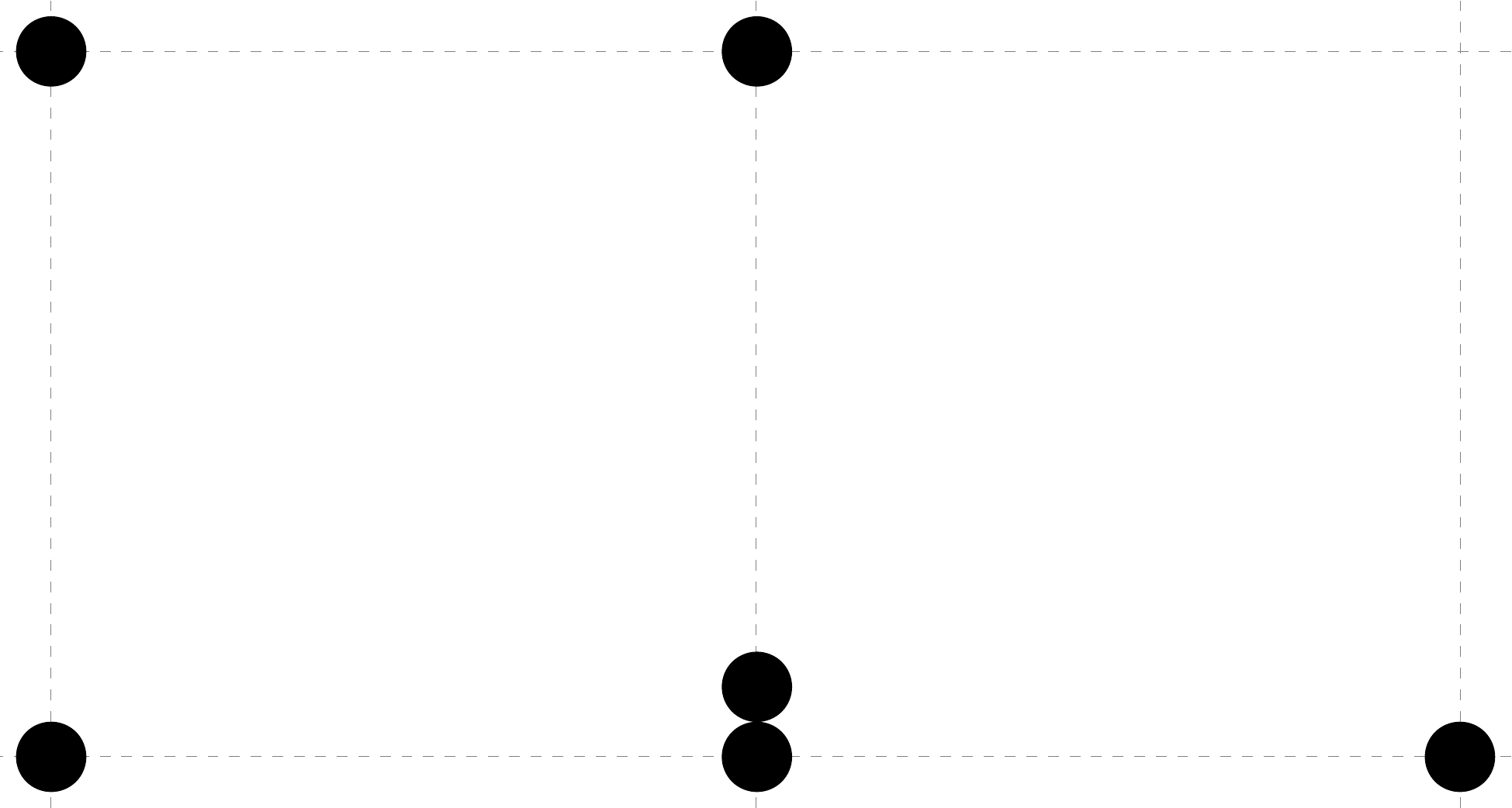}  \\
$ SPP $
\end{tabular}
&
\begin{tabular}[b]{c} 
$\phi _1^{}.X_{12}^{}.X_{21}^{}$ \\
$-\phi _1^{}.X_{13}^{}.X_{31}^{}$ \\
$-X_{12}^{}.X_{23}^{}.X_{32}^{}.X_{21}^{}$ \\
$+X_{13}^{}.X_{32}^{}.X_{23}^{}.X_{31}^{}$ 
\end{tabular}
\\ \hline

(2.3) &
\includegraphics[width=2.5cm]{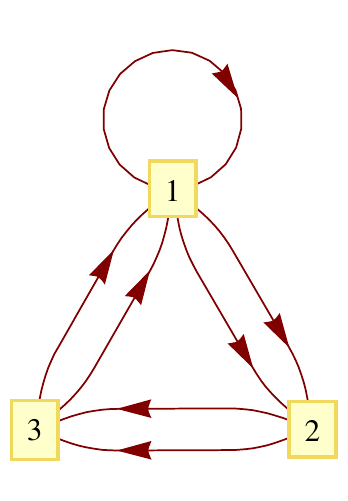} &
\includegraphics[height=3.5cm]{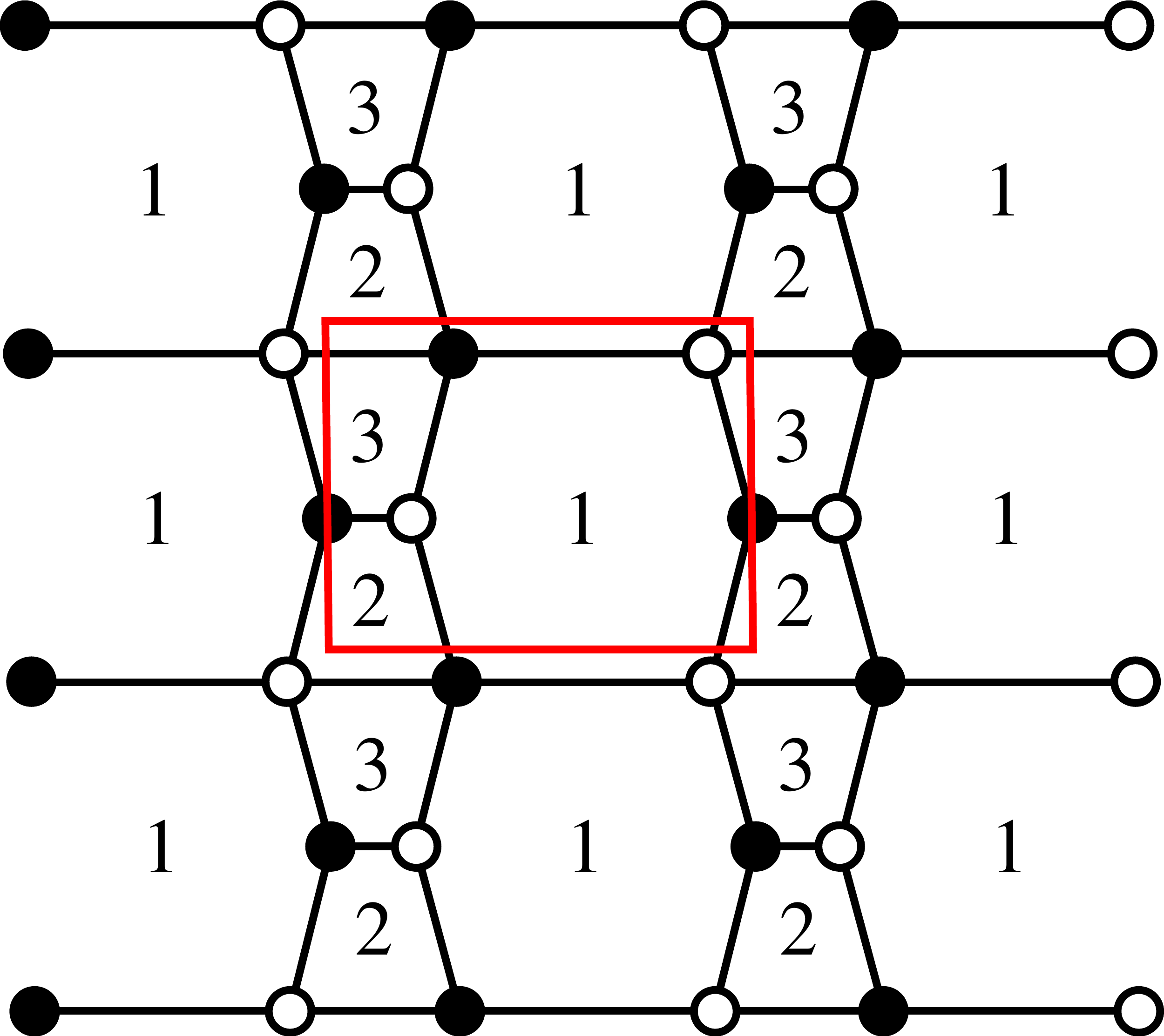} 
&
\begin{tabular}[b]{c}
\includegraphics[height=2.4cm,angle=90]{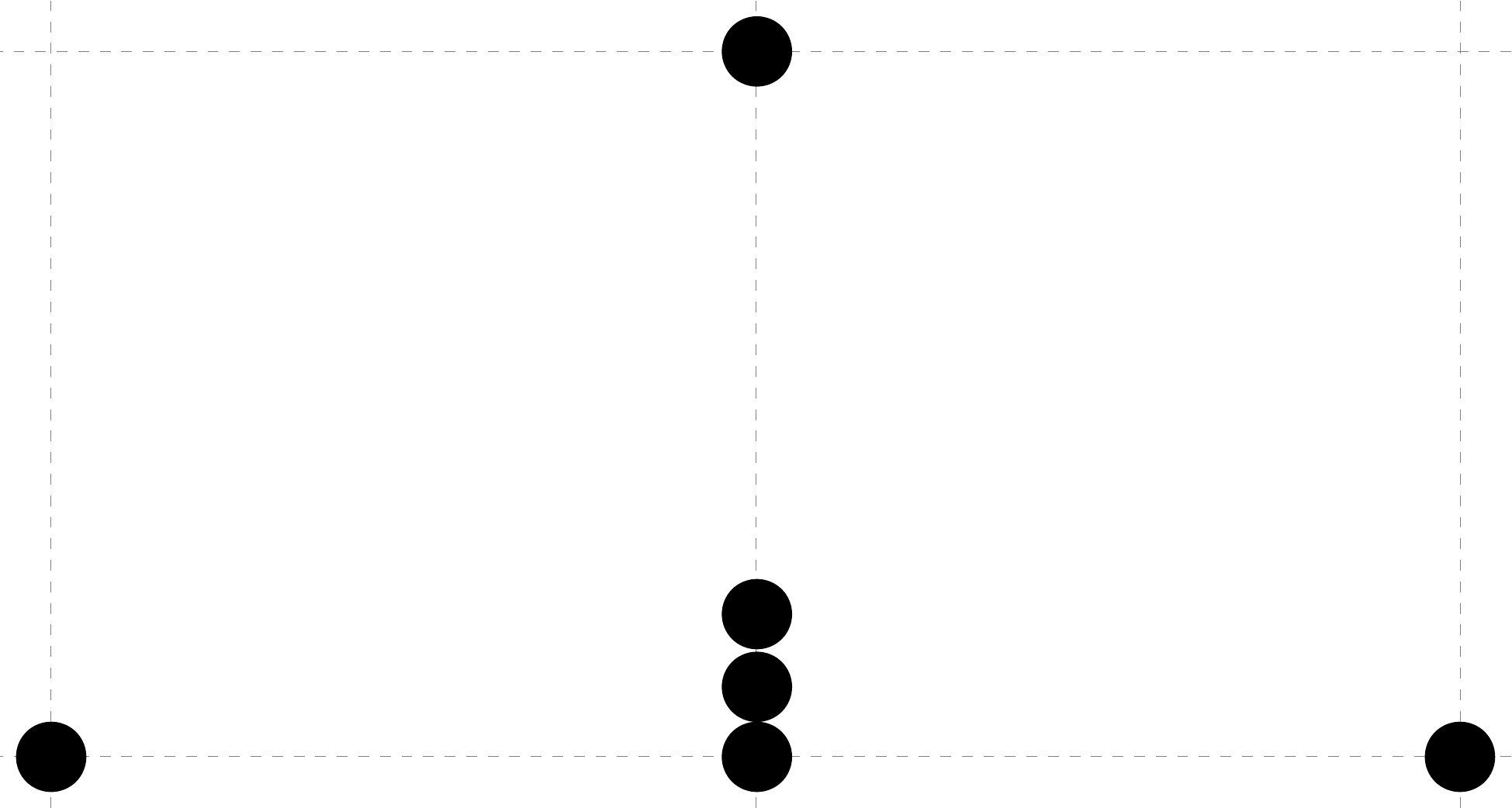} \\
$\BC^2/\BZ_2 \times \BC$ \\
(inc.)
\end{tabular}
&
\begin{tabular}[b]{c} 
$-X_{12}^1.X_{23}^2.X_{31}^1$ \\
$+X_{12}^2.X_{23}^2.X_{31}^2$ \\
$+\phi _1^{}.X_{12}^1.X_{23}^1.X_{31}^1$ \\
$-\phi _1^{}.X_{12}^2.X_{23}^1.X_{31}^2$ 
\end{tabular}

\end{tabular}
\end{center}

\caption{Tilings with 4 superpotential terms and 3 gauge groups}
\label{t:tilings4-3}
\end{table}


\begin{table}[h]

\begin{center}
\begin{tabular}{c|c|c|c|c}
\# & Quiver & Tiling & Toric Diagram & Superpotential\\
\hline \hline

(2.4) &
\includegraphics[width=2.5cm]{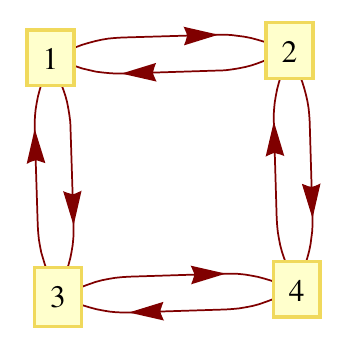} &
\includegraphics[height=3.5cm]{N4-G4-6-tiling} &
\begin{tabular}[b]{c}
\includegraphics[height=2.4cm,angle=90]{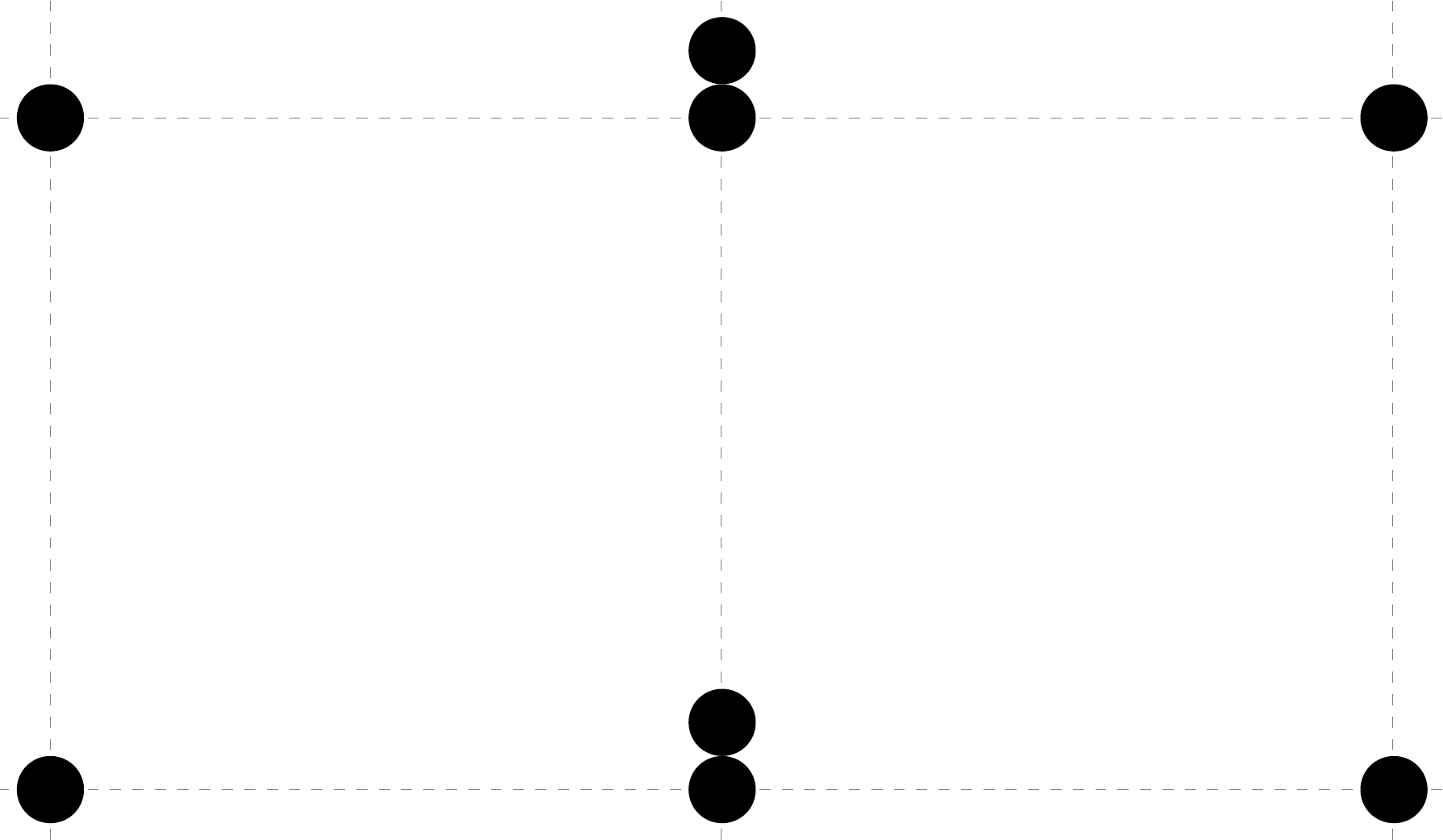}  \\
$L^{222}$ (I)
\end{tabular}
&
\begin{tabular}[b]{c} 
$X_{12}^{}.X_{21}^{}.X_{13}^{}.X_{31}^{}$ \\
$-X_{12}^{}.X_{24}^{}.X_{42}^{}.X_{21}^{}$ \\
$-X_{13}^{}.X_{34}^{}.X_{43}^{}.X_{31}^{}$ \\
$+X_{24}^{}.X_{43}^{}.X_{34}^{}.X_{42}^{}$
\end{tabular}
\\ \hline

(2.5) &
\includegraphics[width=2.5cm]{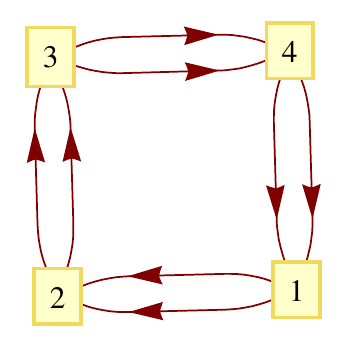} &
\includegraphics[height=3.5cm]{N4-G4-4-tiling.pdf} &
\begin{tabular}[b]{c}
\includegraphics[height=2.0cm]{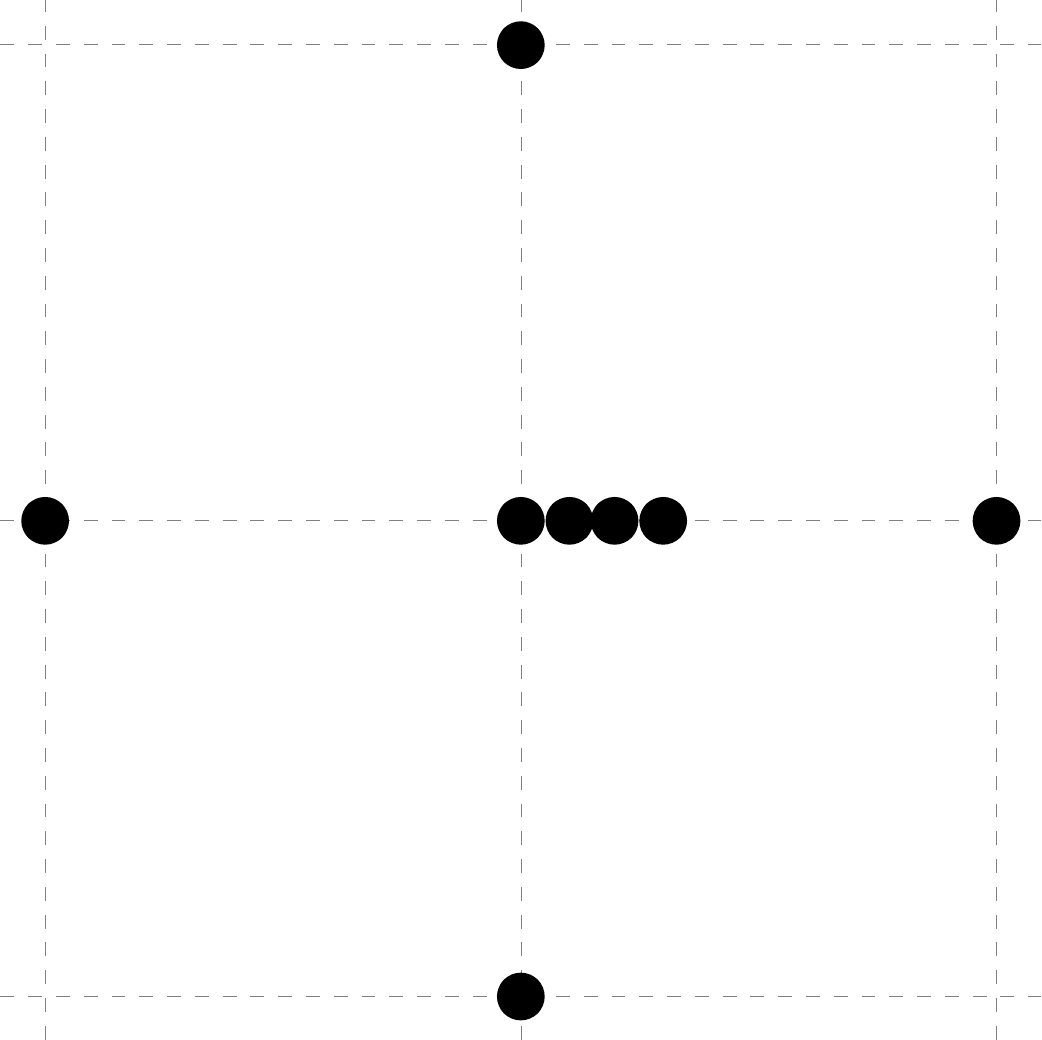} \\
$\BF_0$ (I)
\end{tabular}
&
\begin{tabular}[b]{c} 
$X_{12}^1.X_{23}^1.X_{34}^1.X_{41}^1$ \\
$-X_{12}^1.X_{23}^2.X_{34}^1.X_{41}^2$ \\
$-X_{12}^2.X_{23}^1.X_{34}^2.X_{41}^1$ \\
$+X_{12}^2.X_{23}^2.X_{34}^2.X_{41}^2$
\end{tabular}
\\ \hline

(2.6) &
\includegraphics[width=2.5cm]{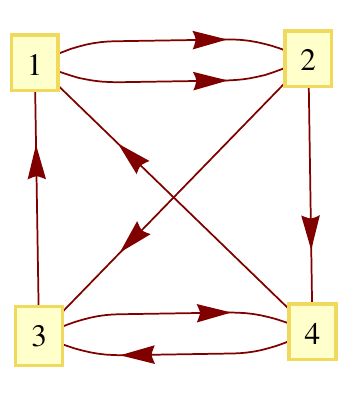} &
\includegraphics[height=3.5cm]{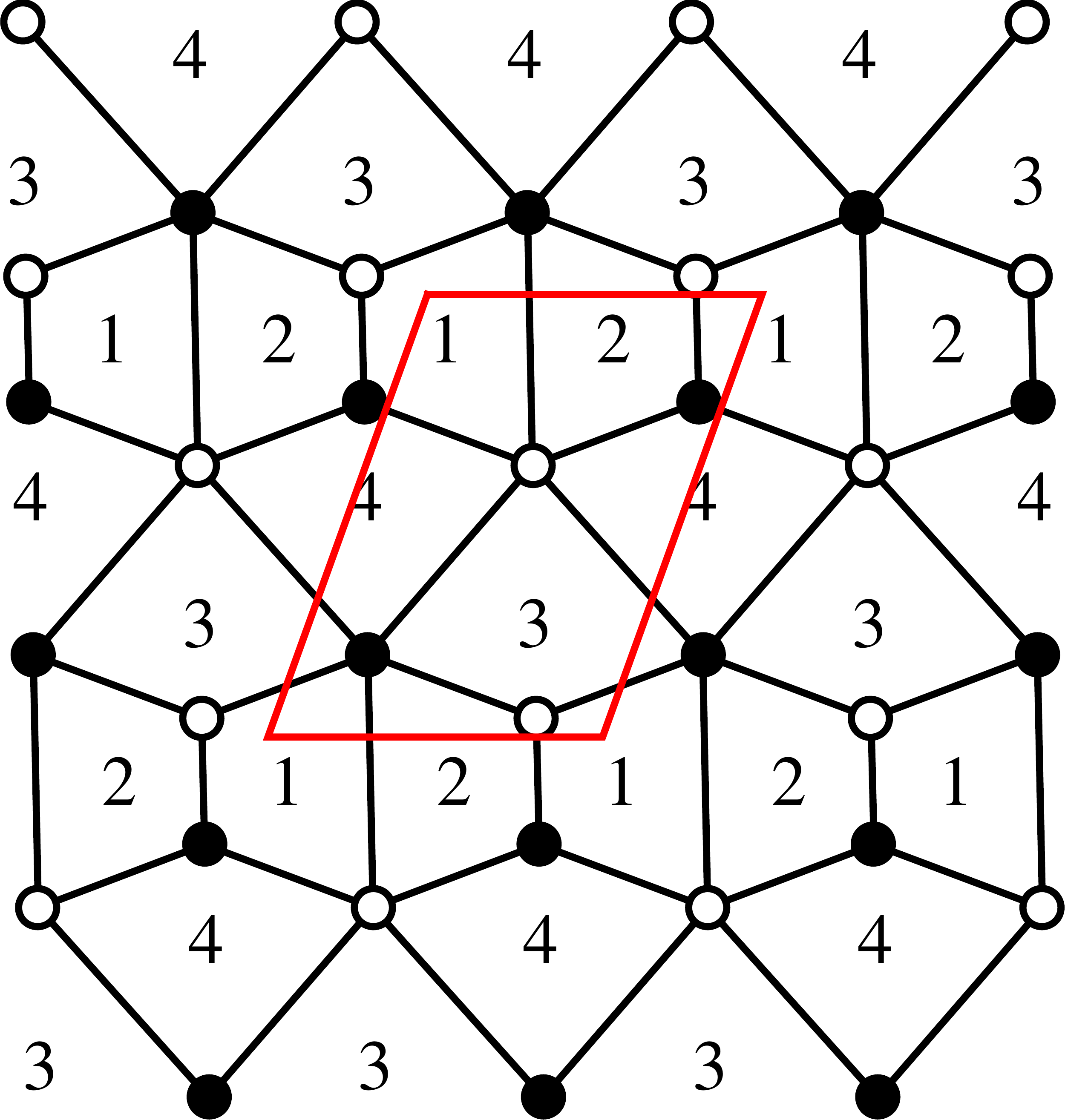} 
&
\begin{tabular}[b]{c}
\includegraphics[height=2.4cm,angle=90]{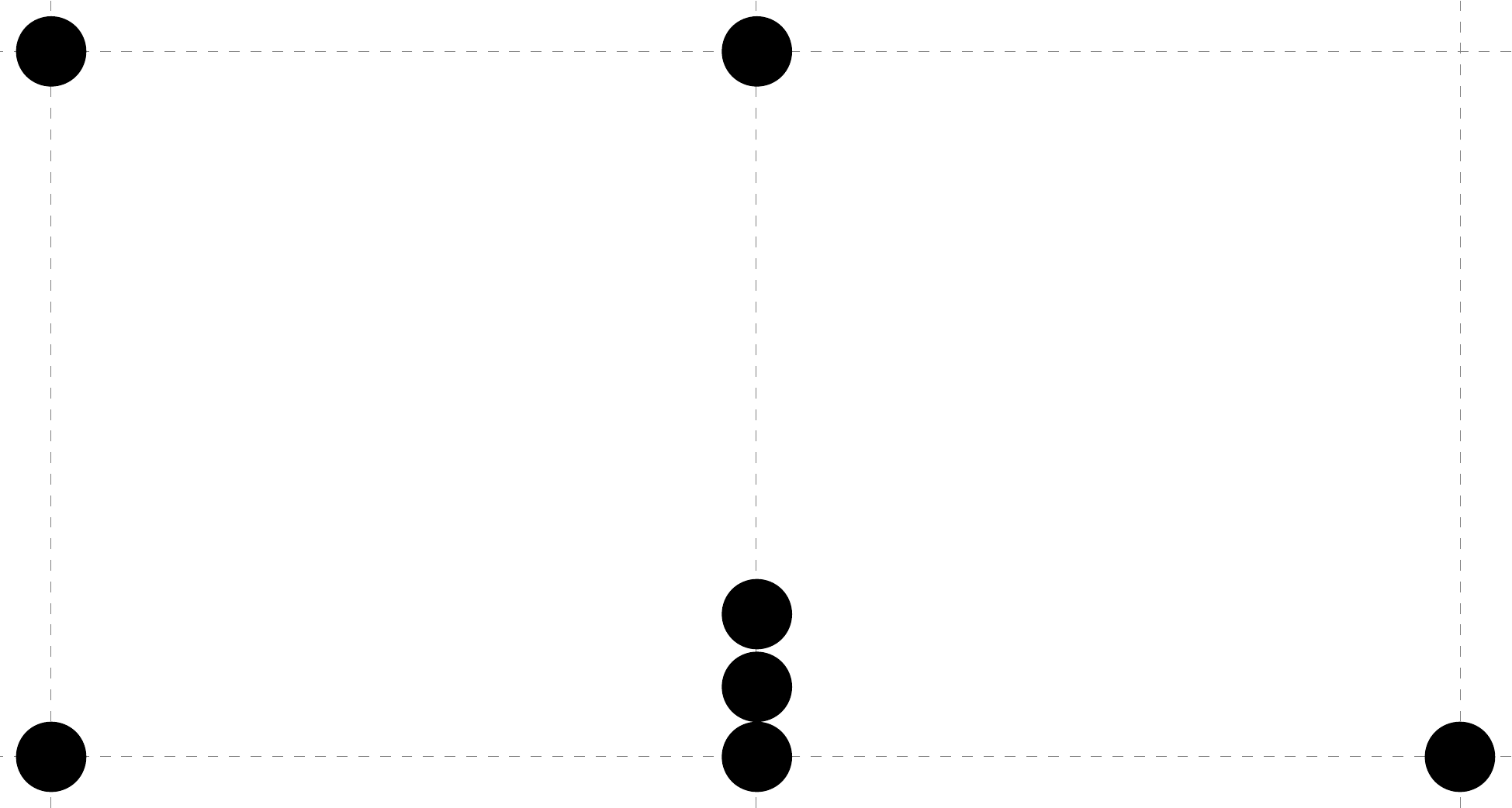}  \\
$SPP$ (inc.)
\end{tabular}
&
\begin{tabular}[b]{c} 
$X_{12}^1.X_{23}^{}.X_{31}^{}$ \\
$-X_{12}^1.X_{24}^{}.X_{41}^{}$ \\
$-X_{12}^2.X_{23}^{}.X_{34}^{}.X_{43}^{}.X_{31}^{}$ \\
$+X_{12}^2.X_{24}^{}.X_{43}^{}.X_{34}^{}.X_{41}^{}$
\end{tabular}

\end{tabular}
\end{center}

\caption{Tilings with 4 superpotential terms and 4 gauge groups}
\label{t:tilings4-4}
\end{table}

\ \clearpage


\subsection{Six superpotential terms}


\begin{table}[h]

\begin{center}
\begin{tabular}{c|c|c|c|c}
\# & Quiver & Tiling & Toric Diagram & Superpotential\\
\hline \hline

(3.1) &
\includegraphics[width=2.5cm]{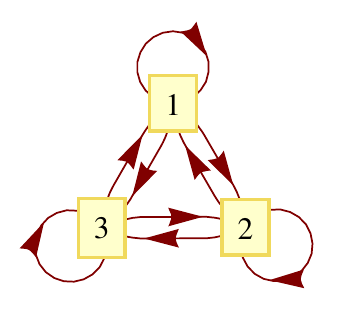} &
\includegraphics*[height=3.5cm]
{N6-G3-3-tiling.pdf} &
\begin{tabular}[b]{c}
\includegraphics[height=2.4cm,angle=90]{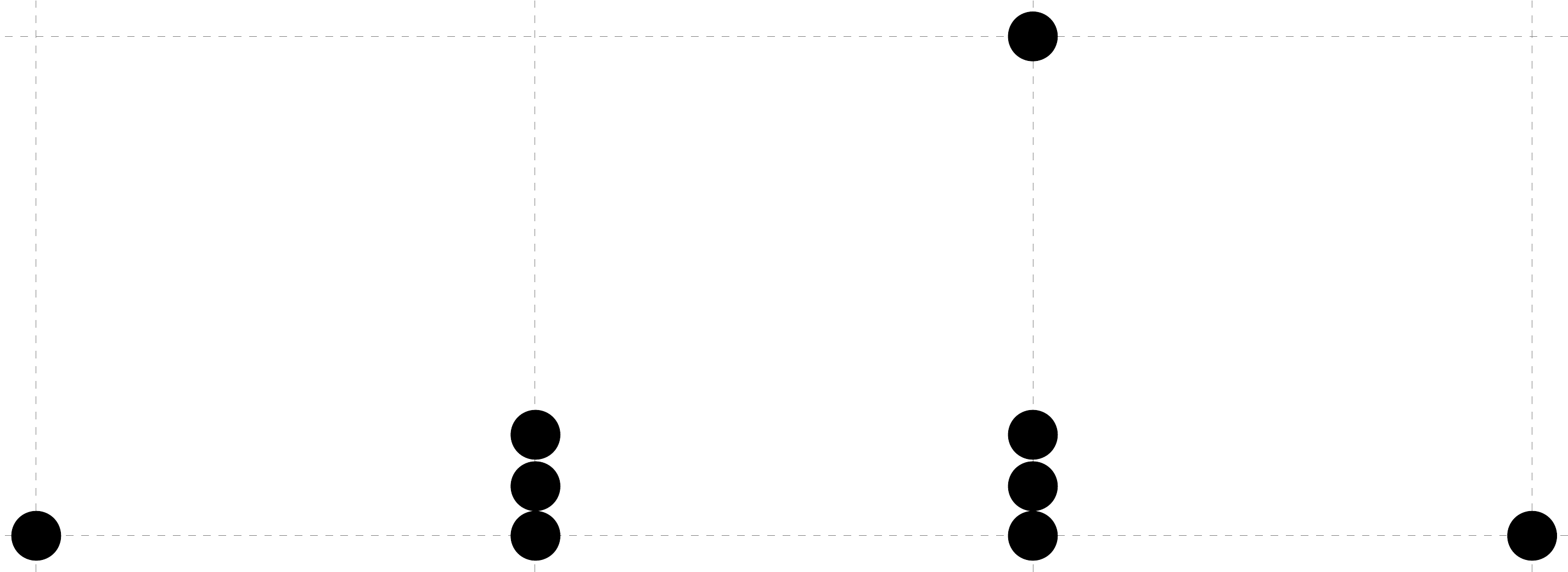}  \\
$\BC^2/\BZ_3 \times \BC$
\end{tabular}
&
\begin{tabular}[b]{c} 
$-X_{12}^{}.\phi _2^{}.X_{21}^{}$ \\
$+X_{13}^{}.\phi _3^{}.X_{31}^{}$ \\
$-X_{23}^{}.\phi _3^{}.X_{32}^{}$ \\
$+\phi _1^{}.X_{12}^{}.X_{21}^{}$ \\
$-\phi _1^{}.X_{13}^{}.X_{31}^{}$ \\
$+\phi _2^{}.X_{23}^{}.X_{32}^{}$
\end{tabular}
\\ \hline

(3.2) &
\includegraphics[width=2.5cm]{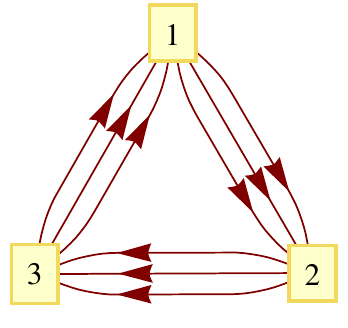} &
\includegraphics*[height=3.5cm]
{N6-G3-1-tiling.pdf} &
\begin{tabular}[b]{c}
\includegraphics[height=2.0cm]{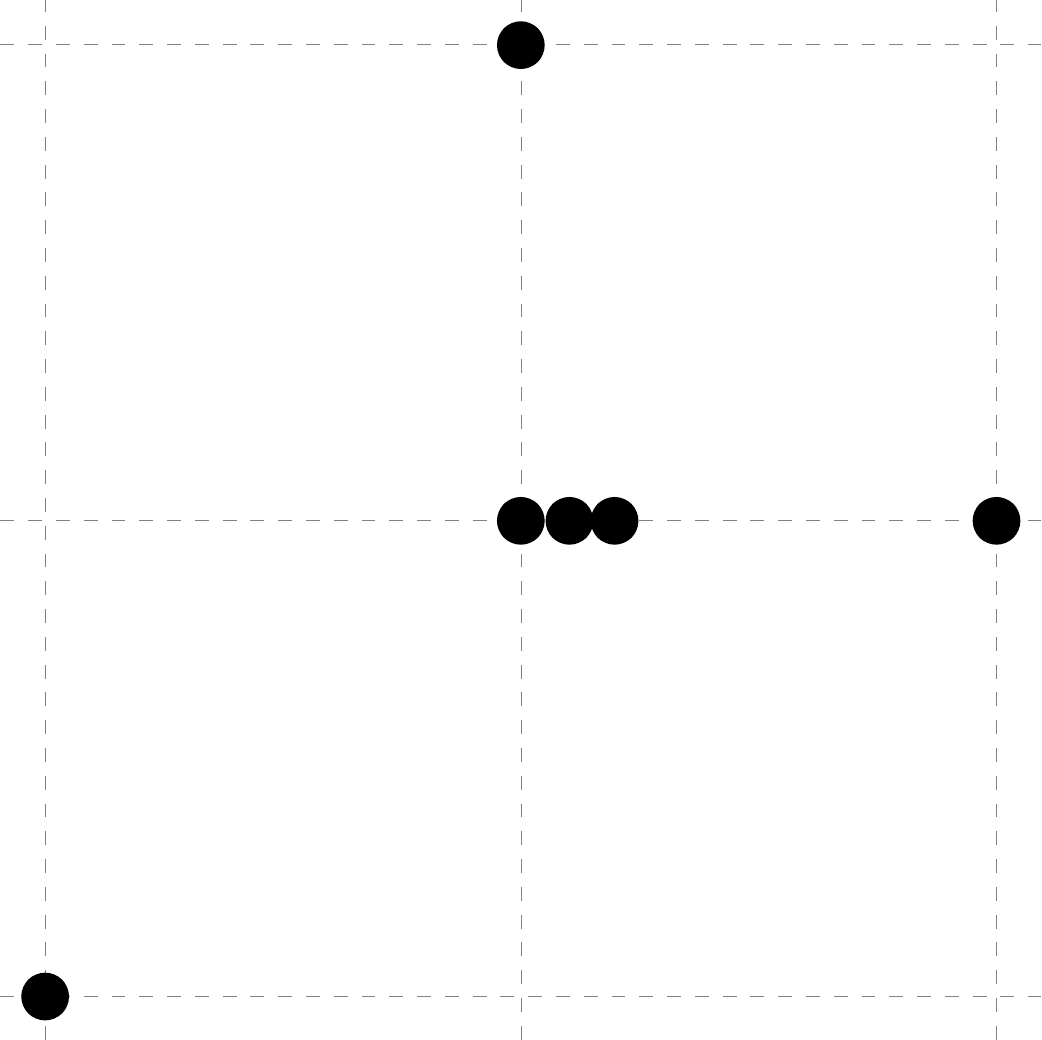}   \\
$\BC^3 / \BZ_3 $
\end{tabular}
&
\begin{tabular}[b]{c} 
$X_{12}^1.X_{23}^1.X_{31}^1$ \\
$-X_{12}^1.X_{23}^3.X_{31}^2$ \\
$-X_{12}^2.X_{23}^1.X_{31}^3$ \\
$+X_{12}^2.X_{23}^2.X_{31}^2$ \\
$-X_{12}^3.X_{23}^2.X_{31}^1$ \\
$+X_{12}^3.X_{23}^3.X_{31}^3$
\end{tabular}
\\ \hline

(3.3) &
\includegraphics[width=2.5cm]{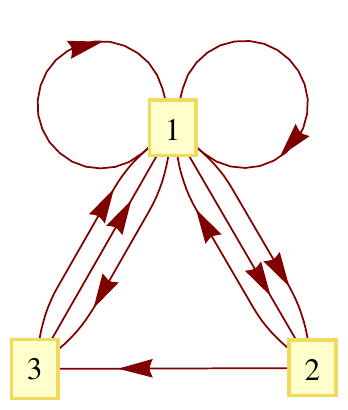} &
\includegraphics*[height=3.5cm]
{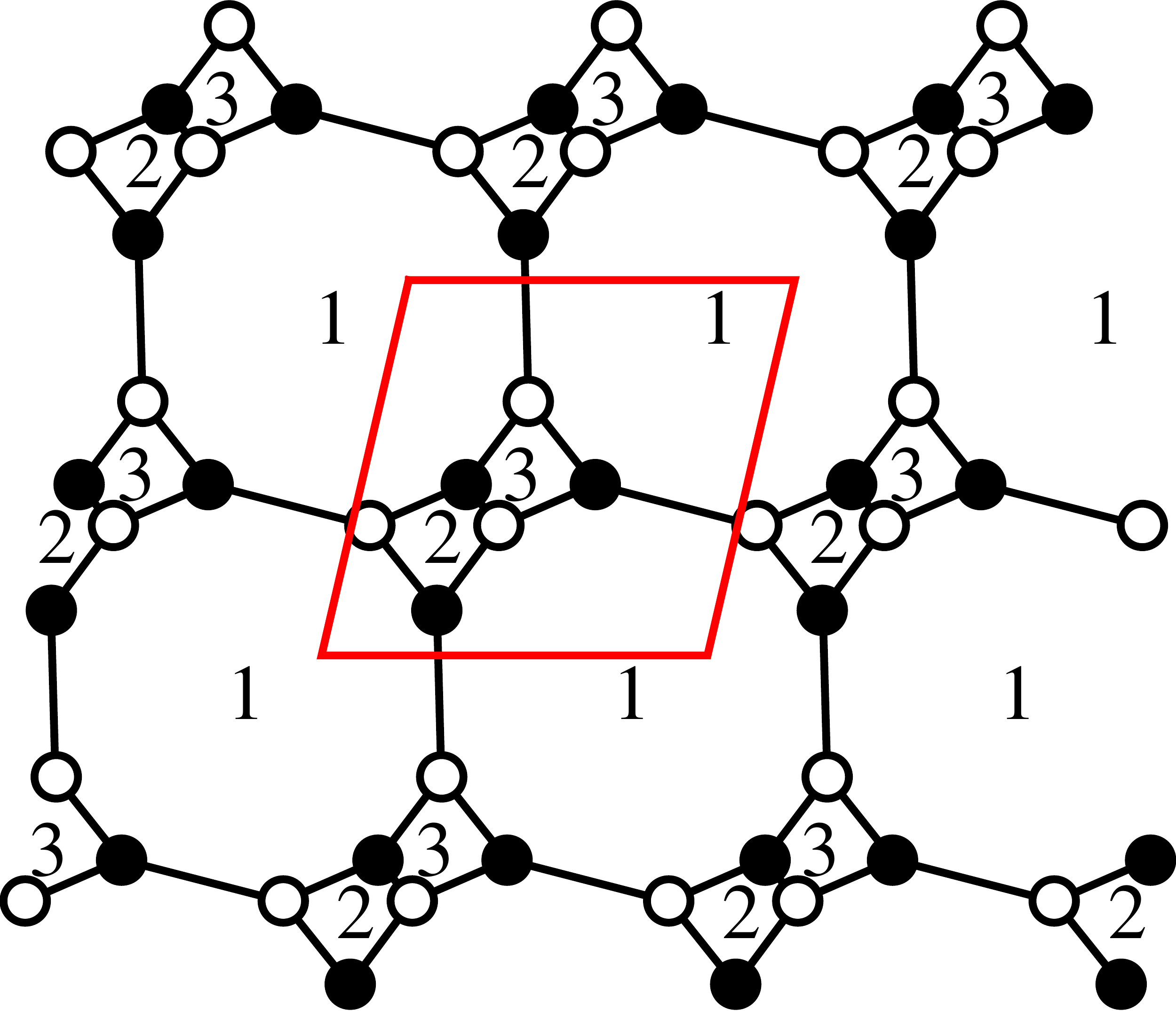} &
\begin{tabular}[b]{c}
\includegraphics[height=2.0cm]{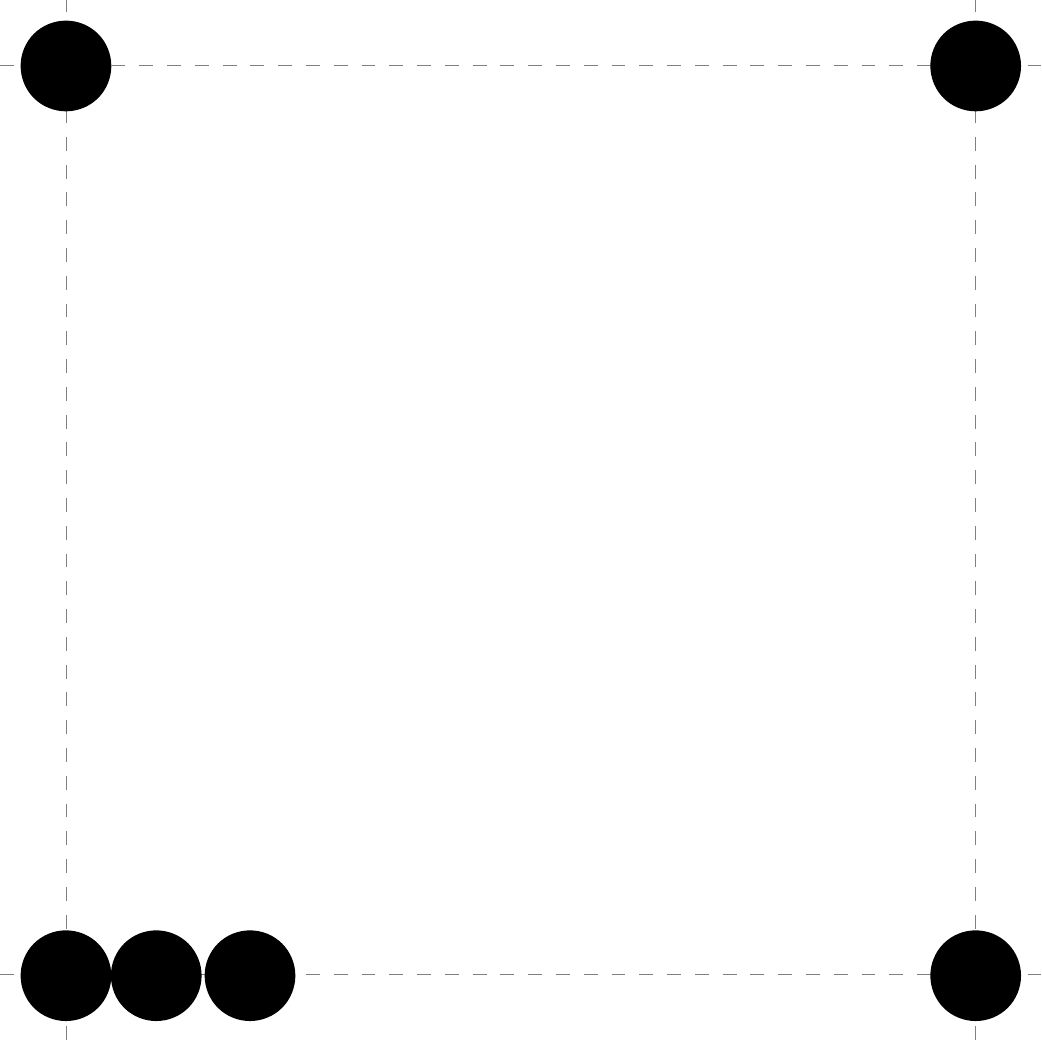}  \\
$\cal{C}$ (inc.)
\end{tabular}
&
\begin{tabular}[b]{c} 
$-X_{12}^1.X_{23}^{}.X_{31}^1$ \\
$+X_{12}^2.X_{23}^{}.X_{31}^2$ \\
$+\phi _1^1.X_{12}^1.X_{21}^{}$ \\
$-\phi _1^1.X_{13}^{}.X_{31}^2$ \\
$-\phi _1^2.X_{12}^2.X_{21}^{}$ \\
$+\phi _1^2.X_{13}^{}.X_{31}^1$
\end{tabular}

\end{tabular}
\end{center}

\caption{Tilings with 6 superpotential terms and 3 gauge groups}
\label{t:tilings6-3}
\end{table}


\begin{table}[h]

\begin{center}
\begin{tabular}{c|c|c|c|c}
\# & Quiver & Tiling & Toric Diagram & Superpotential\\
\hline \hline

(3.4) &
\includegraphics[width=3.0cm]{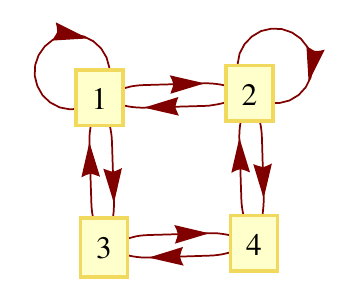} &
\includegraphics*[height=3.5cm]{N6-G4-6-tiling.pdf} &
\begin{tabular}[b]{c}
\includegraphics[height=2.4cm,angle=90]{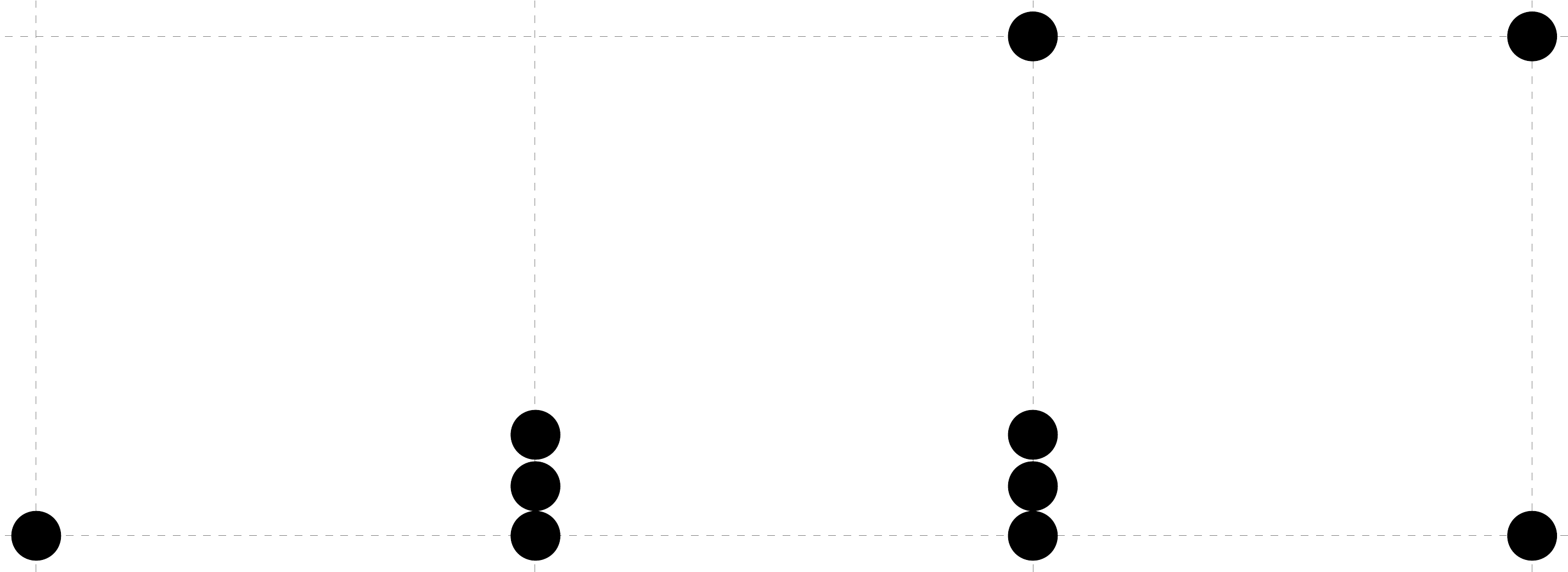}  \\
$ L^{131} $
\end{tabular}
&
\begin{tabular}[b]{c} 
$-X_{12}^{}.\phi _2^{}.X_{21}^{}$ \\
$+\phi _1^{}.X_{12}^{}.X_{21}^{}$ \\
$-\phi _1^{}.X_{13}^{}.X_{31}^{}$ \\
$+\phi _2^{}.X_{24}^{}.X_{42}^{}$ \\
$+X_{13}^{}.X_{34}^{}.X_{43}^{}.X_{31}^{}$ \\
$-X_{24}^{}.X_{43}^{}.X_{34}^{}.X_{42}^{}$ 
\end{tabular}
\\ \hline

(3.5) &
\includegraphics[width=3.0cm]{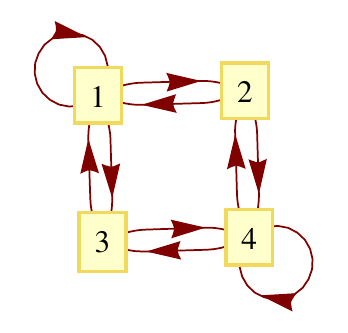} &
\includegraphics*[height=3.5cm]{N6-G4-7-tiling.pdf} & 
\begin{tabular}[b]{c}
\includegraphics[height=2.4cm,angle=90]{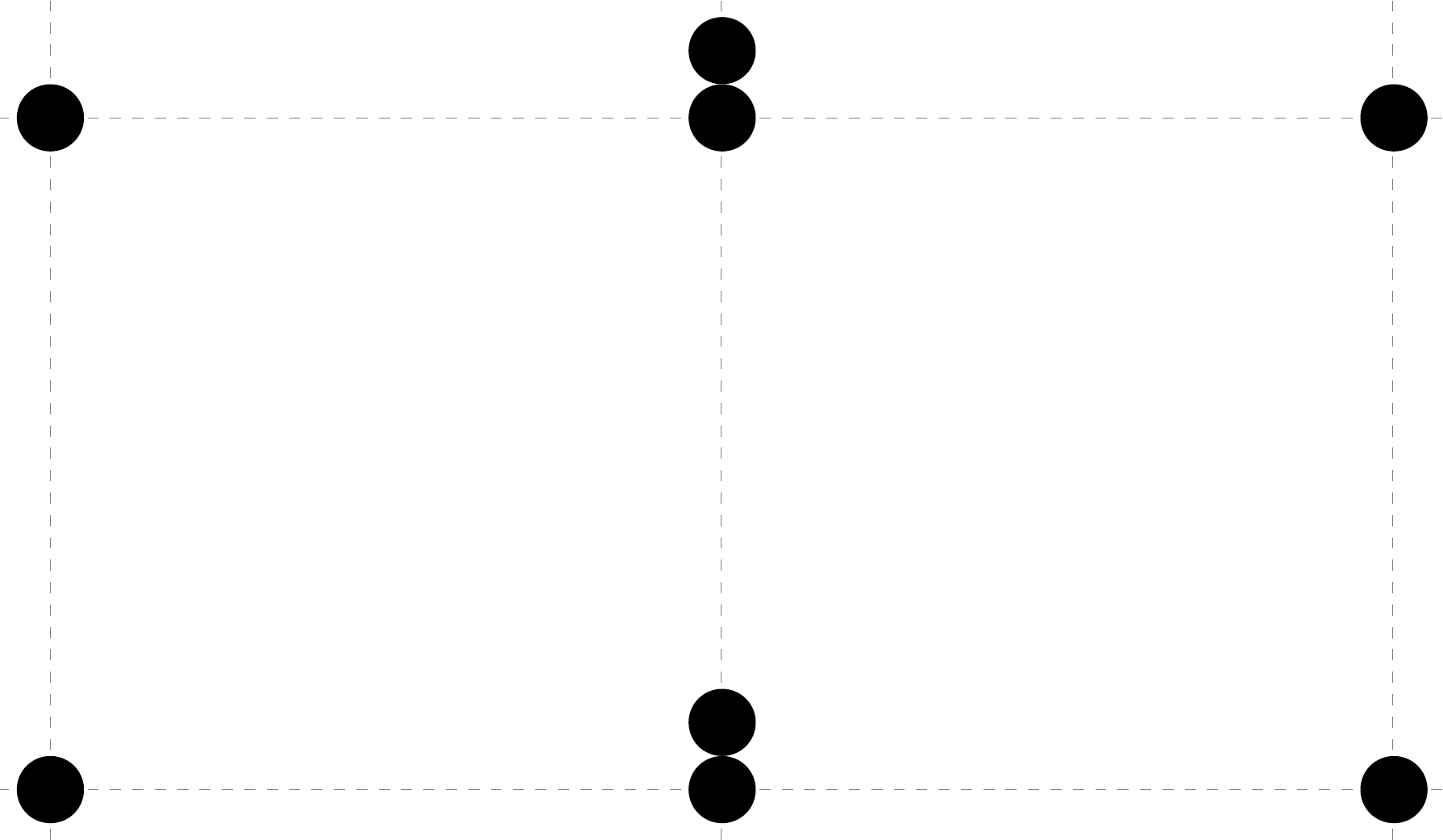} \\
$ L^{222} $ (II)
\end{tabular}
&
\begin{tabular}[b]{c} 
$X_{24}^{}.\phi _4^{}.X_{42}^{}$ \\
$-X_{34}^{}.\phi _4^{}.X_{43}^{}$ \\
$+\phi _1^{}.X_{12}^{}.X_{21}^{}$ \\
$-\phi _1^{}.X_{13}^{}.X_{31}^{}$ \\
$-X_{12}^{}.X_{24}^{}.X_{42}^{}.X_{21}^{}$ \\
$+X_{13}^{}.X_{34}^{}.X_{43}^{}.X_{31}^{}$ 
\end{tabular}
\\ \hline

(3.6) &
\includegraphics[width=3.0cm]{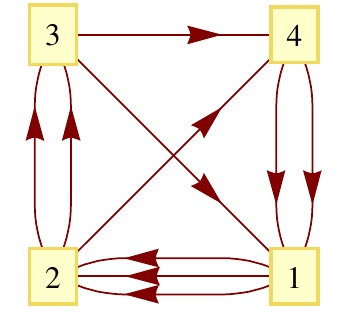} &
\includegraphics*[height=3.5cm]{N6-G4-9-tiling.pdf} &
\begin{tabular}[b]{c}
\includegraphics[height=2.0cm]{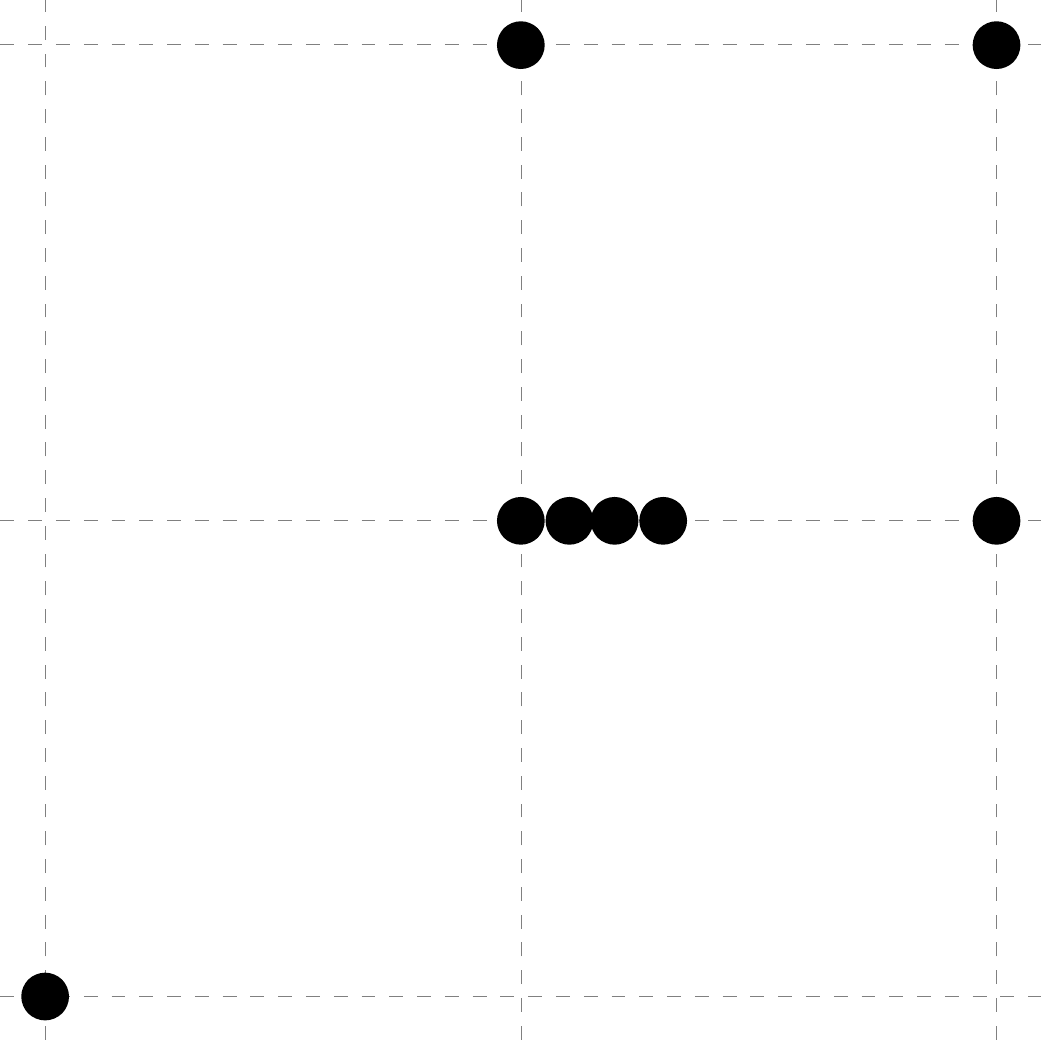}  \\
$ dP_1 $
\end{tabular}
&
\begin{tabular}[b]{c} 
$X_{12}^1.X_{23}^1.X_{31}^{}$ \\
$-X_{12}^1.X_{24}^{}.X_{41}^2$ \\
$-X_{12}^2.X_{23}^2.X_{31}^{}$ \\
$+X_{12}^2.X_{24}^{}.X_{41}^1$ \\
$-X_{12}^3.X_{23}^1.X_{34}^{}.X_{41}^1$ \\
$+X_{12}^3.X_{23}^2.X_{34}^{}.X_{41}^2$ 
\end{tabular}
\\ \hline

(3.7) &
\includegraphics[width=3.0cm]{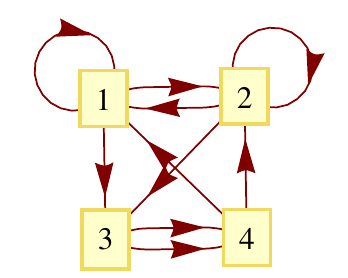} &
\includegraphics*[height=3.5cm]{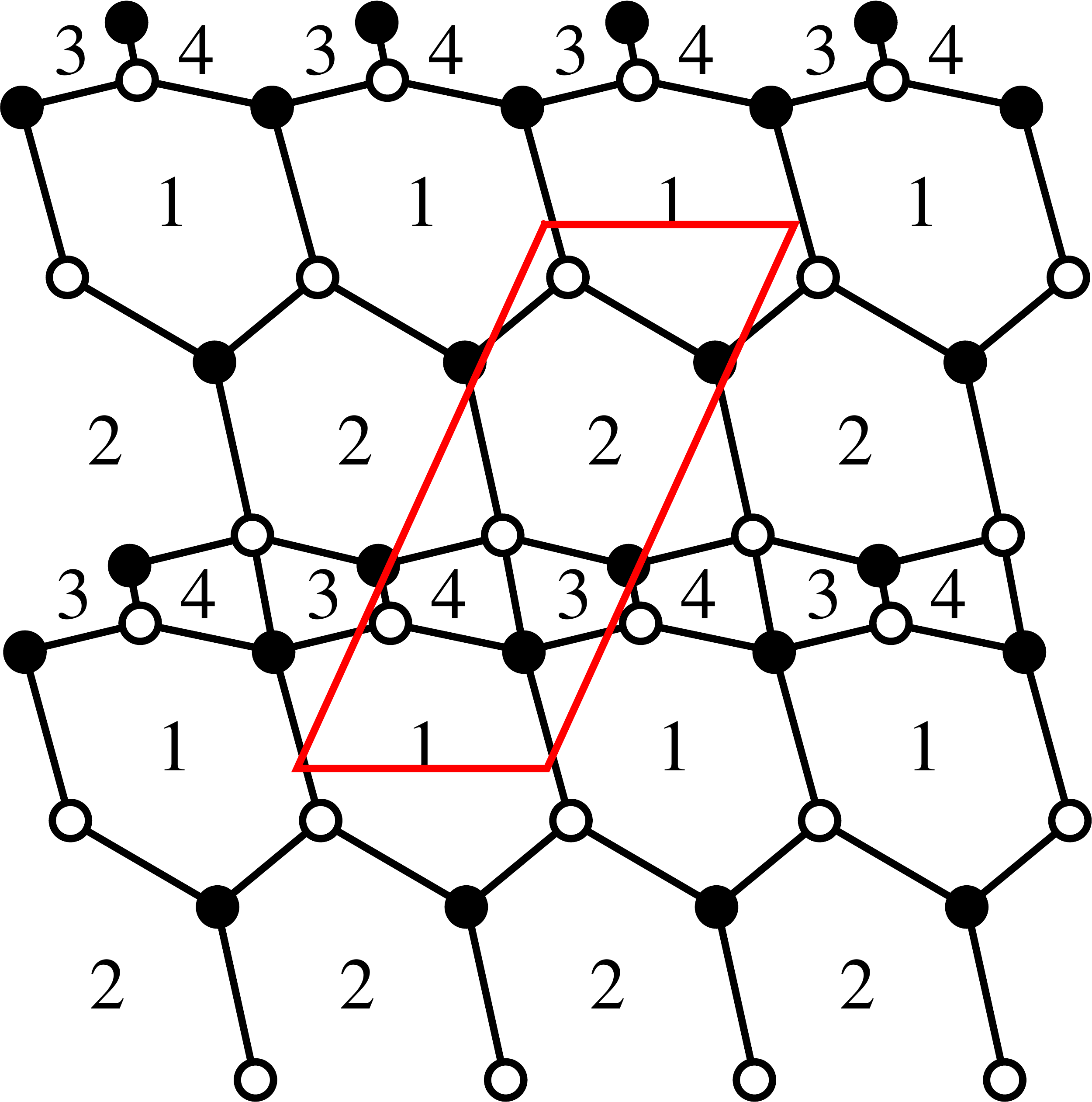} &
\begin{tabular}[b]{c}
\includegraphics[height=2.4cm,angle=90]{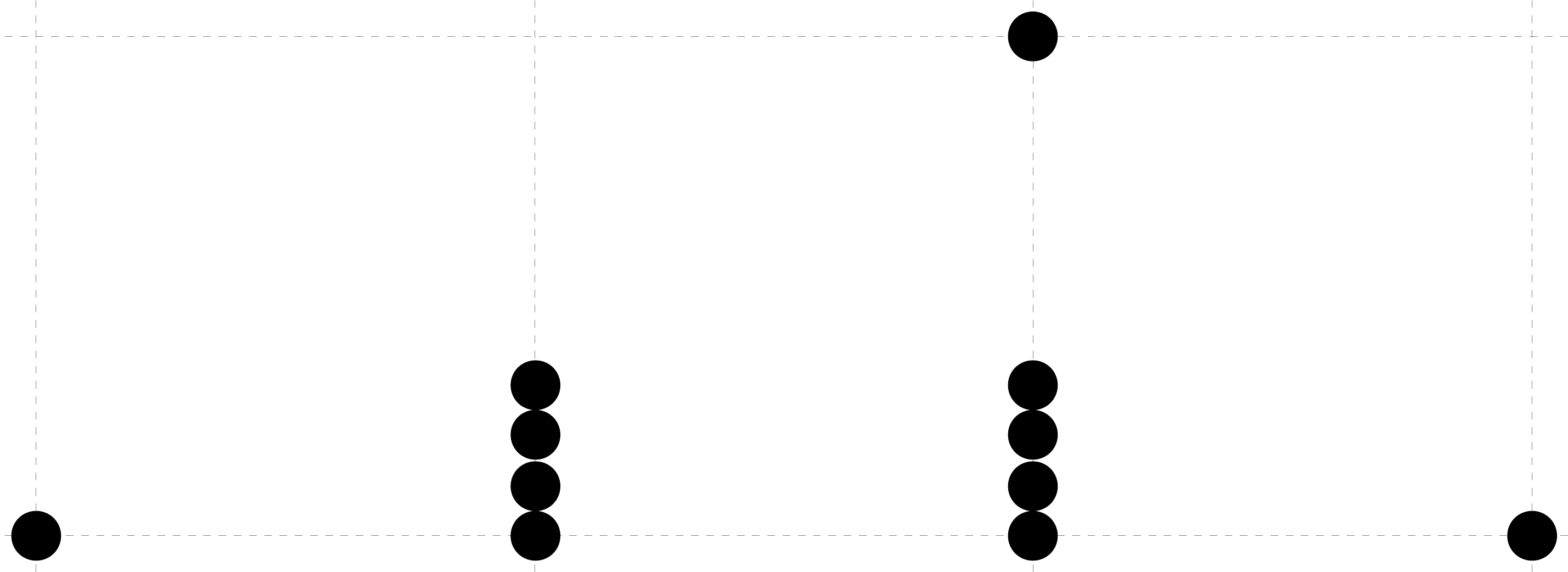}  \\
$\BC^2/\BZ_3 \times \BC$ \\
(inc.)
\end{tabular}
&
\begin{tabular}[b]{c} 
$-X_{12}^{}.\phi _2^{}.X_{21}^{}$ \\
$+X_{13}^{}.X_{34}^1.X_{41}^{}$ \\
$-X_{23}^{}.X_{34}^1.X_{42}^{}$ \\
$+\phi _1^{}.X_{12}^{}.X_{21}^{}$ \\
$-\phi _1^{}.X_{13}^{}.X_{34}^2.X_{41}^{}$ \\
$+\phi _2^{}.X_{23}^{}.X_{34}^2.X_{42}^{}$ 
\end{tabular}
\\ \hline

(3.8) &
\includegraphics[width=3.0cm]{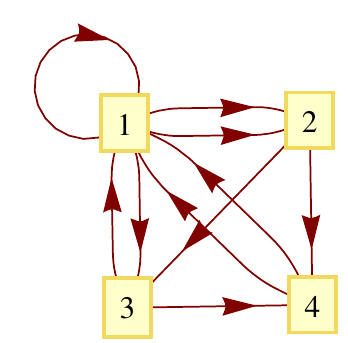} &
\includegraphics*[height=3.5cm]{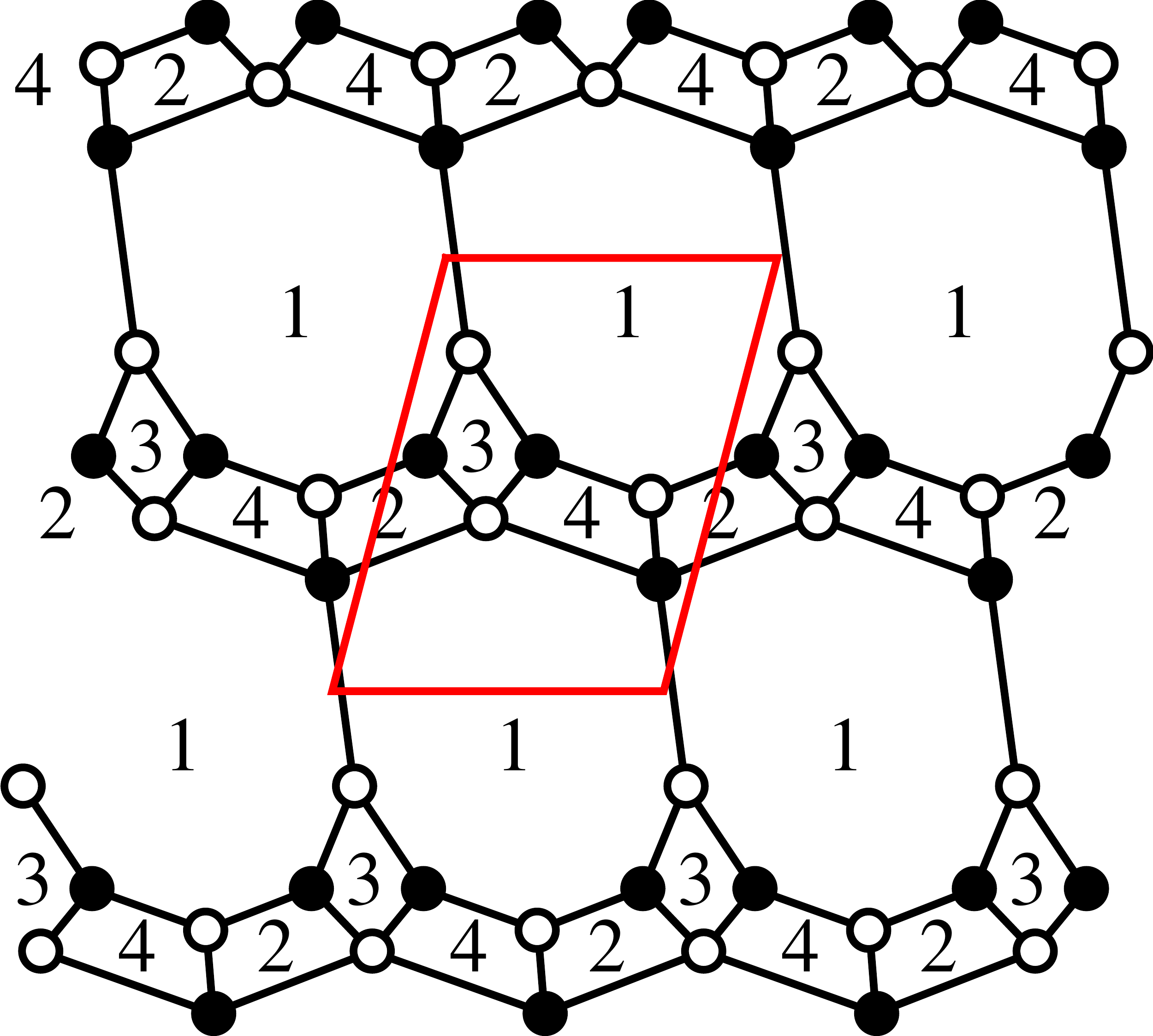} &
\begin{tabular}[b]{c}
\includegraphics[height=2.4cm,angle=90]{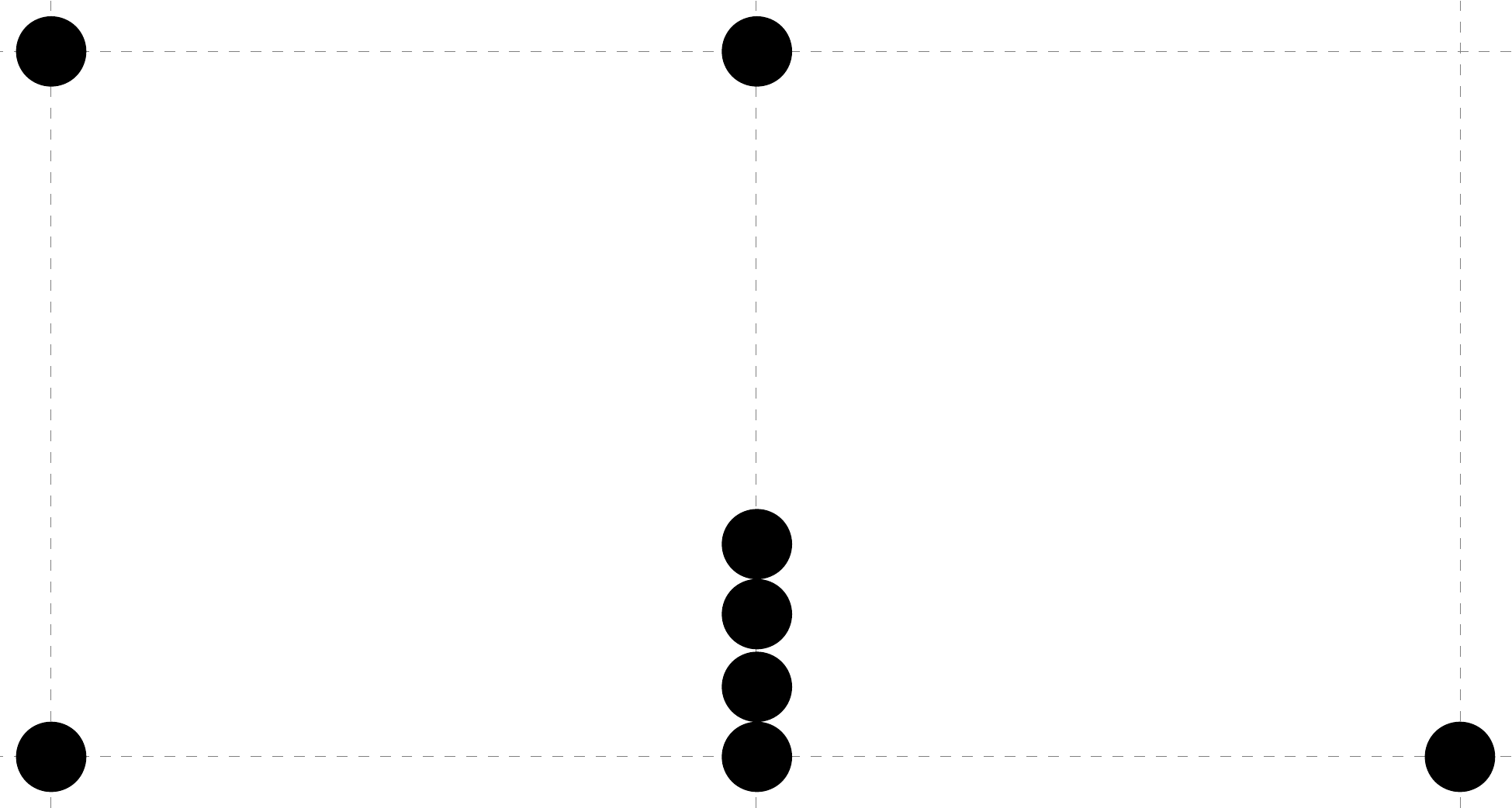}  \\
$SPP$ (inc.)
\end{tabular}
&
\begin{tabular}[b]{c} 
$-X_{12}^1.X_{23}^{}.X_{31}^{}$ \\
$+X_{12}^1.X_{24}^{}.X_{41}^1$ \\
$-X_{13}^{}.X_{34}^{}.X_{41}^1$ \\
$+\phi _1^{}.X_{13}^{}.X_{31}^{}$ \\
$+X_{12}^2.X_{23}^{}.X_{34}^{}.X_{41}^2$ \\
$-\phi _1^{}.X_{12}^2.X_{24}^{}.X_{41}^2$
\end{tabular}
\\ \hline



\end{tabular}
\end{center}

\caption{Tilings with 6 superpotential terms and 4 gauge groups \bf{(page 1/2)}}
\label{t:tilings6-4a}
\end{table}

\begin{table}[h]

\begin{center}
\begin{tabular}{c|c|c|c|c}
\# & Quiver & Tiling & Toric Diagram & Superpotential\\
\hline \hline

(3.9) &
\includegraphics[width=3.0cm]{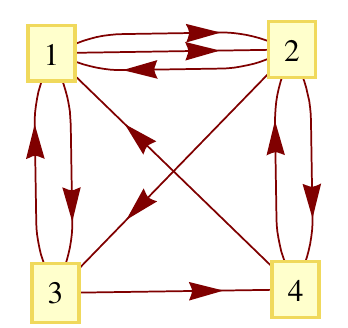} &
\includegraphics*[height=3.5cm]{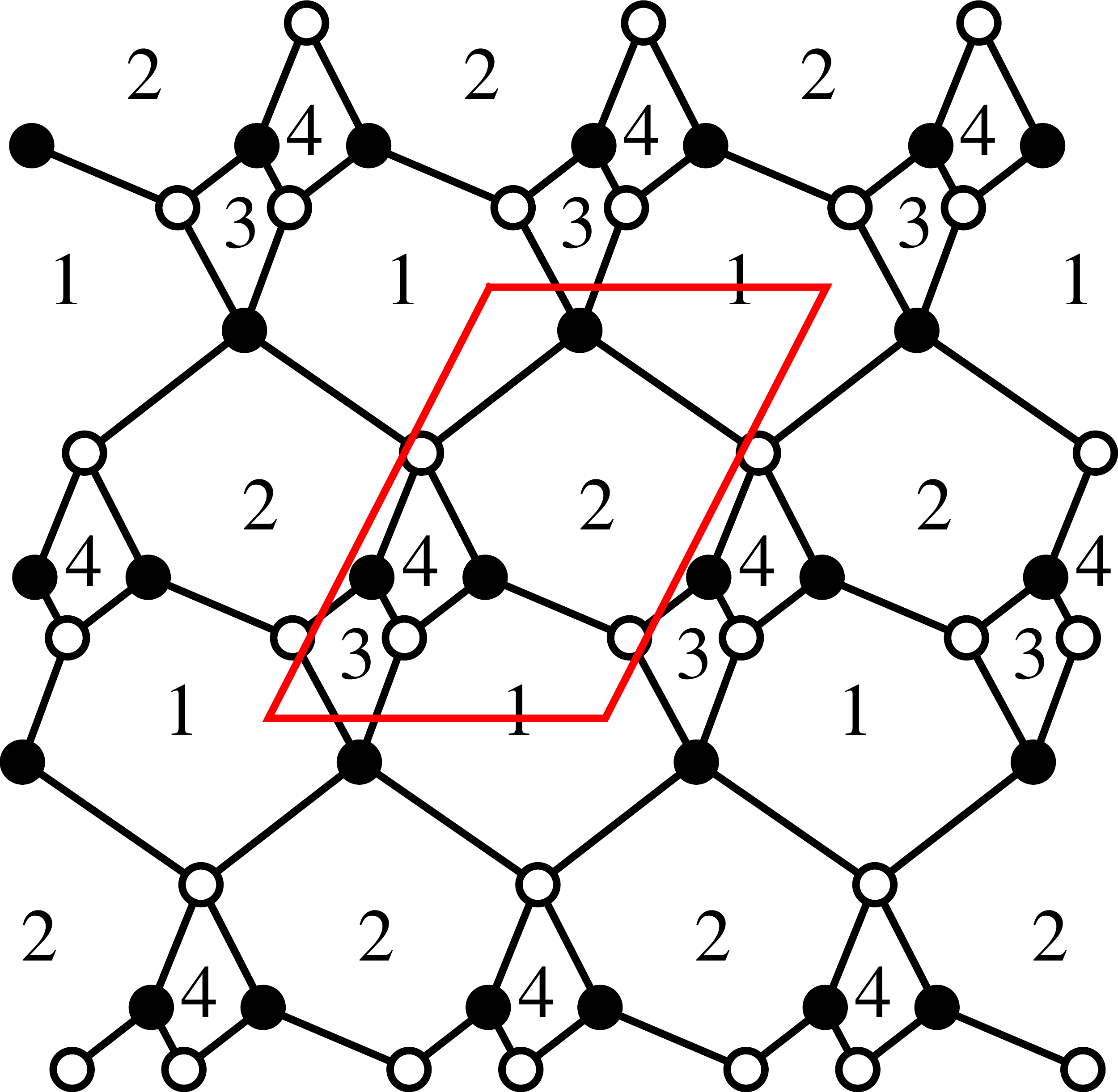} & 
\begin{tabular}[b]{c}
\includegraphics[height=2.4cm,angle=90]{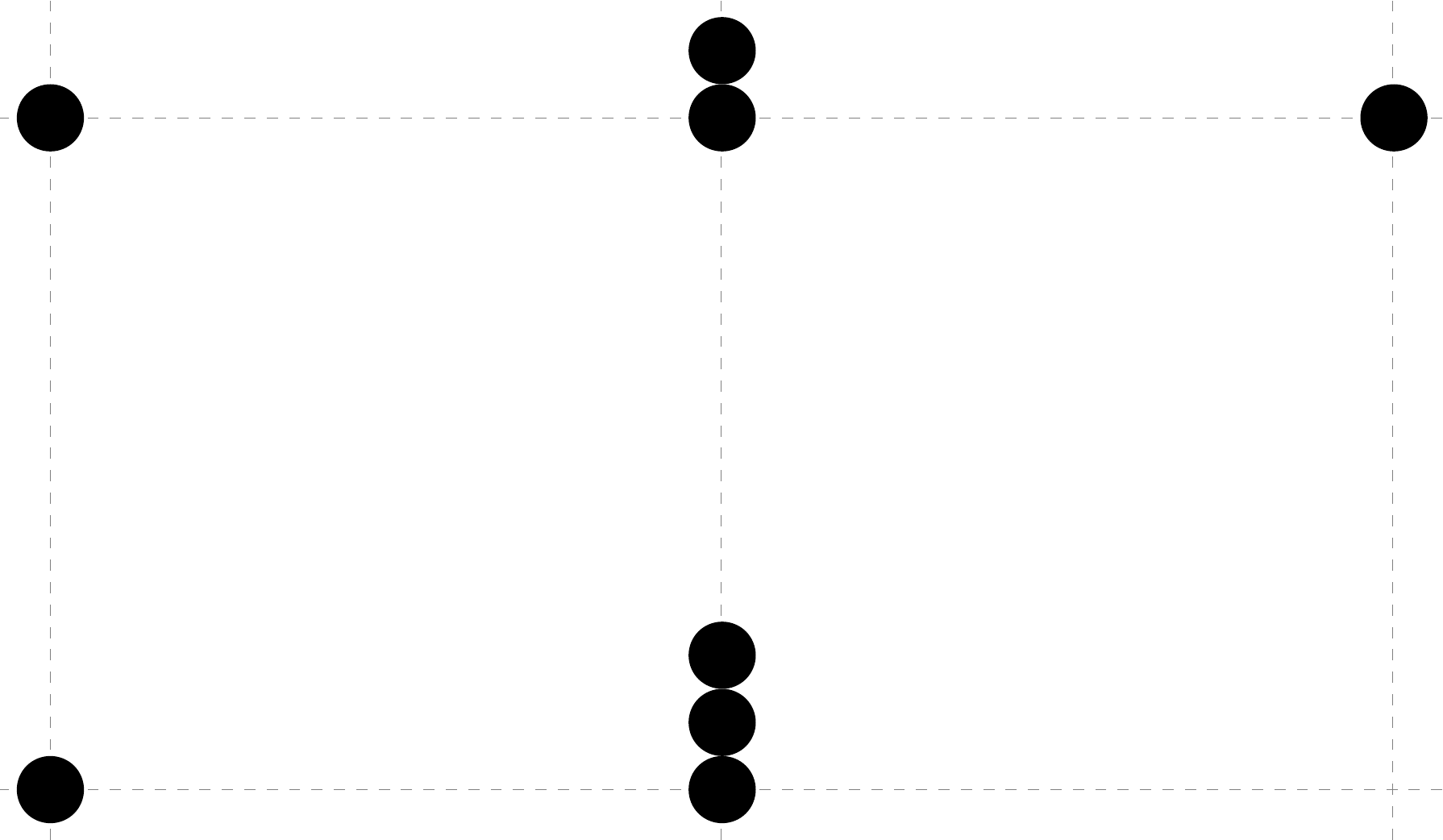} \\
$SPP$ (inc.)
\end{tabular}
&
\begin{tabular}[b]{c} 
$X_{12}^1.X_{23}^{}.X_{31}^{}$ \\
$-X_{12}^1.X_{24}^{}.X_{41}^{}$ \\
$+X_{13}^{}.X_{34}^{}.X_{41}^{}$ \\
$-X_{23}^{}.X_{34}^{}.X_{42}^{}$ \\
$-X_{12}^2.X_{21}^{}.X_{13}^{}.X_{31}^{}$ \\
$+X_{12}^2.X_{24}^{}.X_{42}^{}.X_{21}^{}$ 
\end{tabular}
\\ \hline

(3.10) &
\includegraphics[width=3.0cm]{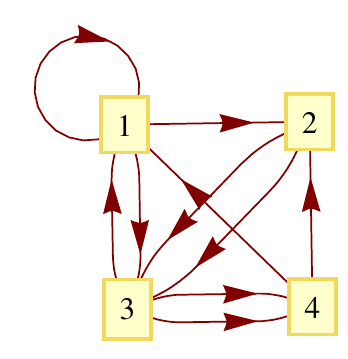} &
\includegraphics*[height=3.5cm]{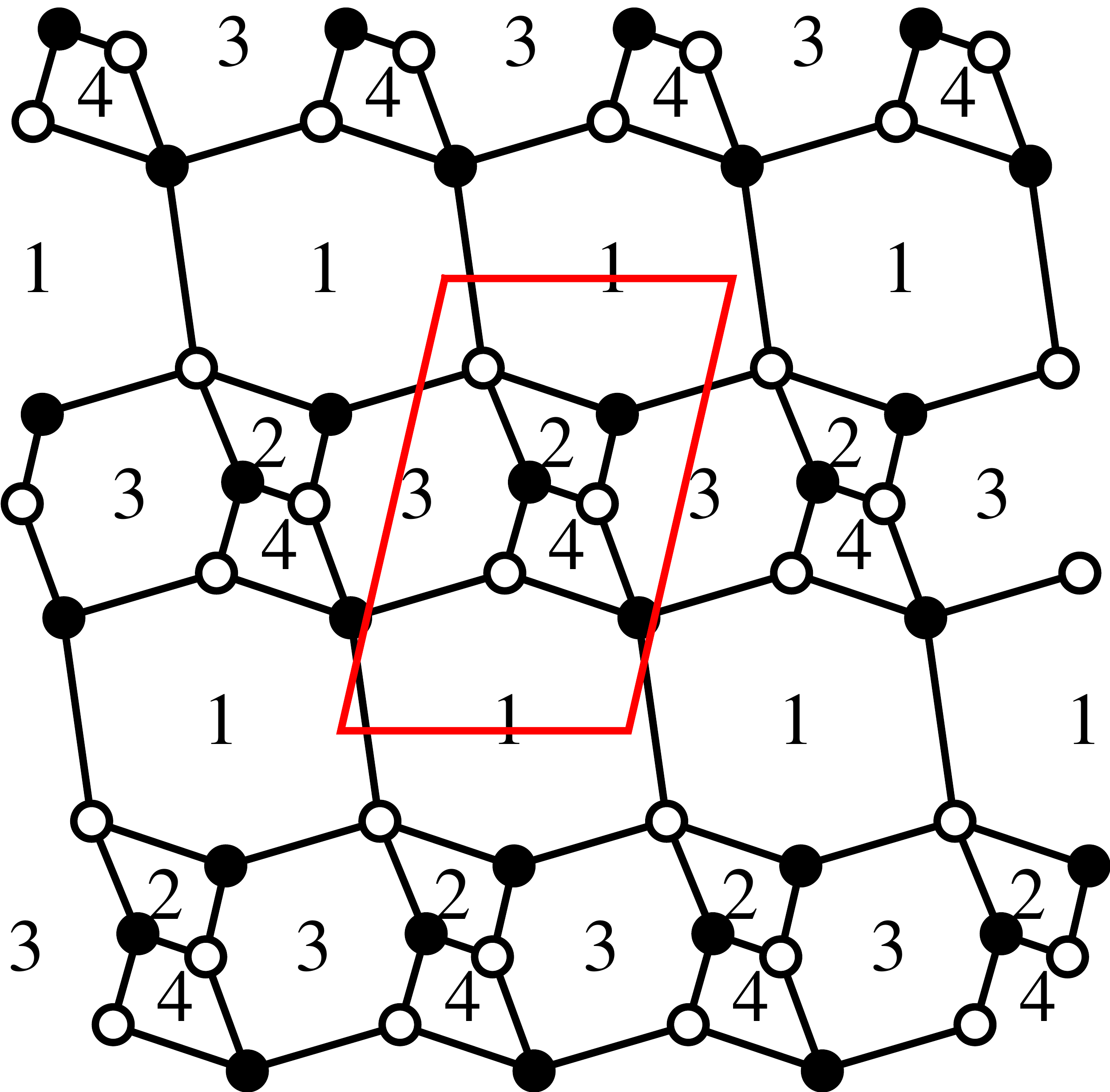} &
\begin{tabular}[b]{c}
\includegraphics[height=2.4cm,angle=90]{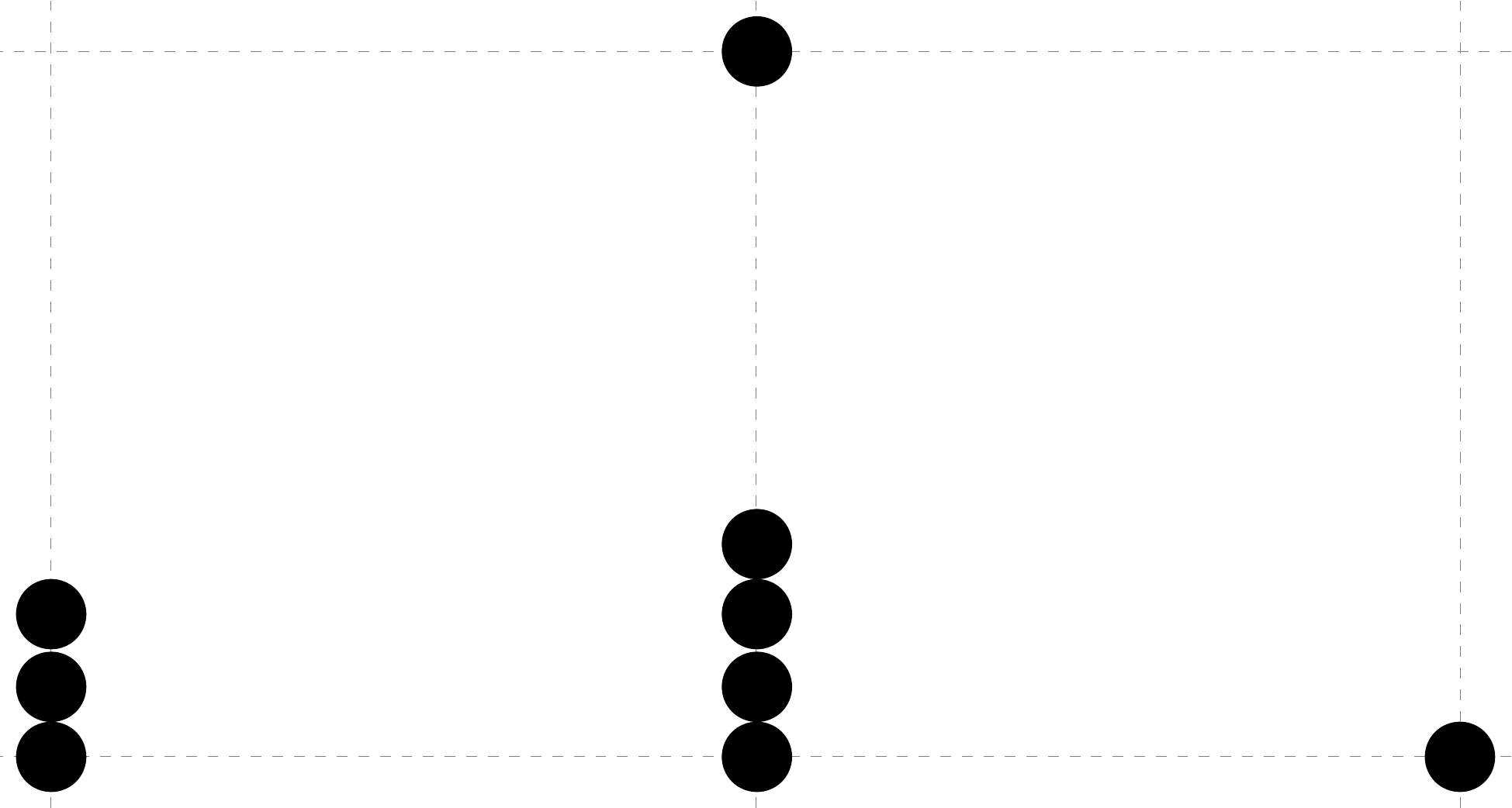}  \\
$ \BC^2/\BZ_2 \times \BC $ \\
(inc.)
\end{tabular}
&
\begin{tabular}[b]{c} 
$-X_{12}^{}.X_{23}^2.X_{31}^{}$ \\
$+X_{13}^{}.X_{34}^1.X_{41}^{}$ \\
$-X_{23}^1.X_{34}^1.X_{42}^{}$ \\
$+X_{23}^2.X_{34}^2.X_{42}^{}$ \\
$+\phi _1^{}.X_{12}^{}.X_{23}^1.X_{31}^{}$ \\
$-\phi _1^{}.X_{13}^{}.X_{34}^2.X_{41}^{}$ 
\end{tabular}
\\ \hline


(3.11) &
\includegraphics[width=3.0cm]{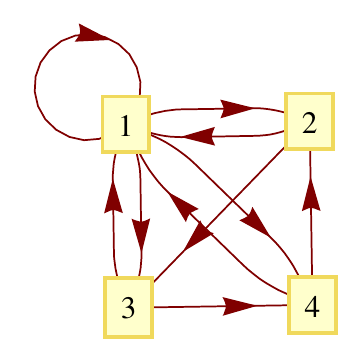} &
\includegraphics*[height=3.5cm]{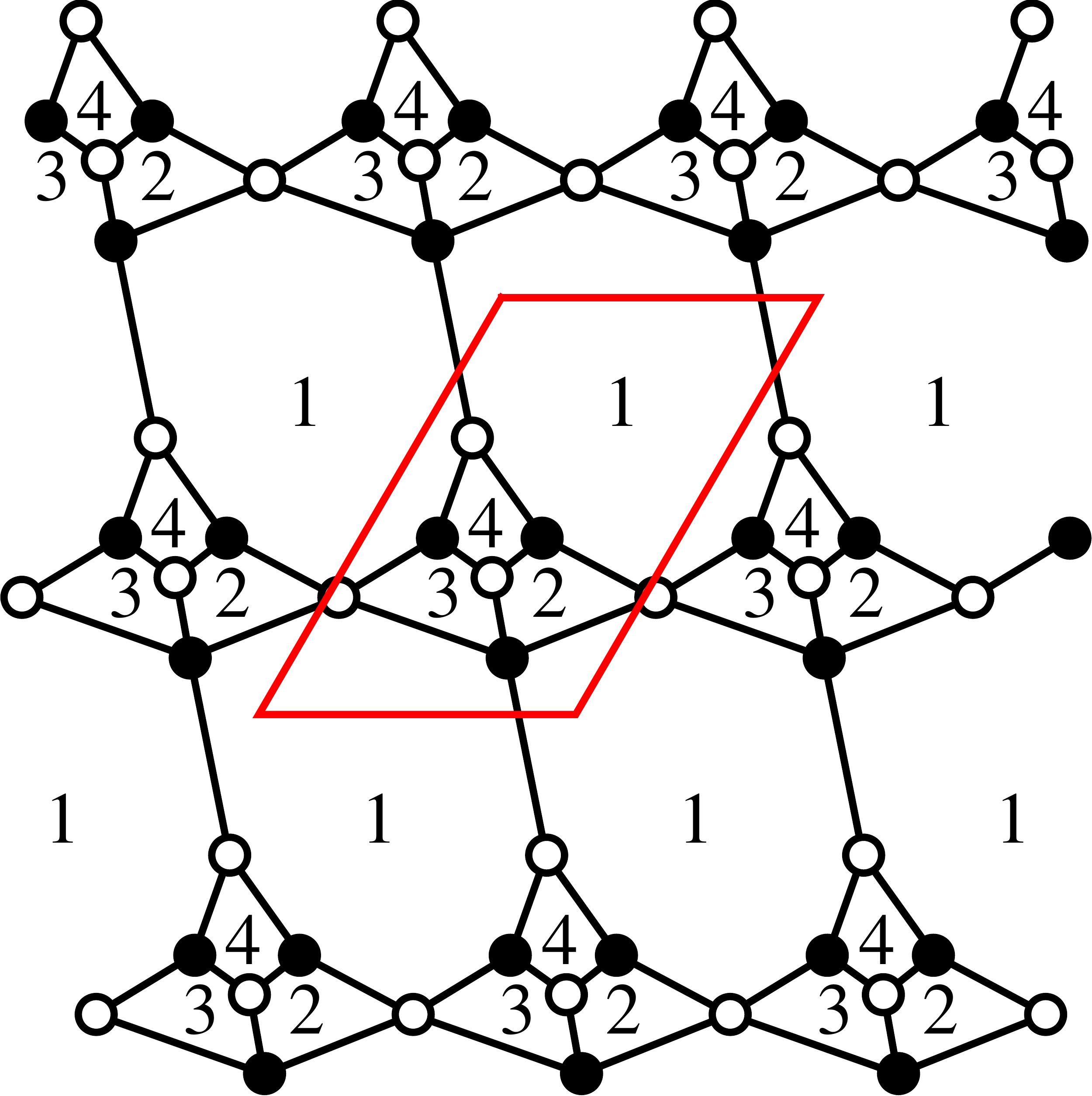} &
\begin{tabular}[b]{c}
\includegraphics[height=2.0cm]{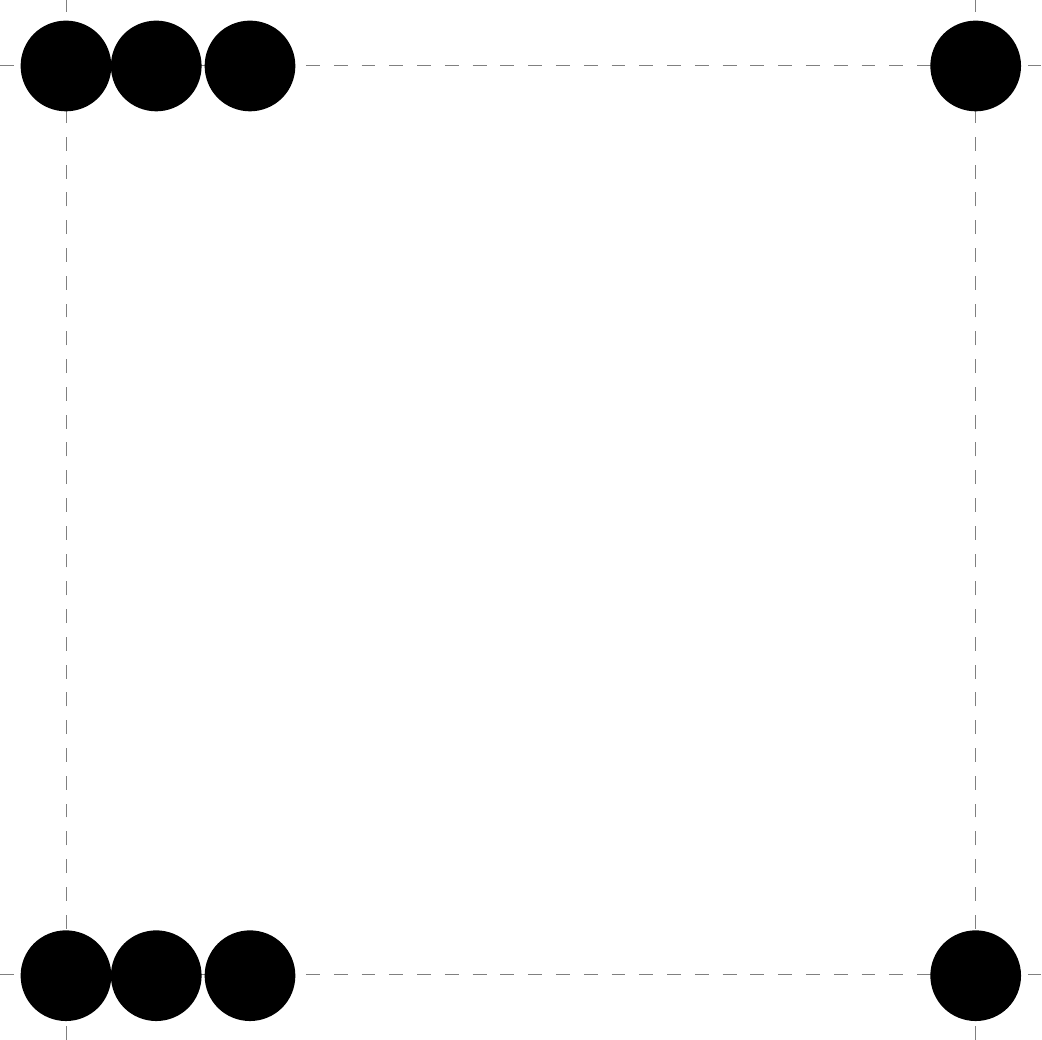}  \\
$ \mathcal{C} $ (inc.)
\end{tabular}
&
\begin{tabular}[b]{c} 
$-X_{13}^{}.X_{34}^{}.X_{41}^{}$ \\
$-X_{14}^{}.X_{42}^{}.X_{21}^{}$ \\
$+X_{23}^{}.X_{34}^{}.X_{42}^{}$ \\
$+\phi _1^{}.X_{14}^{}.X_{41}^{}$ \\
$+X_{12}^{}.X_{21}^{}.X_{13}^{}.X_{31}^{}$ \\
$-\phi _1^{}.X_{12}^{}.X_{23}^{}.X_{31}^{}$ 
\end{tabular}
\\ \hline

(3.12) &
\includegraphics[width=3.0cm]{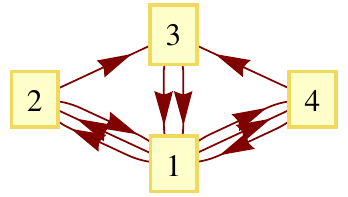} &
\includegraphics*[height=3.5cm]{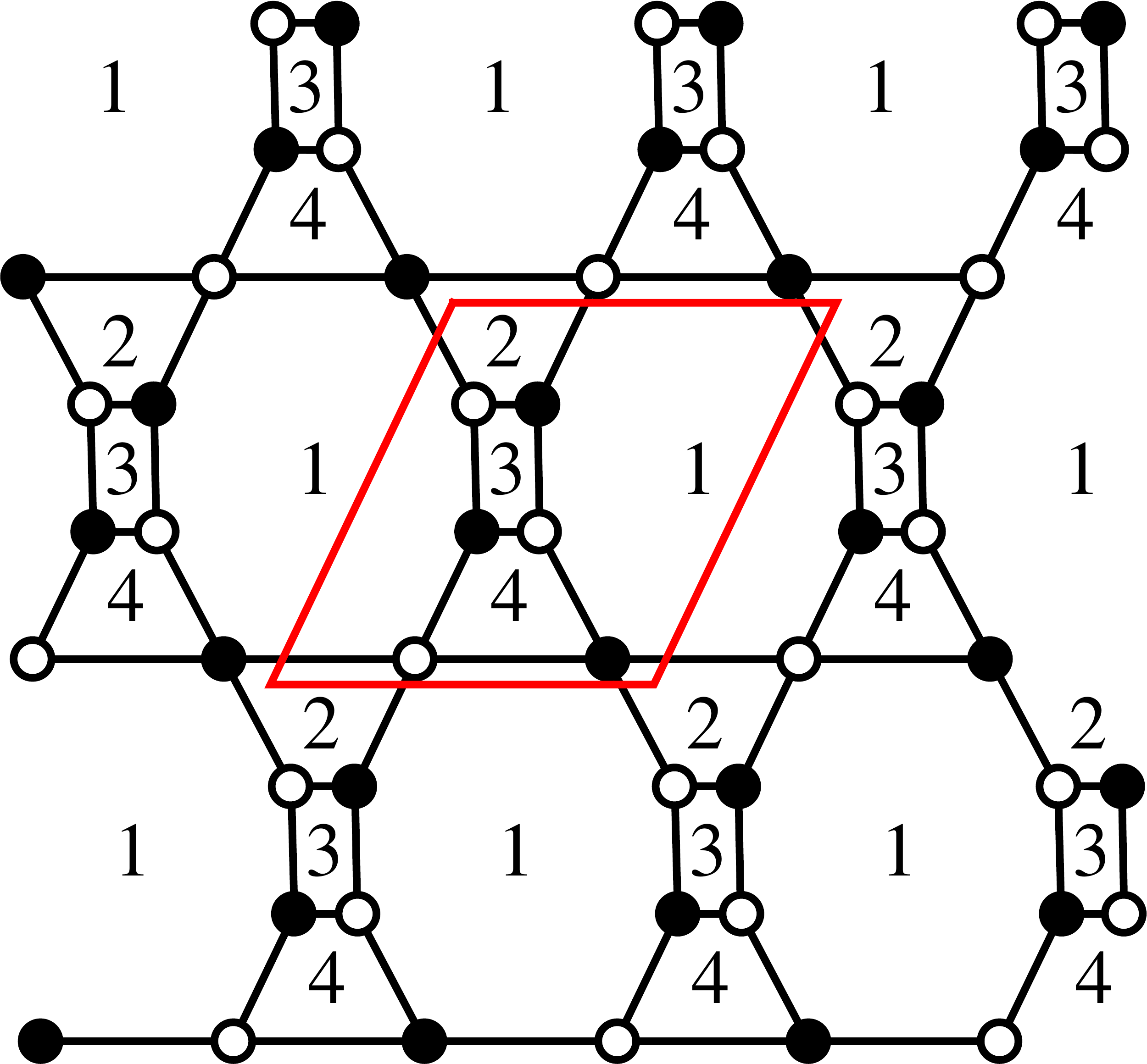} &
\begin{tabular}[b]{c}
\includegraphics[height=2.0cm]{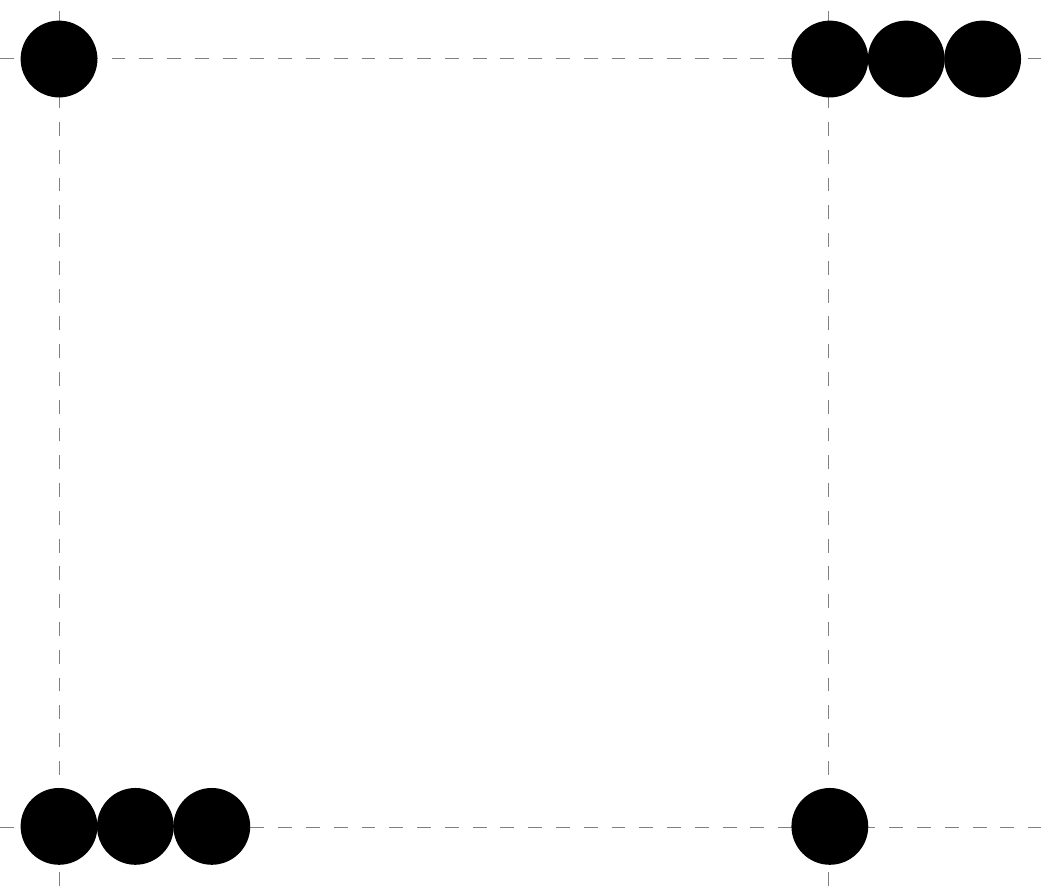} \\  
$ \CC $ (inc.)
\end{tabular}
&
\begin{tabular}[b]{c} 
$X_{12}^1.X_{23}^{}.X_{31}^1$ \\
$-X_{12}^2.X_{23}^{}.X_{31}^2$ \\
$-X_{14}^1.X_{43}^{}.X_{31}^1$ \\
$+X_{14}^2.X_{43}^{}.X_{31}^2$ \\
$-X_{12}^1.X_{21}^{}.X_{14}^2.X_{41}^{}$ \\
$+X_{12}^2.X_{21}^{}.X_{14}^1.X_{41}^{}$
\end{tabular}

\end{tabular}
\end{center}

\caption{Tilings with 6 superpotential terms and 4 gauge groups \bf{(page 2/2)}}
\label{t:tilings6-4}
\end{table}


\begin{table}[h]

\begin{center}
\begin{tabular}{c|c|c|c|c}
\# & Quiver & Tiling & Toric Diagram & Superpotential\\
\hline \hline

(3.13) &
\includegraphics[width=3.0cm]{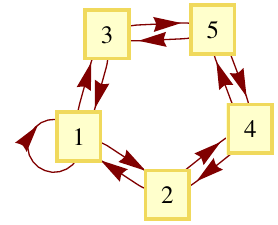} &
\includegraphics*[height=3.5cm]{N6-G5-1-tiling.pdf} &
\begin{tabular}[b]{c}
\includegraphics[height=2.4cm,angle=90]{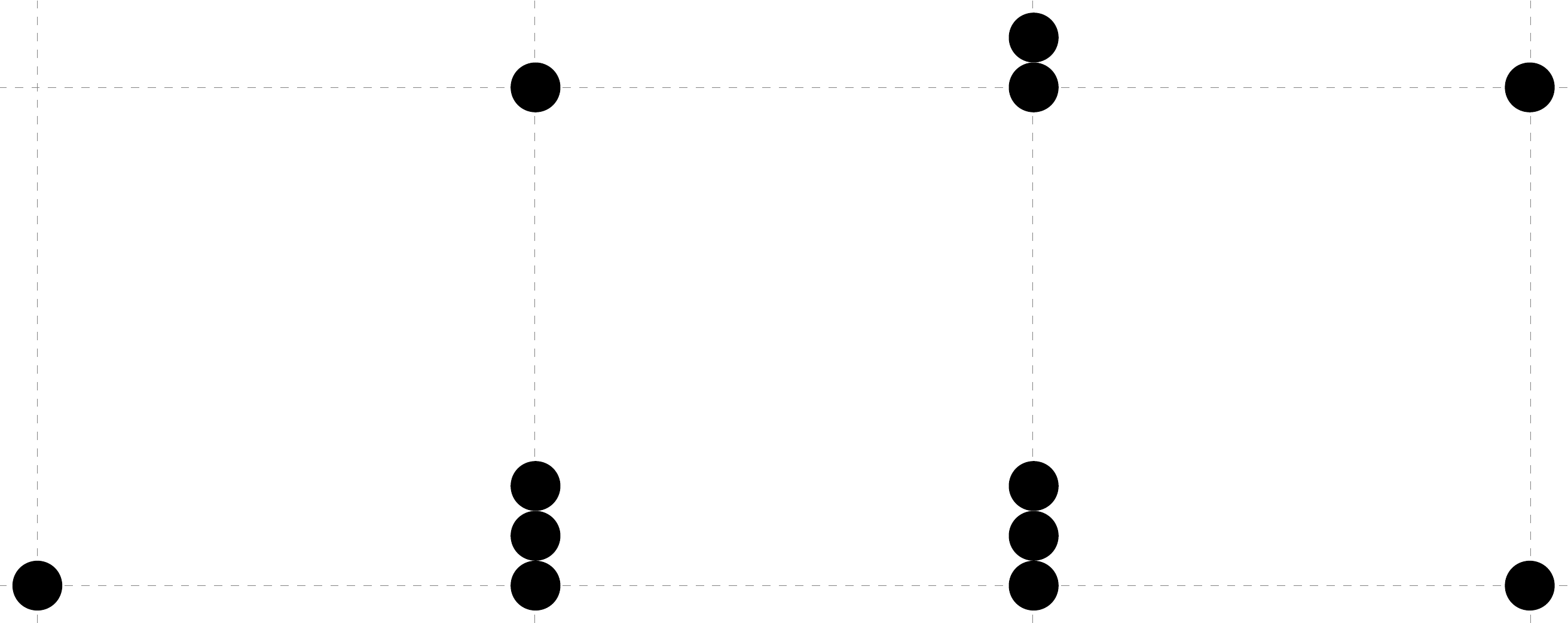}  \\
$ L^{232} $ (I)
\end{tabular}
&
\begin{tabular}[b]{c} 
$\phi _1^{}.X_{12}^{}.X_{21}^{}$ \\
$-\phi _1^{}.X_{13}^{}.X_{31}^{}$ \\
$-X_{12}^{}.X_{24}^{}.X_{42}^{}.X_{21}^{}$ \\
$+X_{13}^{}.X_{35}^{}.X_{53}^{}.X_{31}^{}$ \\
$+X_{24}^{}.X_{45}^{}.X_{54}^{}.X_{42}^{}$ \\
$-X_{35}^{}.X_{54}^{}.X_{45}^{}.X_{53}^{}$
\end{tabular}
\\ \hline

(3.14) &
\includegraphics[width=3.0cm]{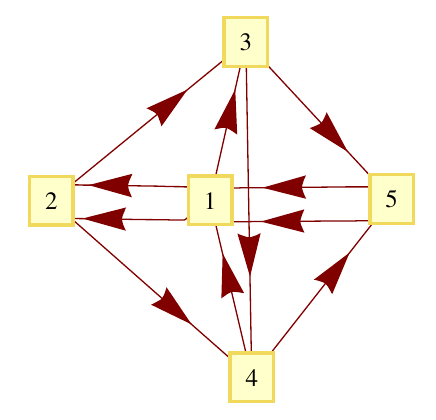} &
\includegraphics*[height=3.5cm]{N6-G5-7-tiling.pdf} &
\begin{tabular}[b]{c}
\includegraphics[height=2.4cm]{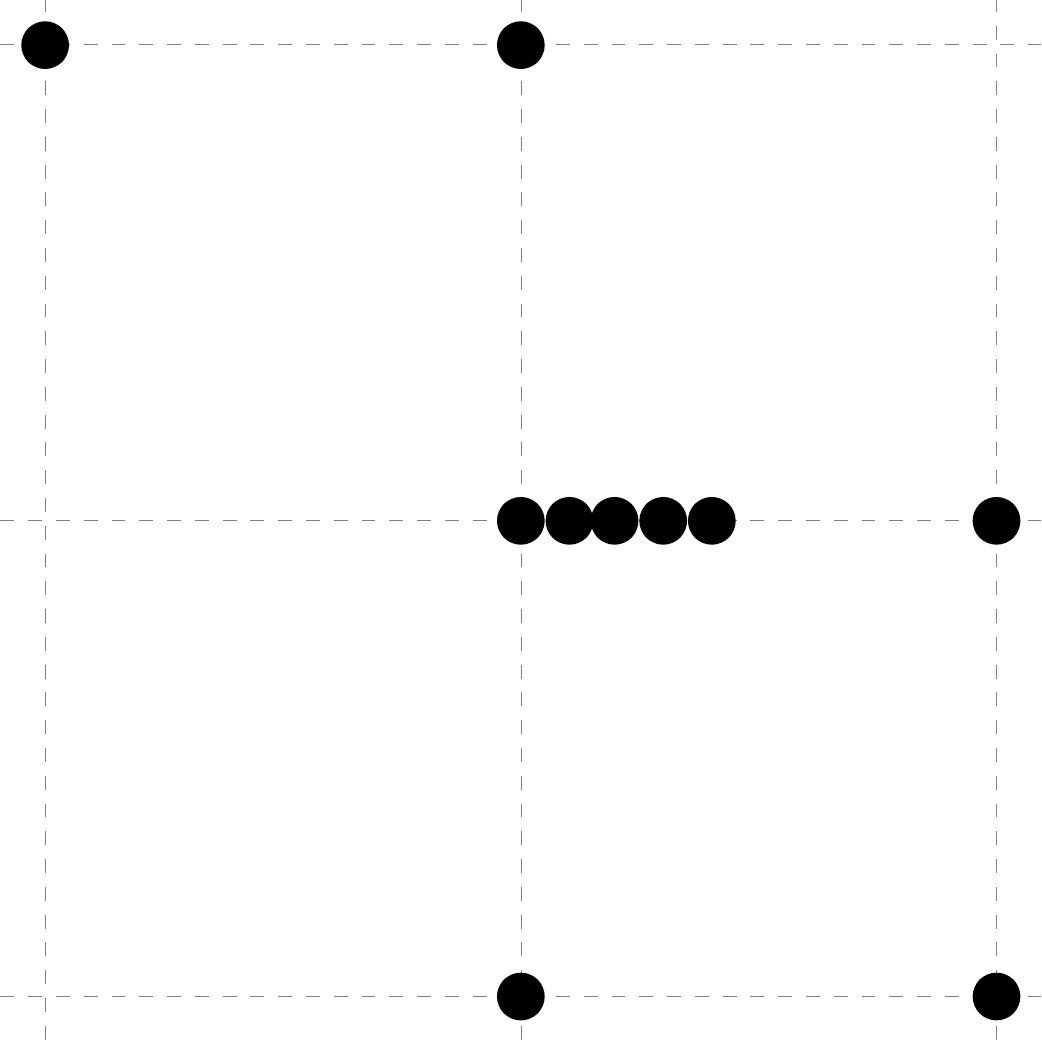} \\
$ dP_2 $ (I)
\end{tabular}
&
\begin{tabular}[b]{c} 
$X_{12}^1.X_{24}^{}.X_{41}^{}$ \\
$-X_{13}^{}.X_{34}^{}.X_{41}^{}$ \\
$+X_{13}^{}.X_{35}^{}.X_{51}^2$ \\
$-X_{12}^1.X_{23}^{}.X_{35}^{}.X_{51}^1$ \\
$-X_{12}^2.X_{24}^{}.X_{45}^{}.X_{51}^2$ \\
$+X_{12}^2.X_{23}^{}.X_{34}^{}.X_{45}^{}.X_{51}^1$
\end{tabular}
\\ \hline

(3.15) &
\includegraphics[width=3.0cm]{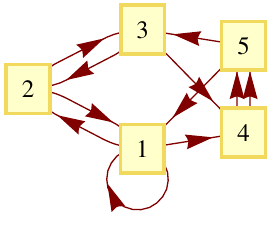} &
\includegraphics*[height=3.5cm]{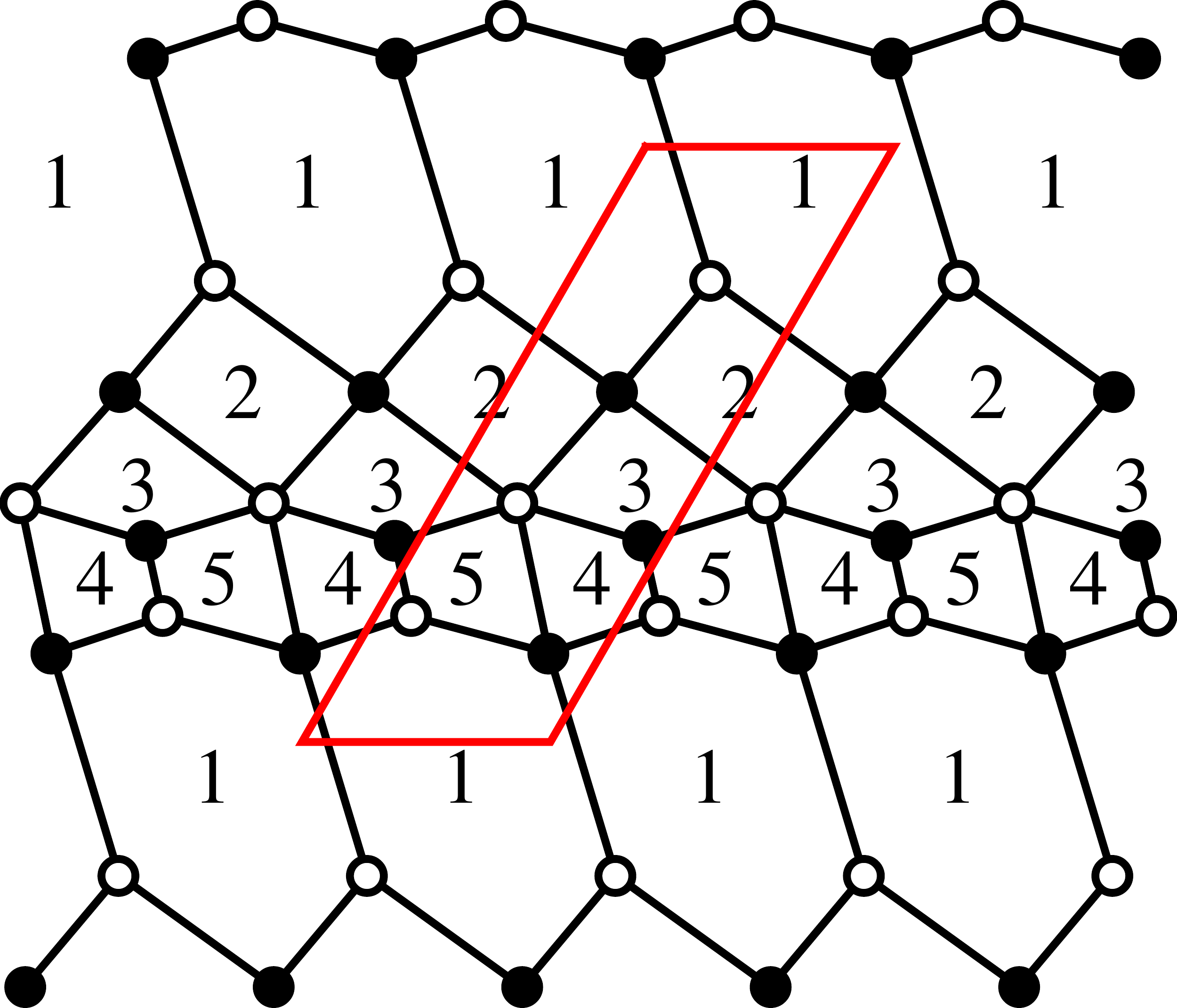} &
\begin{tabular}[b]{c}
\includegraphics[height=2.4cm,angle=90]{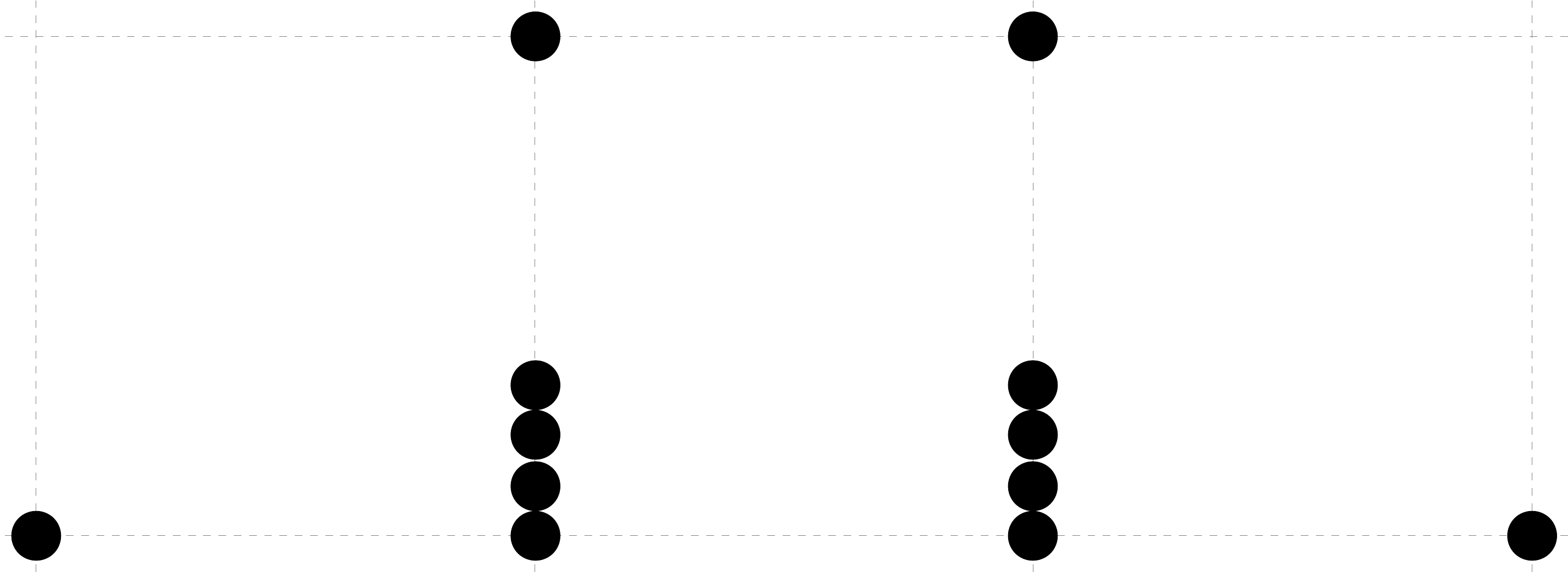}  \\
$ L^{131} $ (inc.)
\end{tabular}
&
\begin{tabular}[b]{c} 
$X_{14}^{}.X_{45}^1.X_{51}^{}$ \\
$-X_{34}^{}.X_{45}^1.X_{53}^{}$ \\
$+\phi _1^{}.X_{12}^{}.X_{21}^{}$ \\
$-X_{12}^{}.X_{23}^{}.X_{32}^{}.X_{21}^{}$ \\
$-\phi _1^{}.X_{14}^{}.X_{45}^2.X_{51}^{}$ \\
$+X_{23}^{}.X_{34}^{}.X_{45}^2.X_{53}^{}.X_{32}^{}$
\end{tabular}
\\ \hline

(3.16) &
\includegraphics[width=3.0cm]{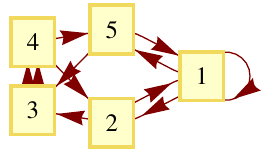} &
\includegraphics*[height=3.5cm]{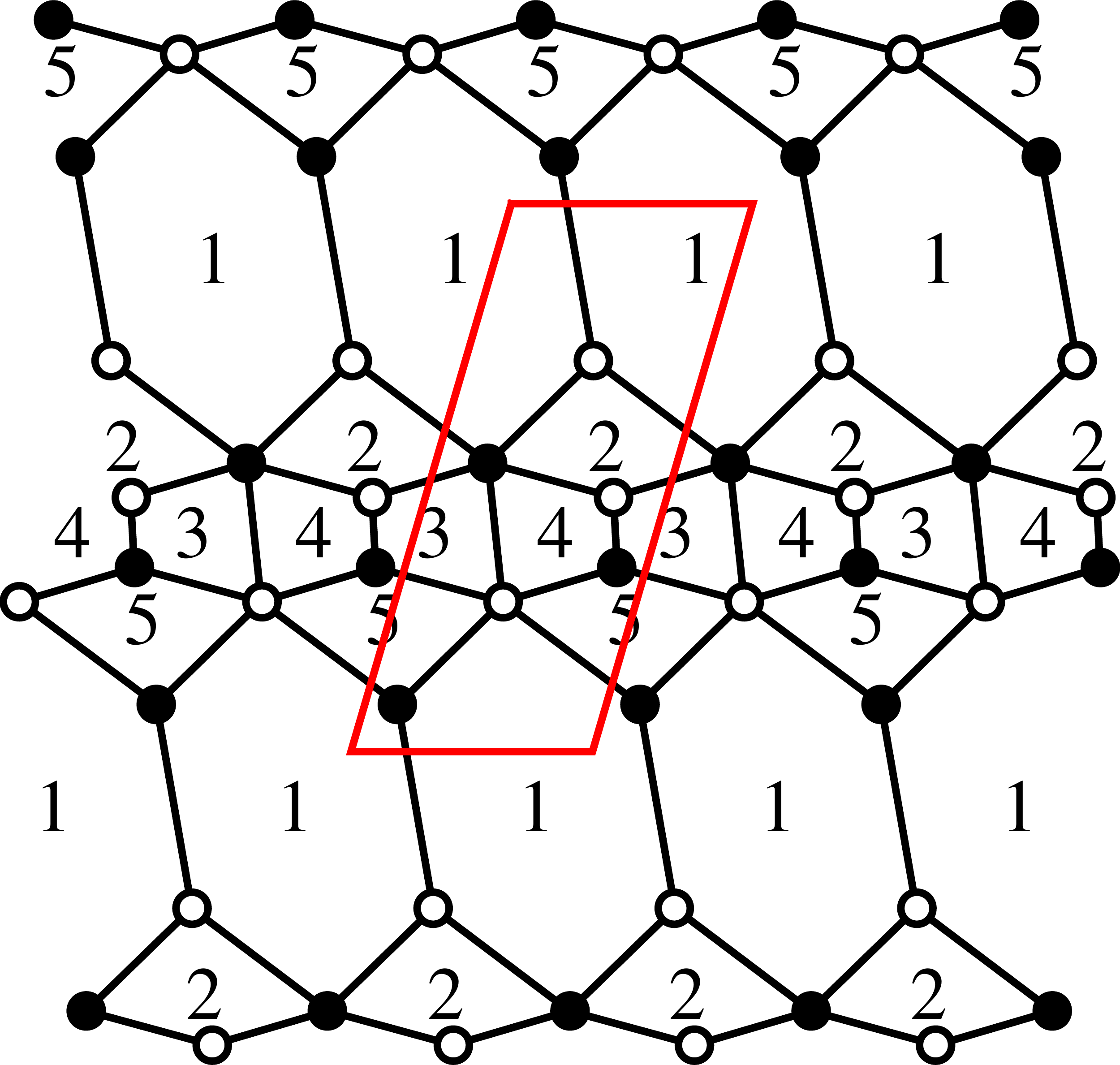} &
\begin{tabular}[b]{c}
\includegraphics[width=2.4cm]{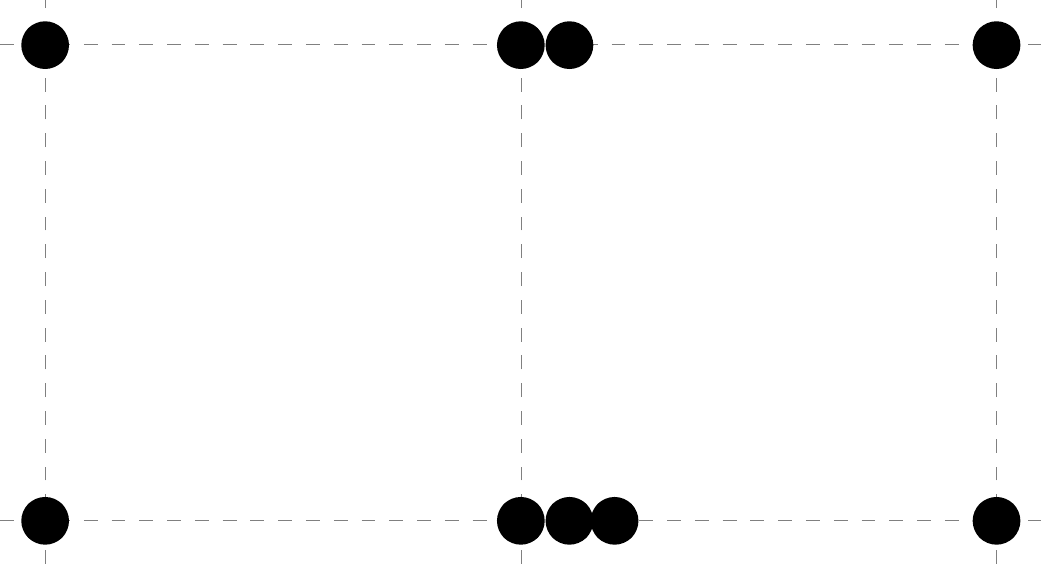}  \\
$ L^{222} $ (inc.)
\end{tabular}
&
\begin{tabular}[b]{c} 
$X_{23}^{}.X_{34}^2.X_{42}^{}$ \\
$-X_{34}^2.X_{45}^{}.X_{53}^{}$ \\
$+\phi _1^{}.X_{12}^{}.X_{21}^{}$ \\
$-\phi _1^{}.X_{15}^{}.X_{51}^{}$ \\
$-X_{12}^{}.X_{23}^{}.X_{34}^1.X_{42}^{}.X_{21}^{}$ \\
$+X_{15}^{}.X_{53}^{}.X_{34}^1.X_{45}^{}.X_{51}^{}$
\end{tabular}
\\ \hline

(3.17) &
\includegraphics[width=3.0cm]{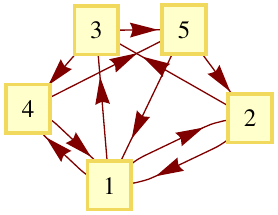} &
\includegraphics*[height=3.5cm]{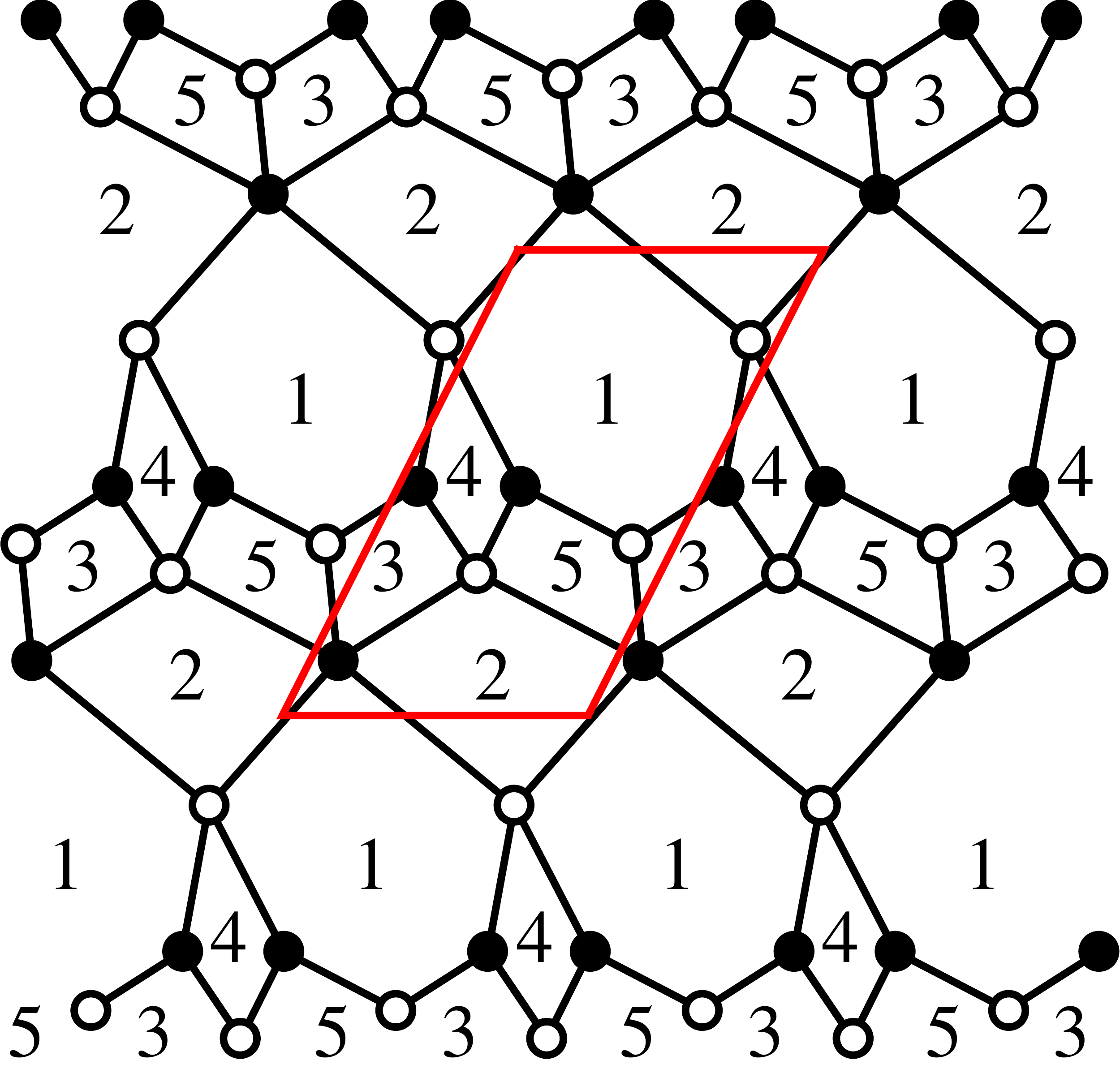} &
\begin{tabular}[b]{c}
\includegraphics[height=2.4cm,angle=90]{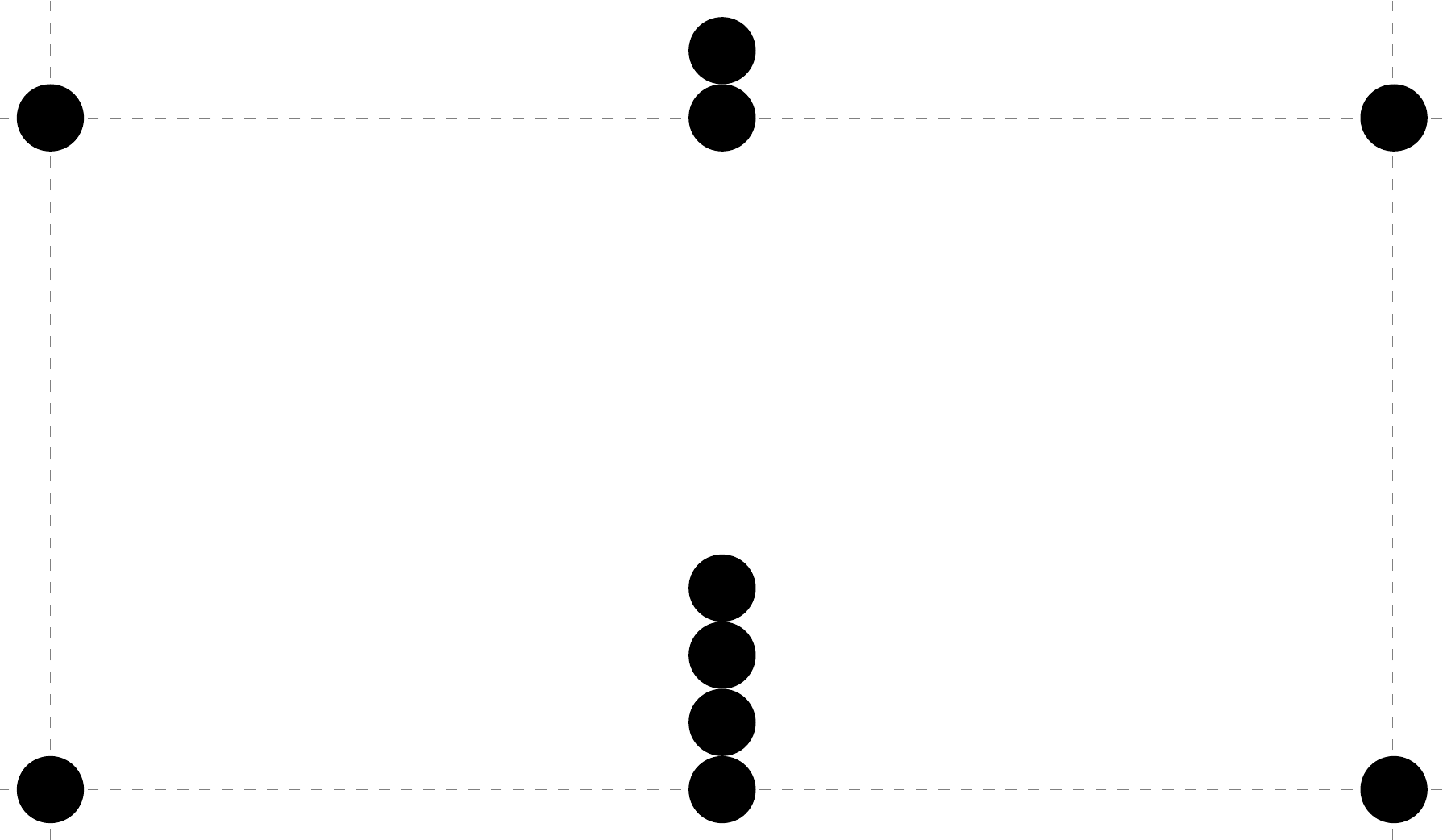} \\
$ L^{222} $ (inc.)
\end{tabular}
 &
\begin{tabular}[b]{c} 
$-X_{13}^{}.X_{34}^{}.X_{41}^{}$ \\
$+X_{13}^{}.X_{35}^{}.X_{51}^{}$ \\
$-X_{14}^{}.X_{45}^{}.X_{51}^{}$ \\
$+X_{12}^{}.X_{21}^{}.X_{14}^{}.X_{41}^{}$ \\
$+X_{23}^{}.X_{34}^{}.X_{45}^{}.X_{52}^{}$ \\
$-X_{12}^{}.X_{23}^{}.X_{35}^{}.X_{52}^{}.X_{21}^{}$
\end{tabular}

\end{tabular}
\end{center}

\caption{Tilings with 6 superpotential terms and 5 gauge groups \bf{(page 1/3)}}
\label{t:tilings6-5a}
\end{table}

\begin{table}[h]

\begin{center}
\begin{tabular}{c|c|c|c|c}
\# & Quiver & Tiling & Toric Diagram & Superpotential\\
\hline \hline

(3.18) &
\includegraphics[width=3.0cm]{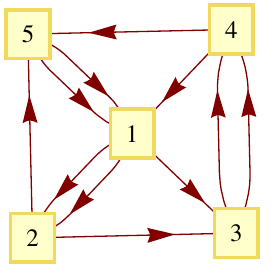} &
\includegraphics*[height=3.5cm]{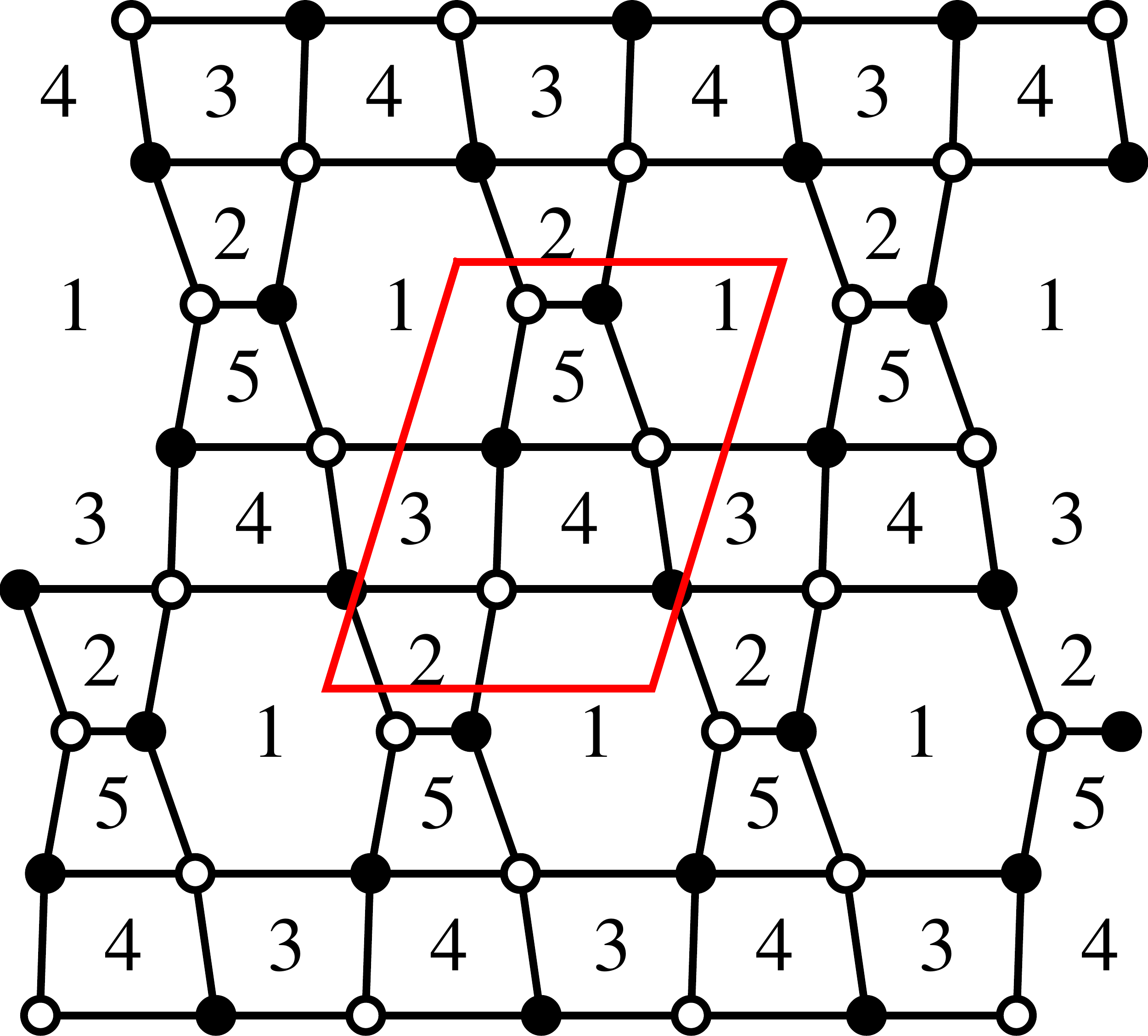} &
\begin{tabular}[b]{c}
\includegraphics[height=2.4cm]{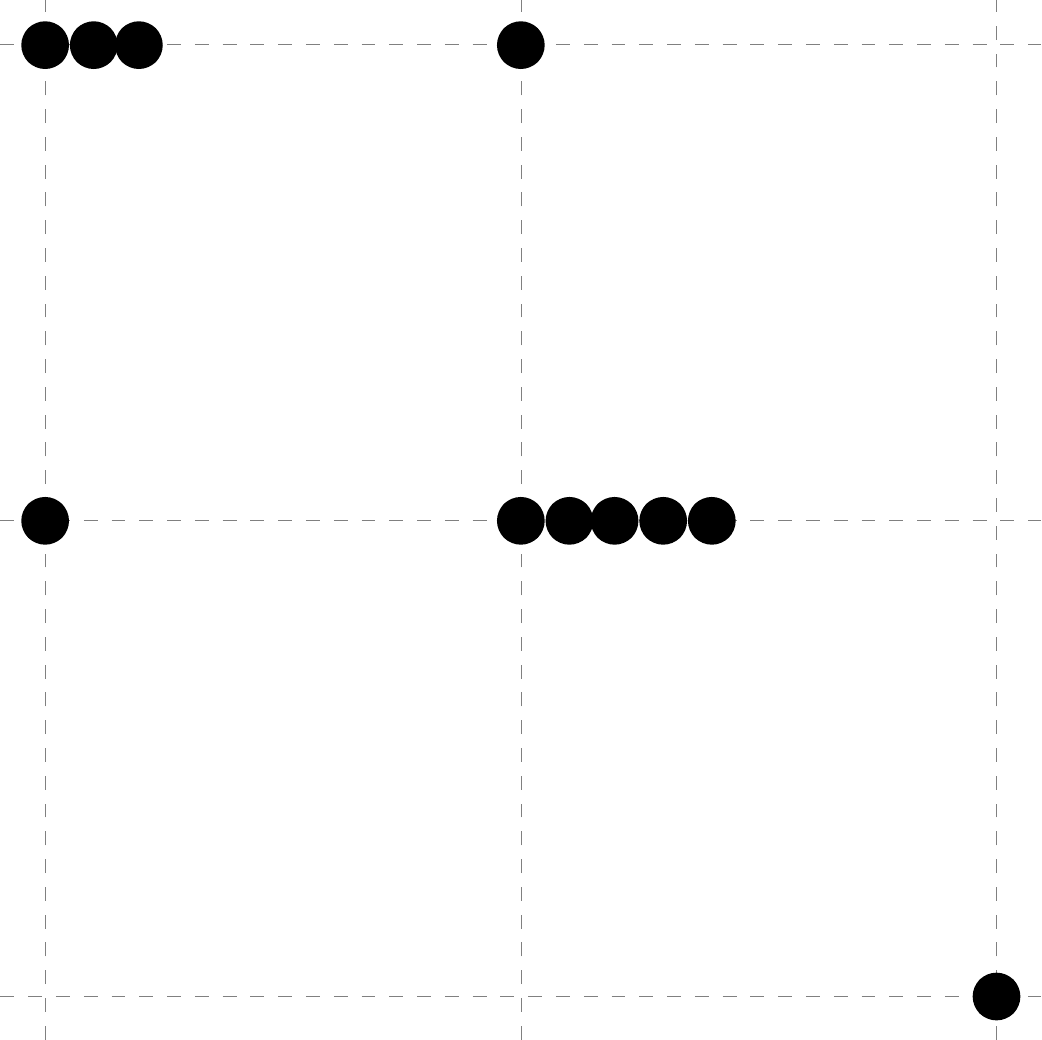} \\
$ dP_1 $ (inc.)
\end{tabular}
 &
\begin{tabular}[b]{c} 
$X_{12}^1.X_{25}^{}.X_{51}^1$ \\
$-X_{12}^2.X_{25}^{}.X_{51}^2$ \\
$-X_{12}^1.X_{23}^{}.X_{34}^2.X_{41}^{}$ \\
$+X_{12}^2.X_{23}^{}.X_{34}^1.X_{41}^{}$ \\
$-X_{13}^{}.X_{34}^1.X_{45}^{}.X_{51}^1$ \\
$+X_{13}^{}.X_{34}^2.X_{45}^{}.X_{51}^2$
\end{tabular}
\\ \hline

(3.19) &
\includegraphics[width=3.0cm]{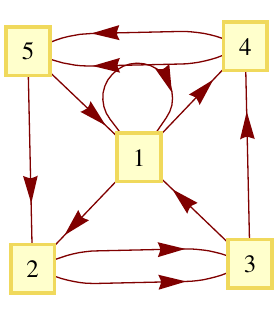} &
\includegraphics*[height=3.5cm]{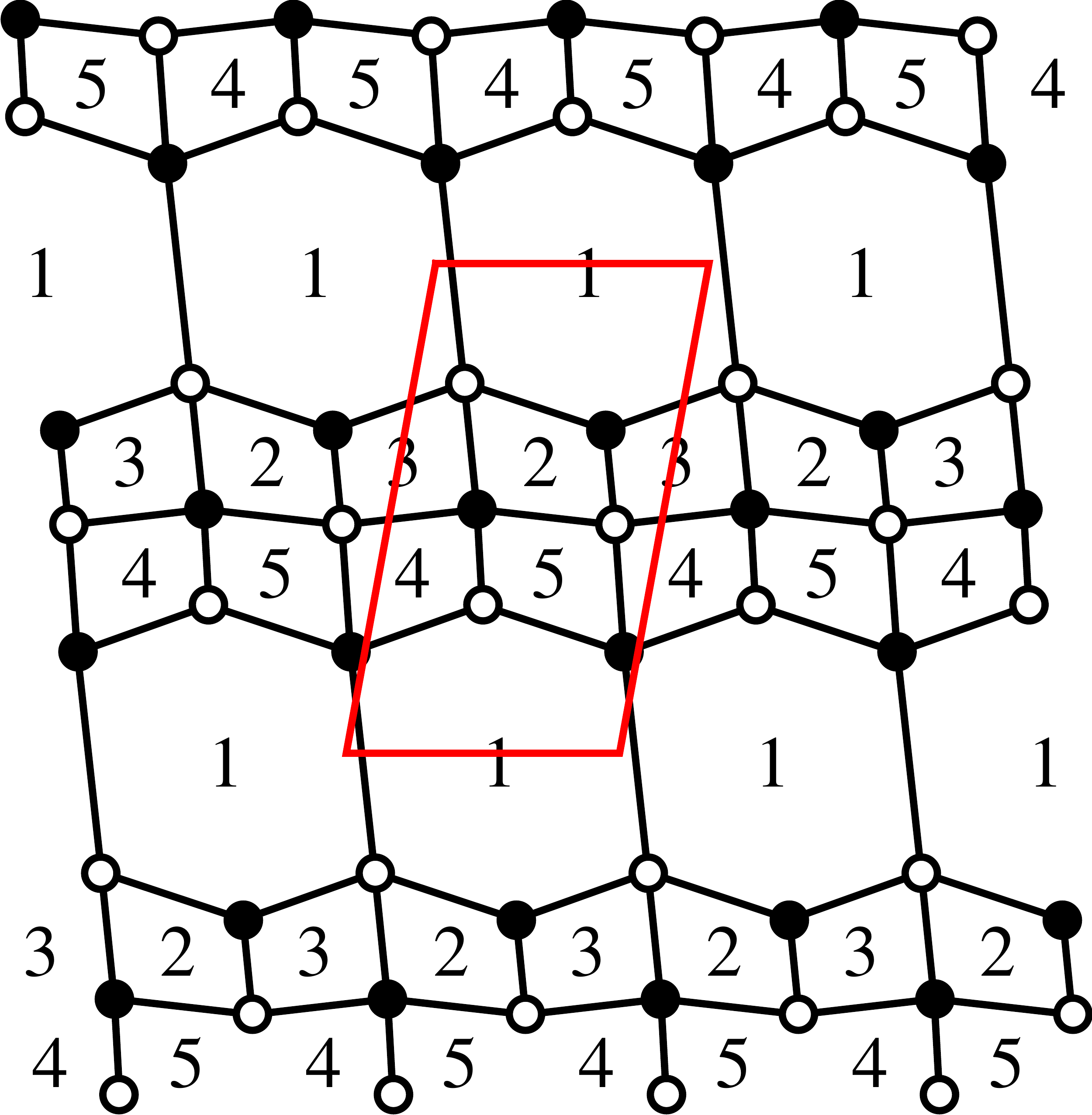} &
\begin{tabular}[b]{c}
\includegraphics[height=2.4cm,angle=90]{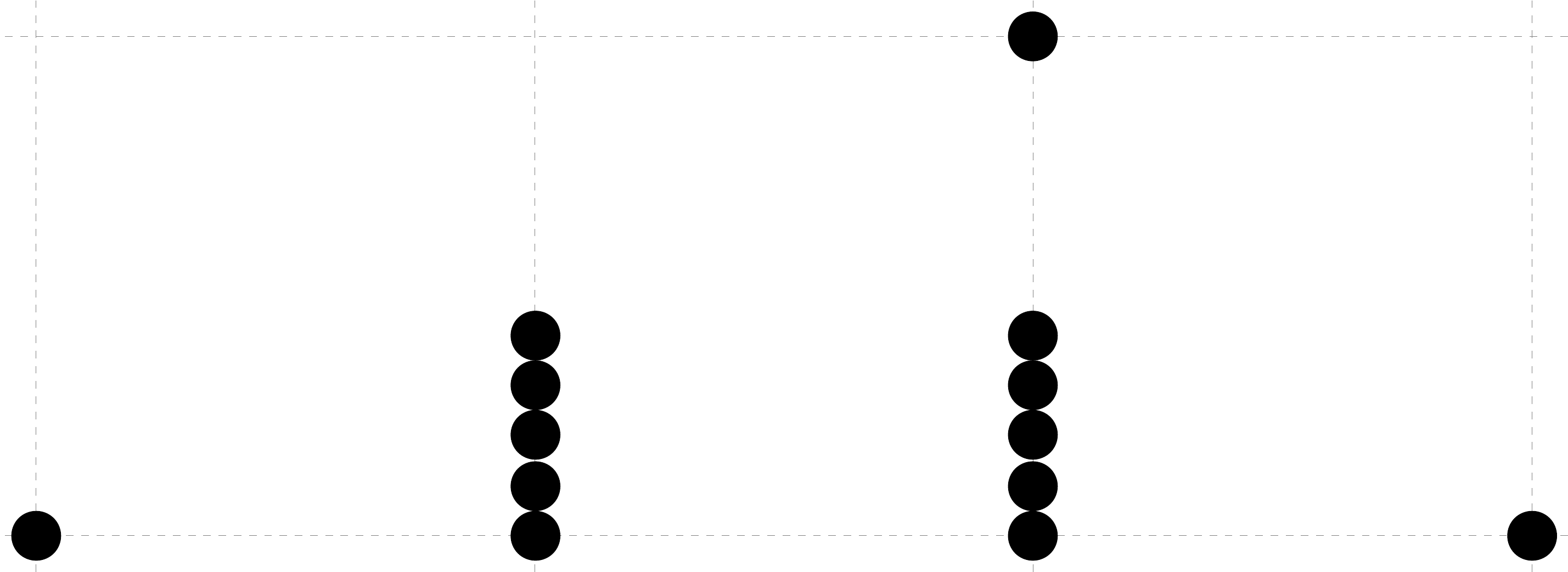}  \\
$ \BC^2 / \BZ_3 \times \BC $ \\
(inc.)
\end{tabular}
&
\begin{tabular}[b]{c} 
$-X_{12}^{}.X_{23}^2.X_{31}^{}$ \\
$+X_{14}^{}.X_{45}^1.X_{51}^{}$ \\
$-X_{23}^1.X_{34}^{}.X_{45}^1.X_{52}^{}$ \\
$+X_{23}^2.X_{34}^{}.X_{45}^2.X_{52}^{}$ \\
$+\phi _1^{}.X_{12}^{}.X_{23}^1.X_{31}^{}$ \\
$-\phi _1^{}.X_{14}^{}.X_{45}^2.X_{51}^{}$
\end{tabular}
\\ \hline

(3.20) &
\includegraphics[width=3.0cm]{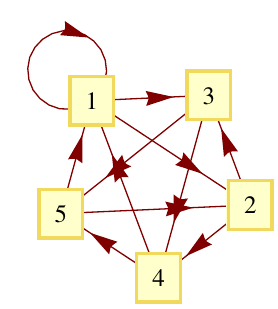} &
\includegraphics*[height=3.5cm]{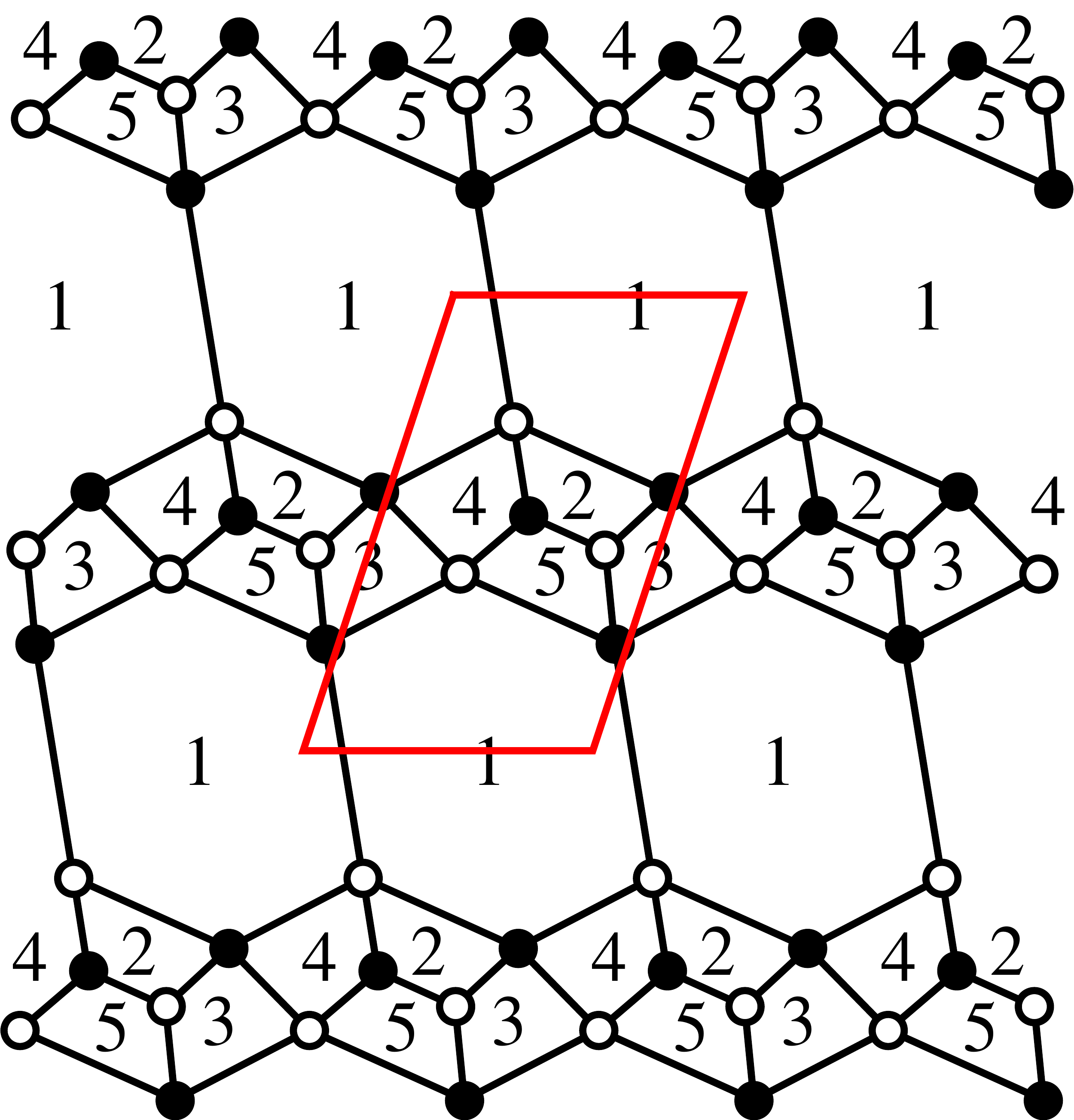} &
\begin{tabular}[b]{c}
\includegraphics[height=2.4cm,angle=90]{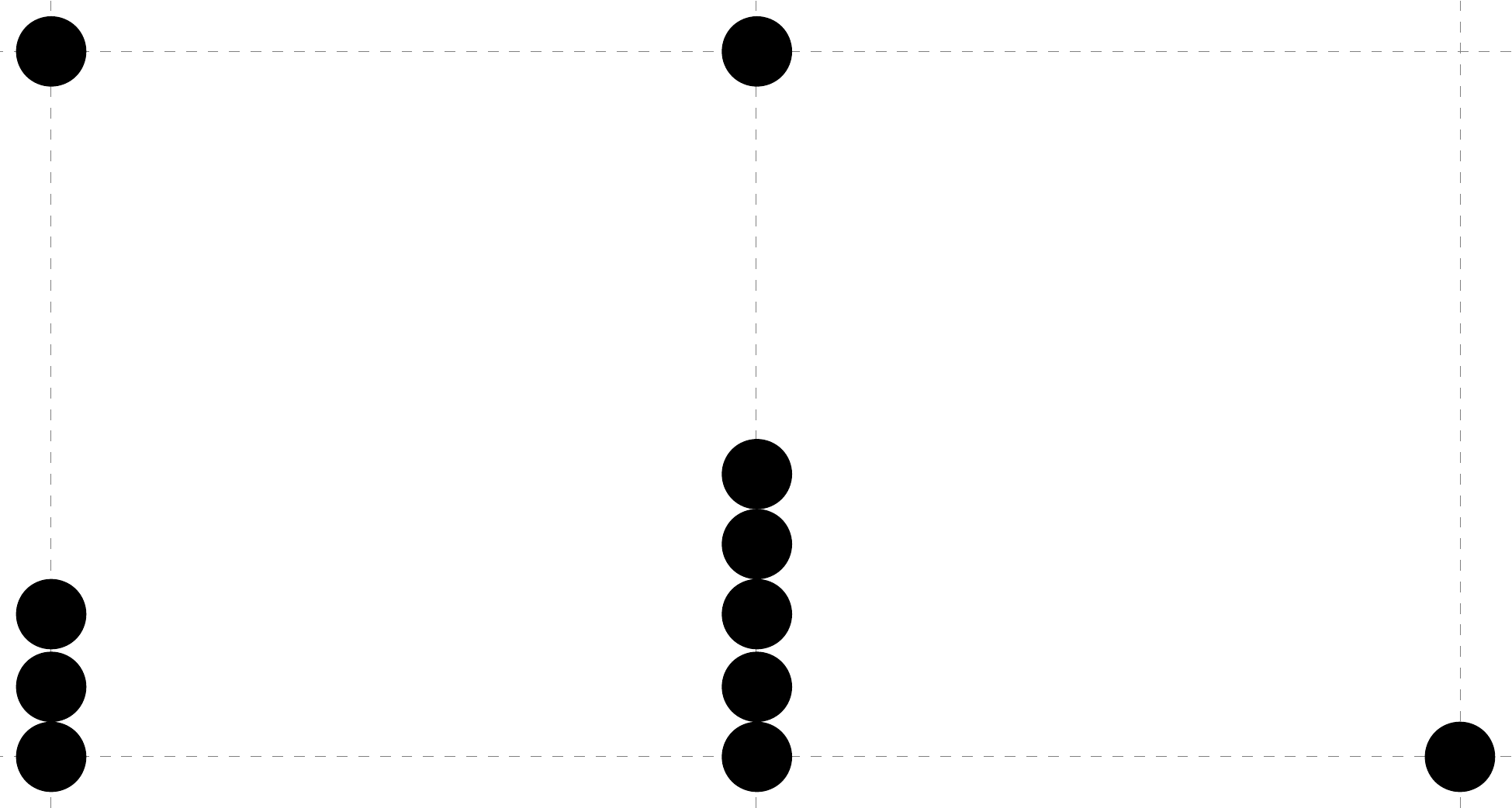}  \\
$ SPP $ (inc.)
\end{tabular}
&
\begin{tabular}[b]{c} 
$X_{23}^{}.X_{35}^{}.X_{52}^{}$ \\
$-X_{24}^{}.X_{45}^{}.X_{52}^{}$ \\
$-X_{12}^{}.X_{23}^{}.X_{34}^{}.X_{41}^{}$ \\
$+X_{13}^{}.X_{34}^{}.X_{45}^{}.X_{51}^{}$ \\
$+\phi _1^{}.X_{12}^{}.X_{24}^{}.X_{41}^{}$ \\
$-\phi _1^{}.X_{13}^{}.X_{35}^{}.X_{51}^{}$
\end{tabular}
\\ \hline

(3.21) &
\includegraphics[width=3.0cm]{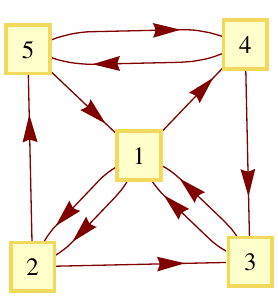} &
\includegraphics*[height=3.5cm]{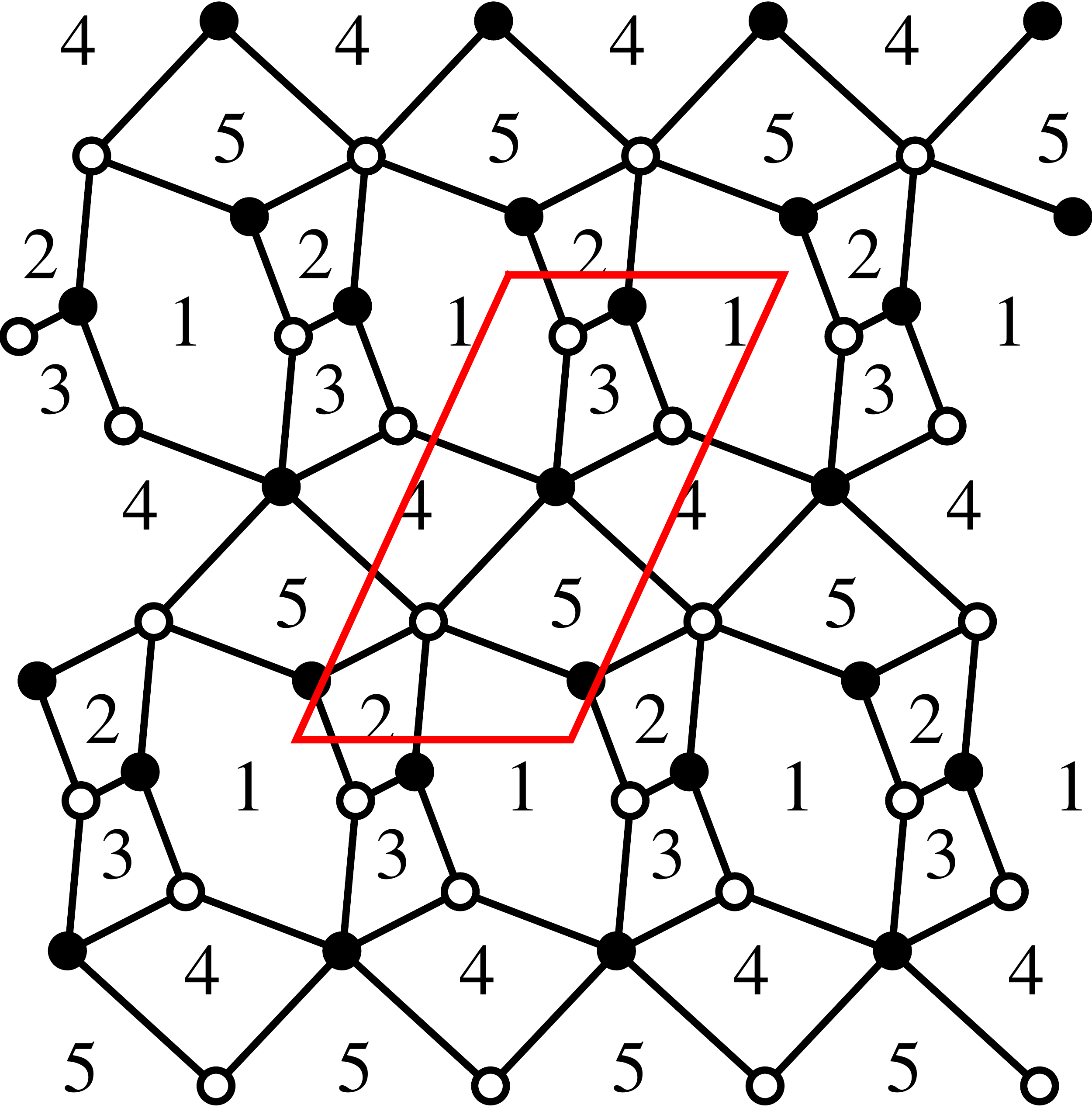} &
\begin{tabular}[b]{c}
\includegraphics[height=2.4cm,angle=90]{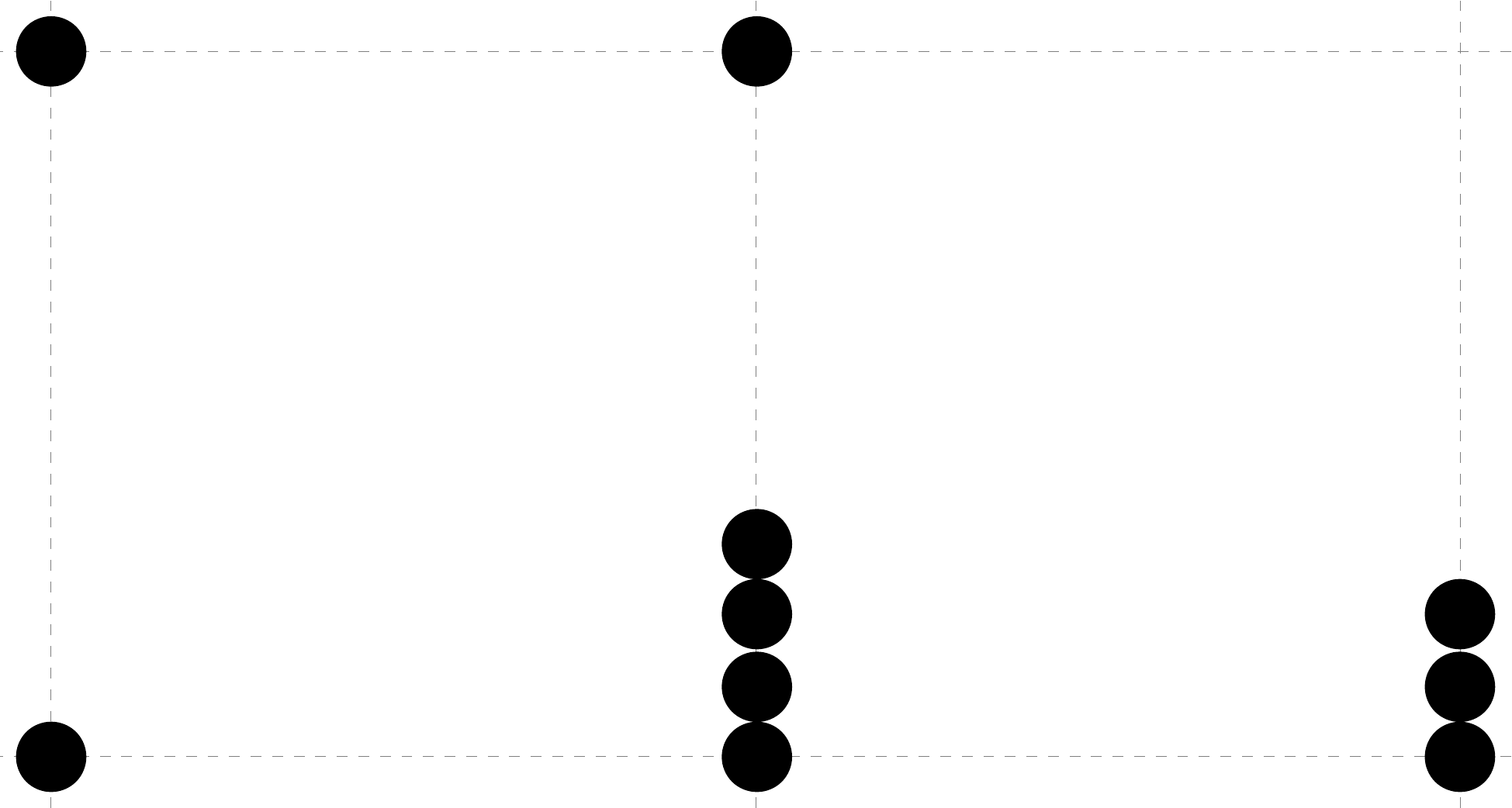} \\
$SPP$ (inc.)
\end{tabular}
 &
\begin{tabular}[b]{c} 
$X_{12}^1.X_{23}^{}.X_{31}^1$ \\
$-X_{12}^1.X_{25}^{}.X_{51}^{}$ \\
$-X_{12}^2.X_{23}^{}.X_{31}^2$ \\
$+X_{14}^{}.X_{43}^{}.X_{31}^2$ \\
$+X_{12}^2.X_{25}^{}.X_{54}^{}.X_{45}^{}.X_{51}^{}$ \\
$-X_{14}^{}.X_{45}^{}.X_{54}^{}.X_{43}^{}.X_{31}^1$
\end{tabular}
\\ \hline

(3.22) &
\includegraphics[width=3.0cm]{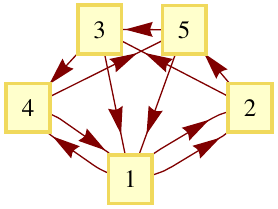} &
\includegraphics*[height=3.5cm]{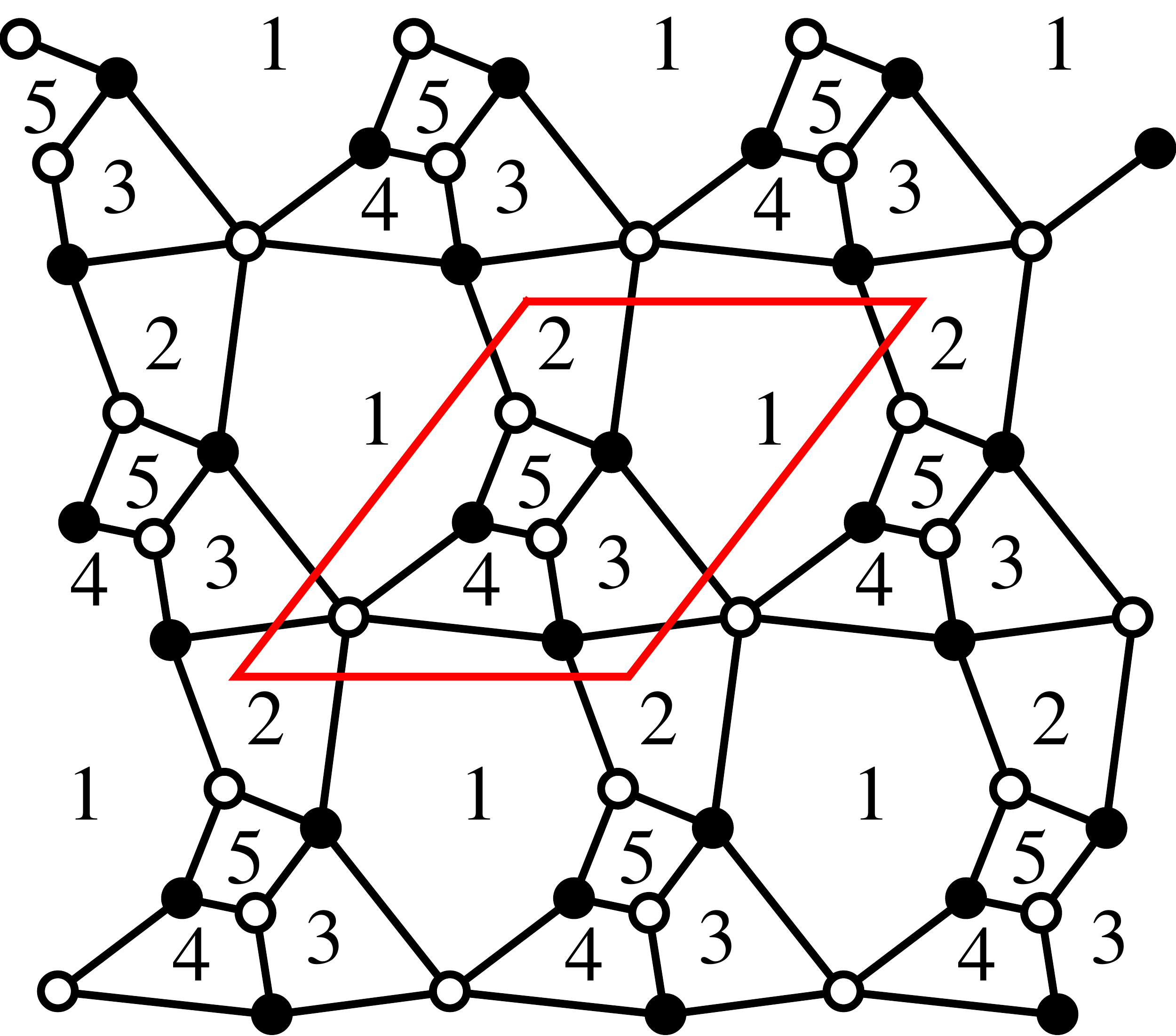} &
\begin{tabular}[b]{c}
\includegraphics[height=2.4cm,angle=90]{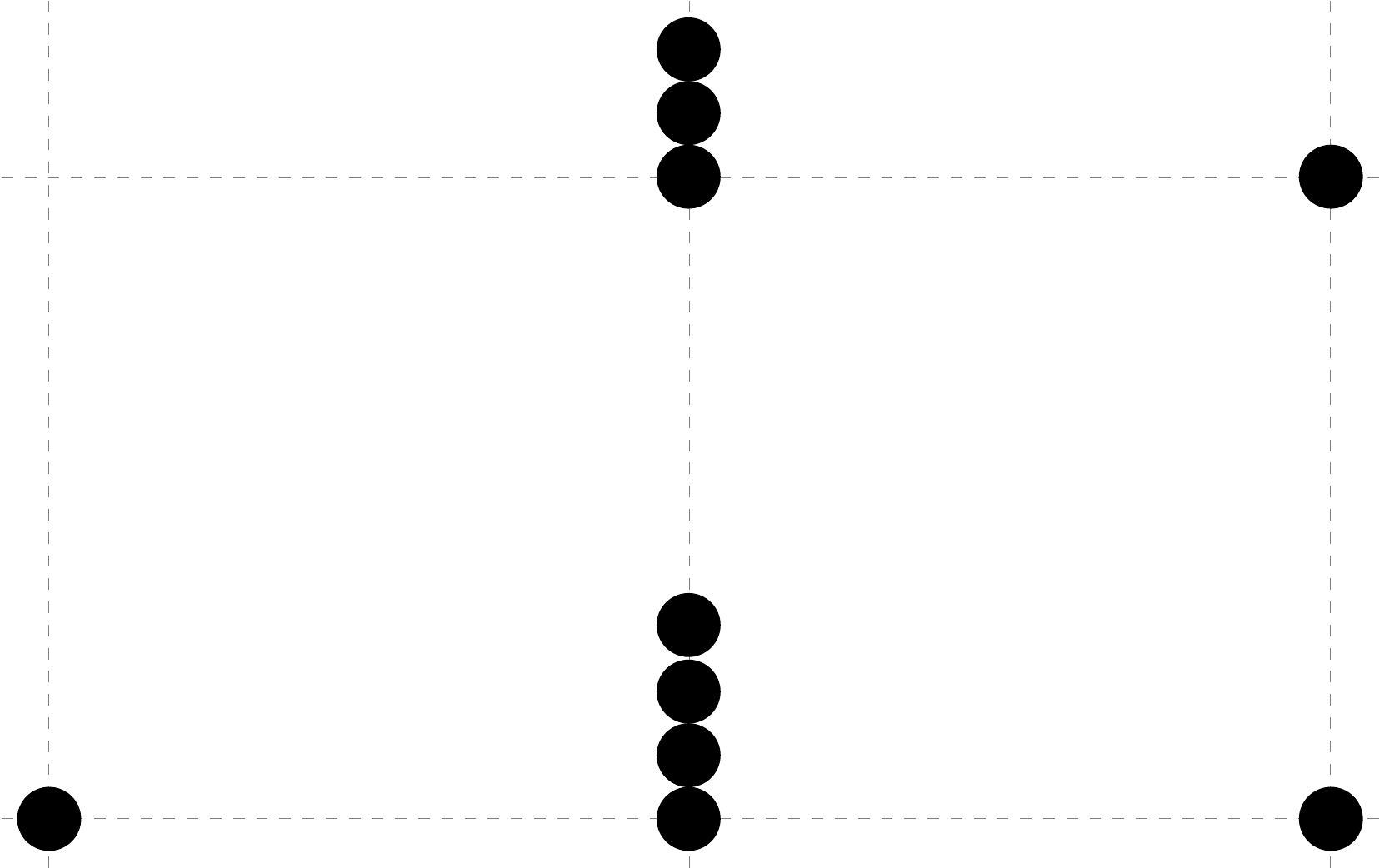} \\
$SPP$ (inc.)
\end{tabular}
 &
\begin{tabular}[b]{c} 
$X_{12}^1.X_{25}^{}.X_{51}^{}$ \\
$-X_{14}^{}.X_{45}^{}.X_{51}^{}$ \\
$+X_{34}^{}.X_{45}^{}.X_{53}^{}$ \\
$-X_{12}^1.X_{23}^{}.X_{34}^{}.X_{41}^{}$ \\
$-X_{12}^2.X_{25}^{}.X_{53}^{}.X_{31}^{}$ \\
$+X_{12}^2.X_{23}^{}.X_{31}^{}.X_{14}^{}.X_{41}^{}$
\end{tabular}

\end{tabular}
\end{center}

\caption{Tilings with 6 superpotential terms and 5 gauge groups \bf{(page 2/3)}}
\label{t:tilings6-5b}
\end{table}

\begin{table}[h]

\begin{center}
\begin{tabular}{c|c|c|c|c}
\# & Quiver & Tiling & Toric Diagram & Superpotential\\
\hline \hline

(3.23) &
\includegraphics[width=3.0cm]{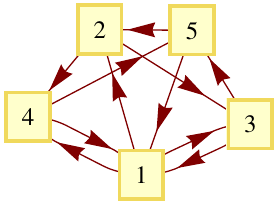} &
\includegraphics*[height=3.5cm]{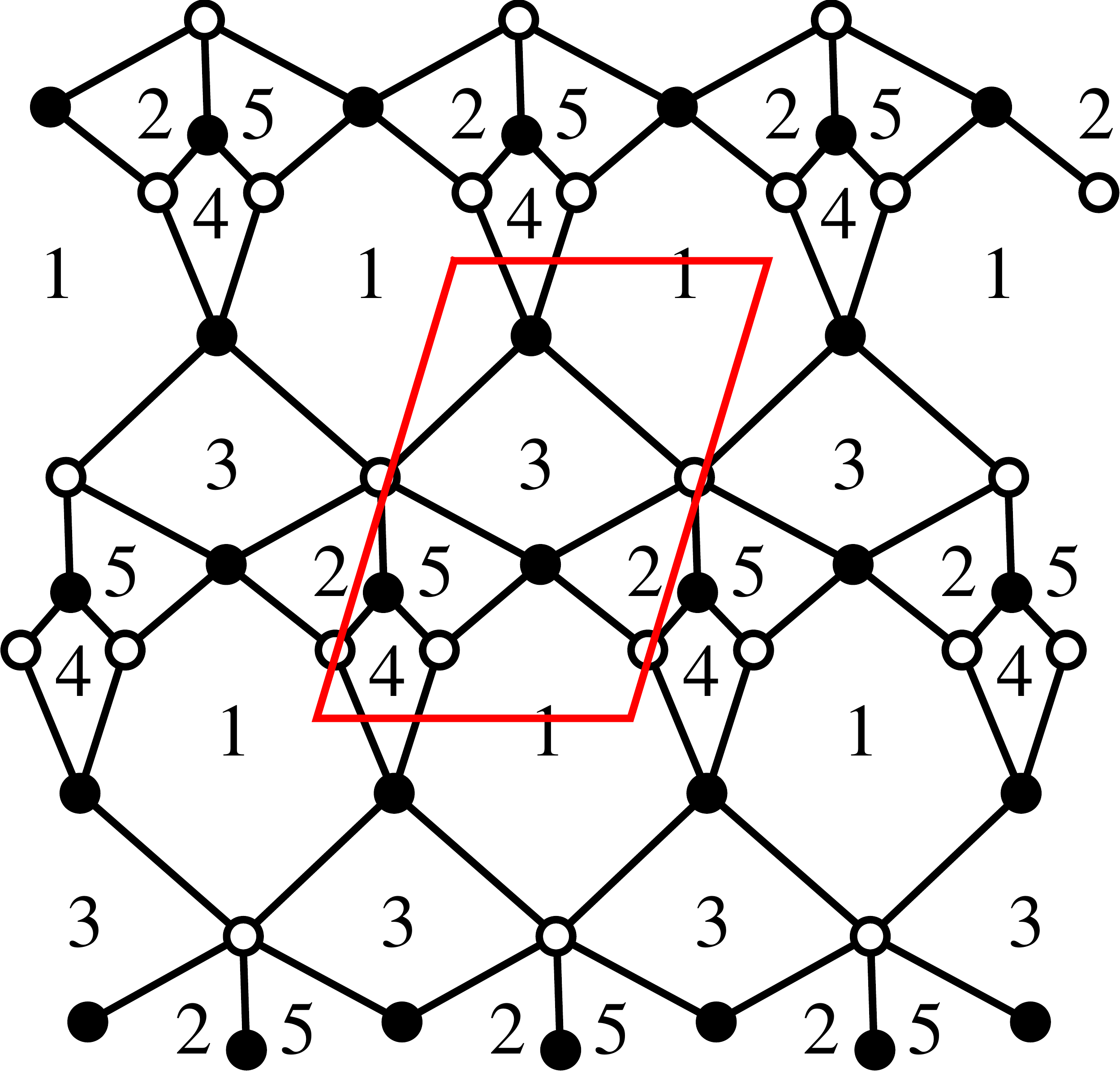} &
\begin{tabular}[b]{c}
\includegraphics[height=2.4cm,angle=90]{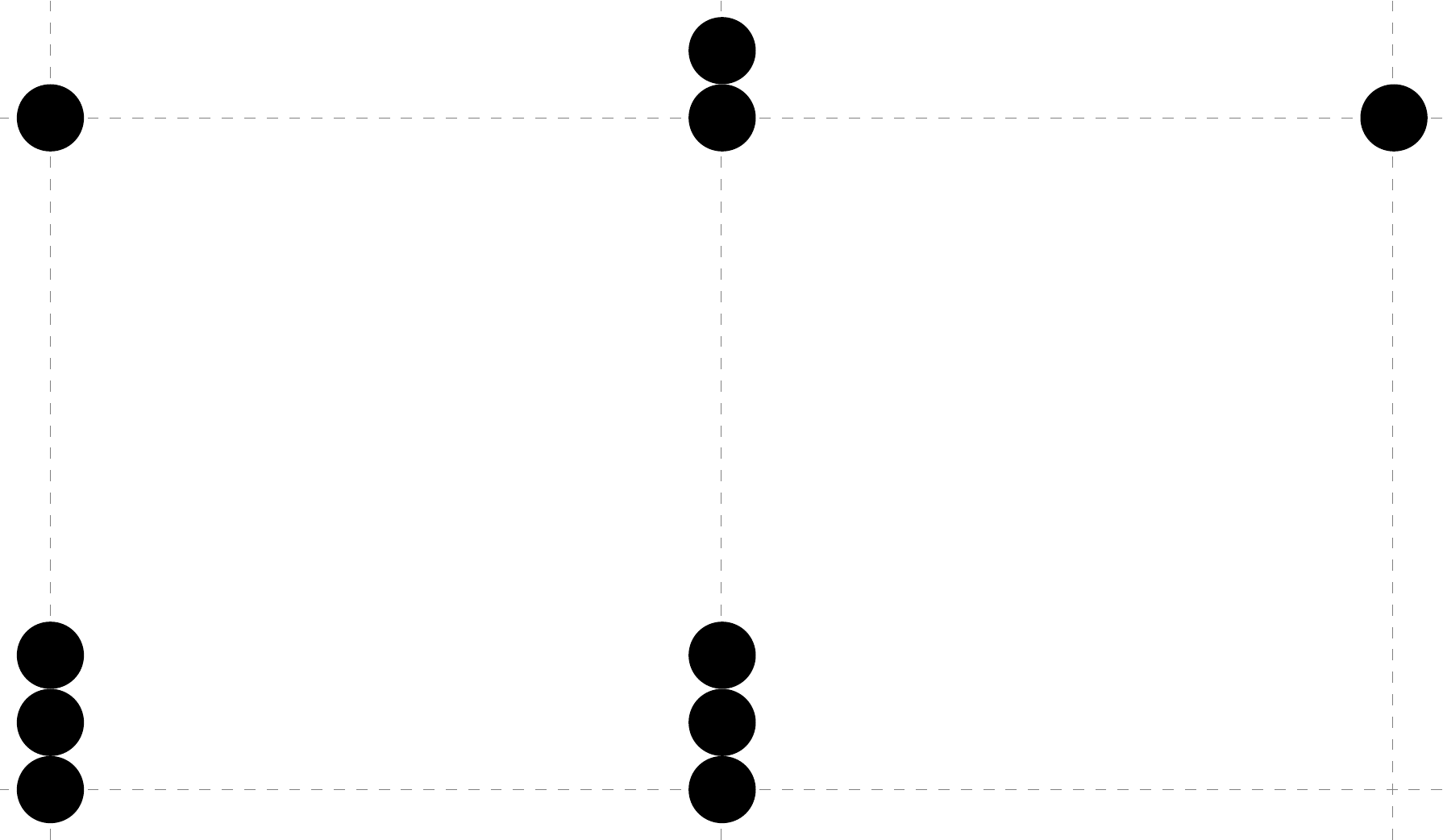} \\
$SPP$ (inc.)
\end{tabular}
 &
\begin{tabular}[b]{c} 
$X_{12}^{}.X_{24}^{}.X_{41}^{}$ \\
$+X_{14}^{}.X_{45}^{}.X_{51}^{}$ \\
$-X_{24}^{}.X_{45}^{}.X_{52}^{}$ \\
$-X_{12}^{}.X_{23}^{}.X_{35}^{}.X_{51}^{}$ \\
$-X_{13}^{}.X_{31}^{}.X_{14}^{}.X_{41}^{}$ \\
$+X_{13}^{}.X_{35}^{}.X_{52}^{}.X_{23}^{}.X_{31}^{}$
\end{tabular}
\\ \hline

(3.24) &
\includegraphics[width=3.0cm]{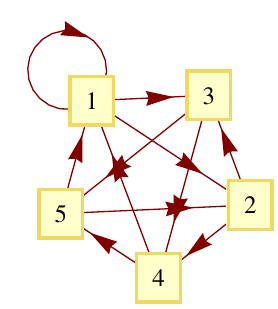} &
\includegraphics*[height=3.5cm]{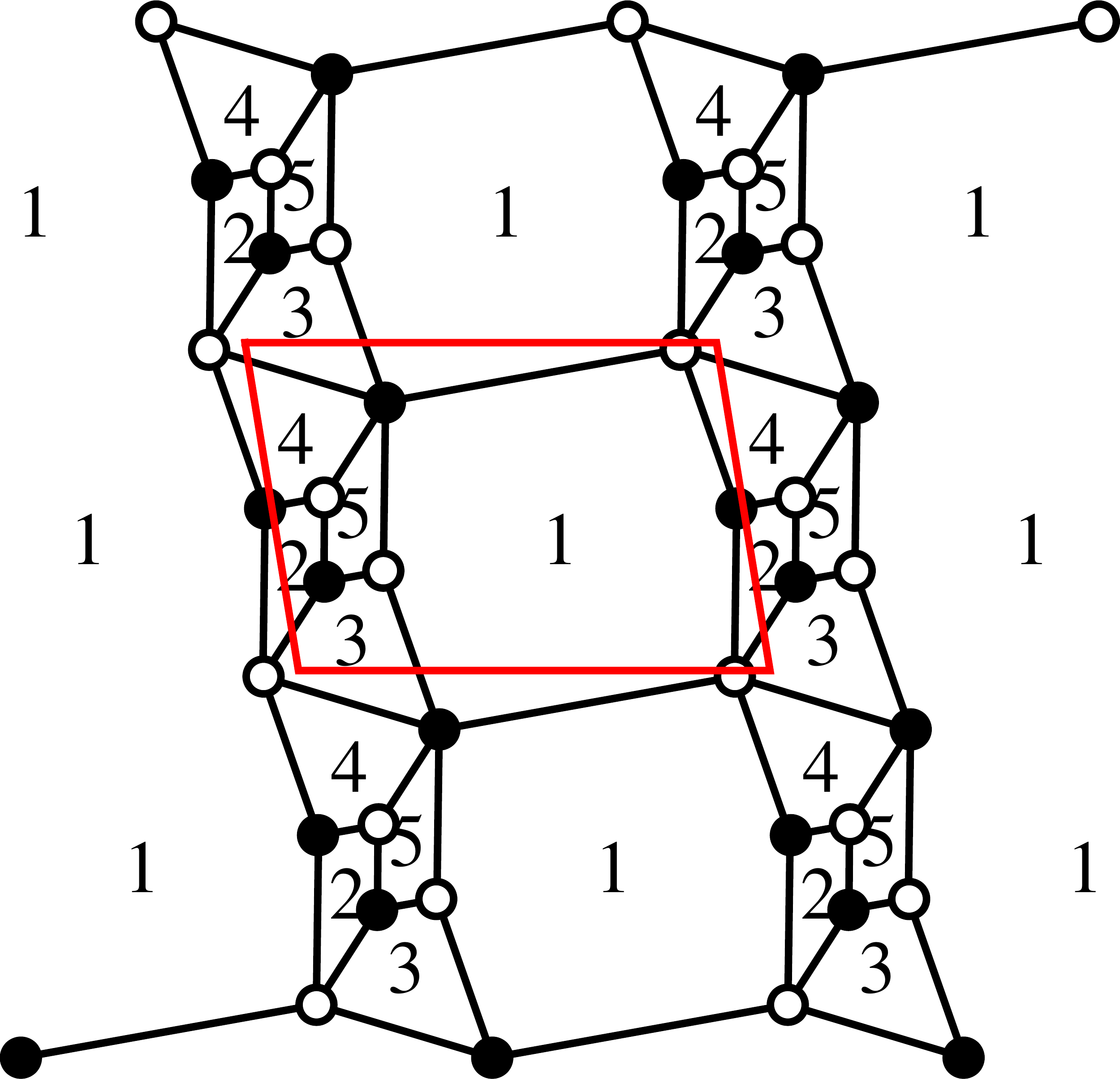} &
\begin{tabular}[b]{c}
\includegraphics[height=2.4cm,angle=90]{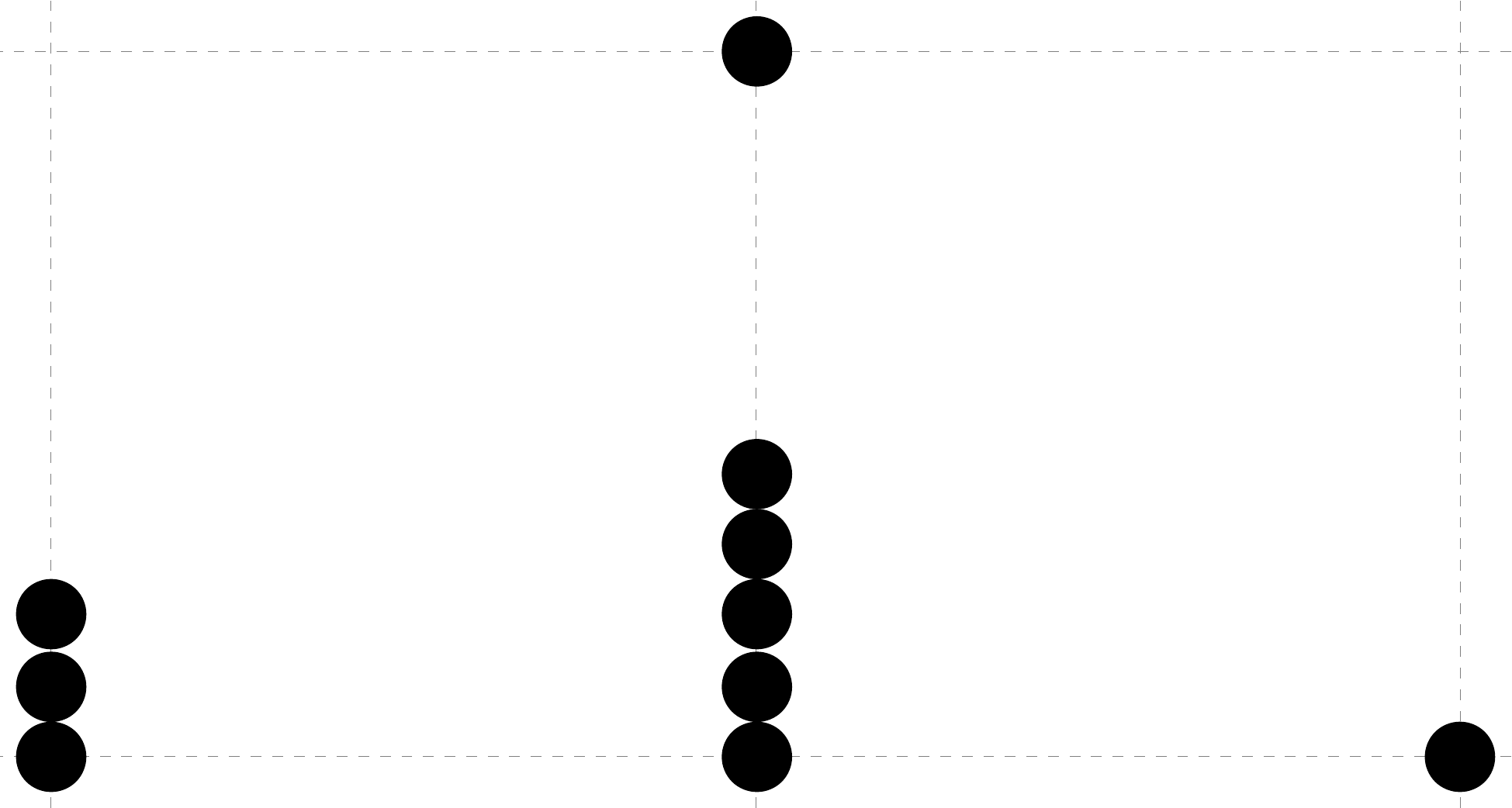}  \\
$ \BC^2 / \BZ_2 \times \BC $ \\
(inc.)
\end{tabular}
&
\begin{tabular}[b]{c} 
$-X_{12}^{}.X_{24}^{}.X_{41}^{}$ \\
$+X_{13}^{}.X_{35}^{}.X_{51}^{}$ \\
$-X_{23}^{}.X_{35}^{}.X_{52}^{}$ \\
$+X_{24}^{}.X_{45}^{}.X_{52}^{}$ \\
$+\phi _1^{}.X_{12}^{}.X_{23}^{}.X_{34}^{}.X_{41}^{}$ \\
$-\phi _1^{}.X_{13}^{}.X_{34}^{}.X_{45}^{}.X_{51}^{}$
\end{tabular}
\\ \hline

(3.25) &
\includegraphics[width=3.0cm]{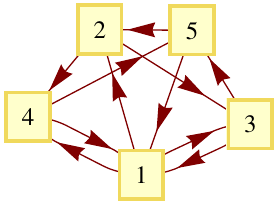} &
\includegraphics*[height=3.5cm]{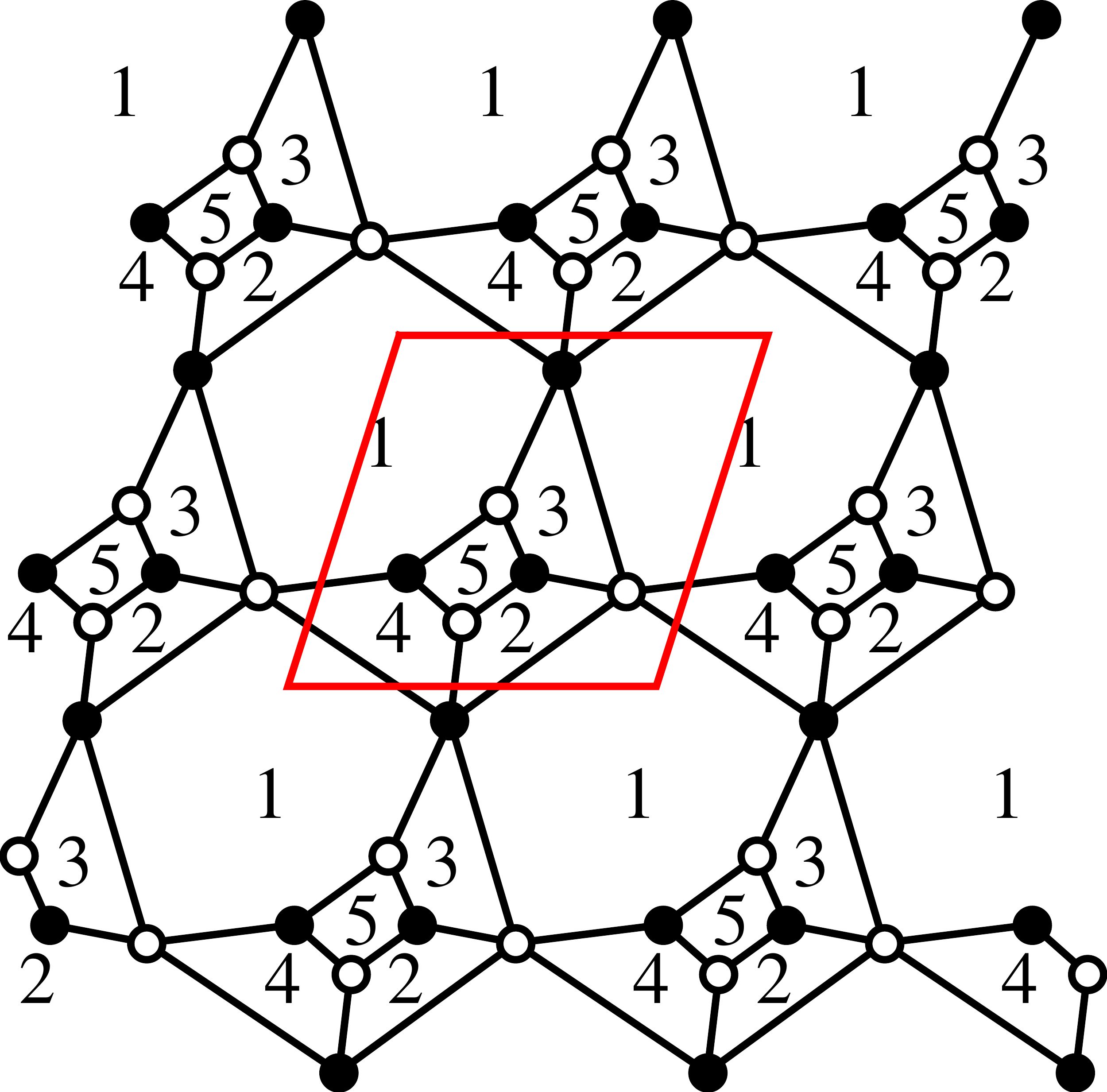} &
\begin{tabular}[b]{c}
\includegraphics[width=2.0cm]{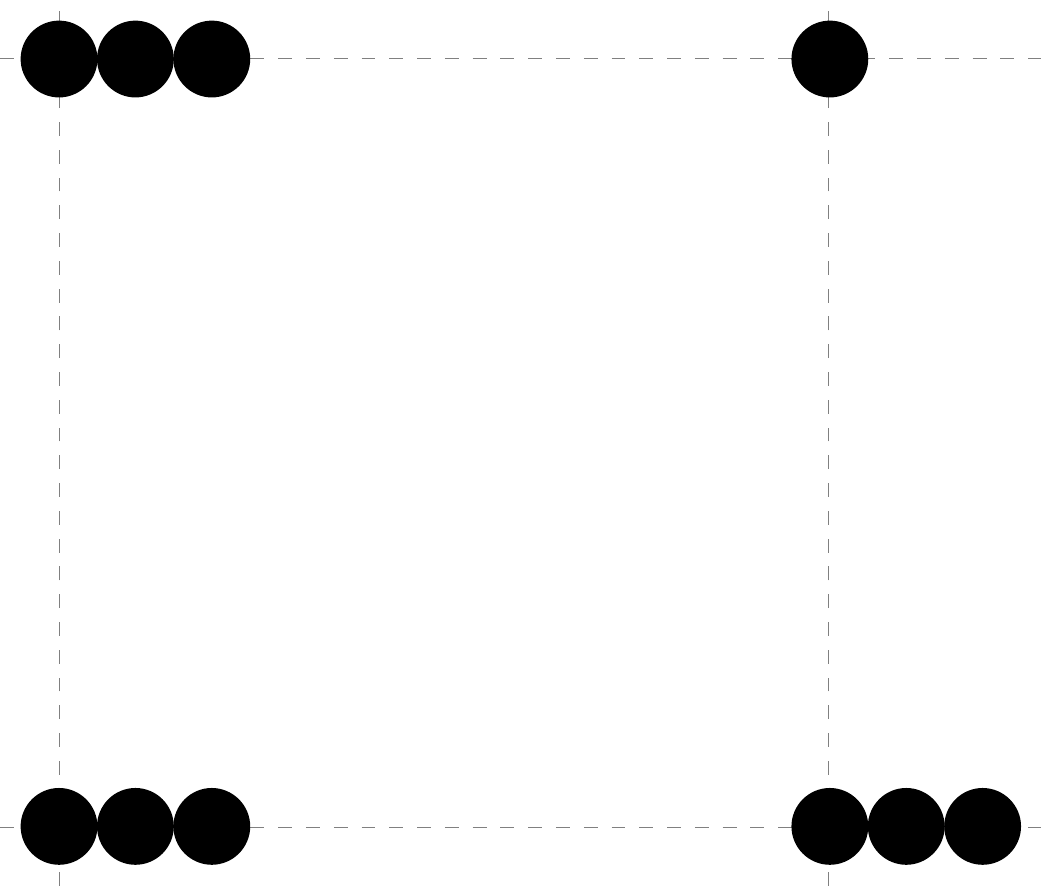} \\
$\CC$ (inc.)
\end{tabular}
 &
\begin{tabular}[b]{c} 
$X_{13}^{}.X_{35}^{}.X_{51}^{}$ \\
$-X_{14}^{}.X_{45}^{}.X_{51}^{}$ \\
$-X_{23}^{}.X_{35}^{}.X_{52}^{}$ \\
$+X_{24}^{}.X_{45}^{}.X_{52}^{}$ \\
$+X_{12}^{}.X_{23}^{}.X_{31}^{}.X_{14}^{}.X_{41}^{}$ \\
$-X_{12}^{}.X_{24}^{}.X_{41}^{}.X_{13}^{}.X_{31}^{}$
\end{tabular}

\end{tabular}
\end{center}

\caption{Tilings with 6 superpotential terms and 5 gauge groups \bf{(page 3/3)}}
\label{t:tilings6-5}
\end{table}


\begin{table}[h]

\begin{center}
\begin{tabular}{c|c|c|c|c}
\# & Quiver & Tiling & Toric Diagram & Superpotential\\
\hline \hline

(3.26) &
\includegraphics[width=3.0cm]{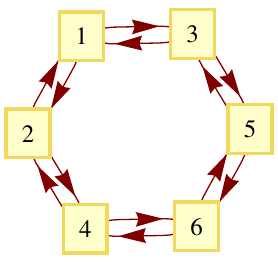} &
\includegraphics*[height=3.5cm]{N6-G6-6-tiling.pdf} &
\begin{tabular}[b]{c}
\includegraphics[height=2.4cm,angle=90]{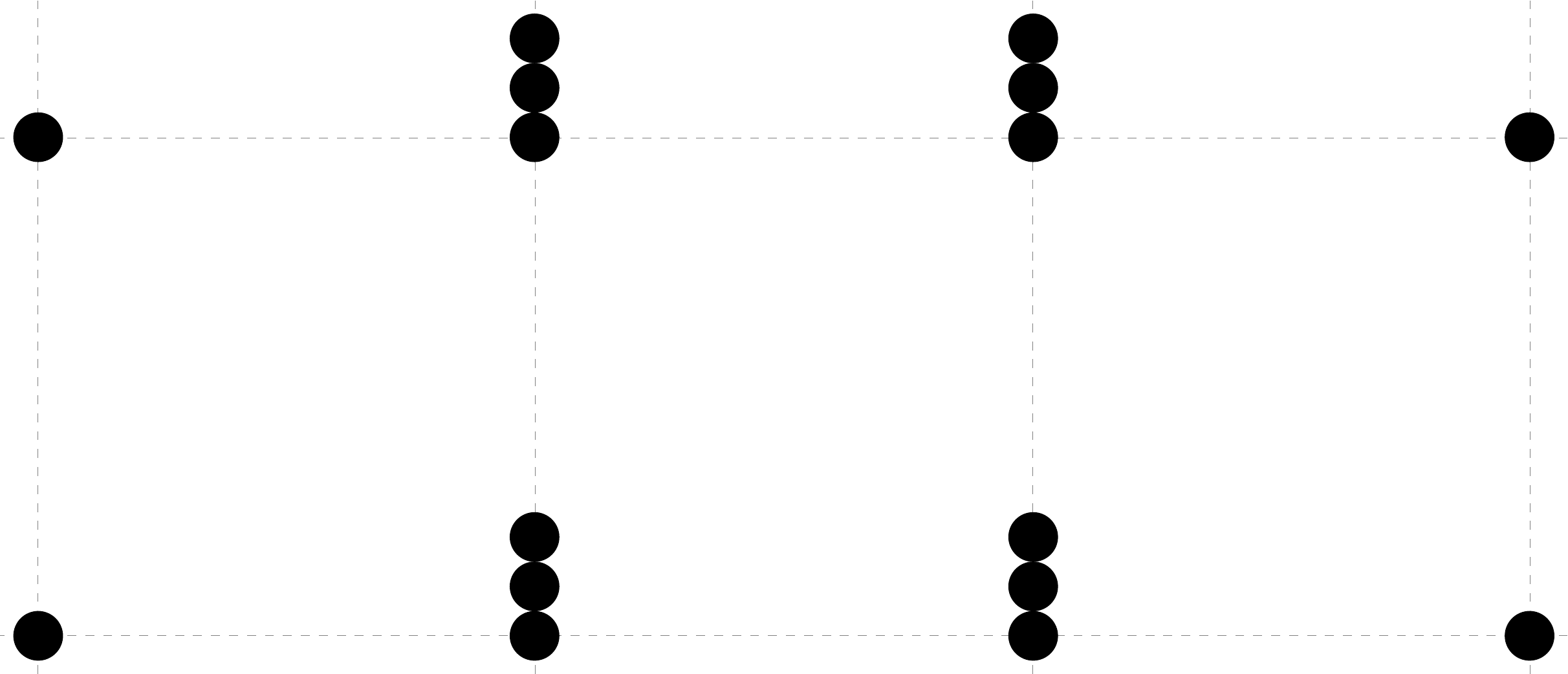} \\
$ L^{333} $ (I)
\end{tabular}
 &
\begin{tabular}[b]{c} 
$X_{12}^{}.X_{21}^{}.X_{13}^{}.X_{31}^{}$ \\
$-X_{12}^{}.X_{24}^{}.X_{42}^{}.X_{21}^{}$ \\
$-X_{13}^{}.X_{35}^{}.X_{53}^{}.X_{31}^{}$ \\
$+X_{24}^{}.X_{46}^{}.X_{64}^{}.X_{42}^{}$ \\
$+X_{35}^{}.X_{56}^{}.X_{65}^{}.X_{53}^{}$ \\
$-X_{46}^{}.X_{65}^{}.X_{56}^{}.X_{64}^{}$
\end{tabular}
\\ \hline

(3.27) &
\includegraphics[width=3.0cm]{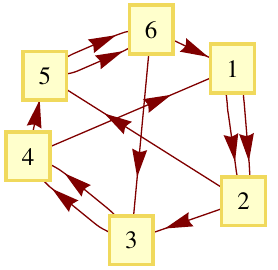} &
\includegraphics*[height=3.5cm]{N6-G6-5-tiling.pdf} &
\begin{tabular}[b]{c}
\includegraphics[height=2.4cm,angle=90]{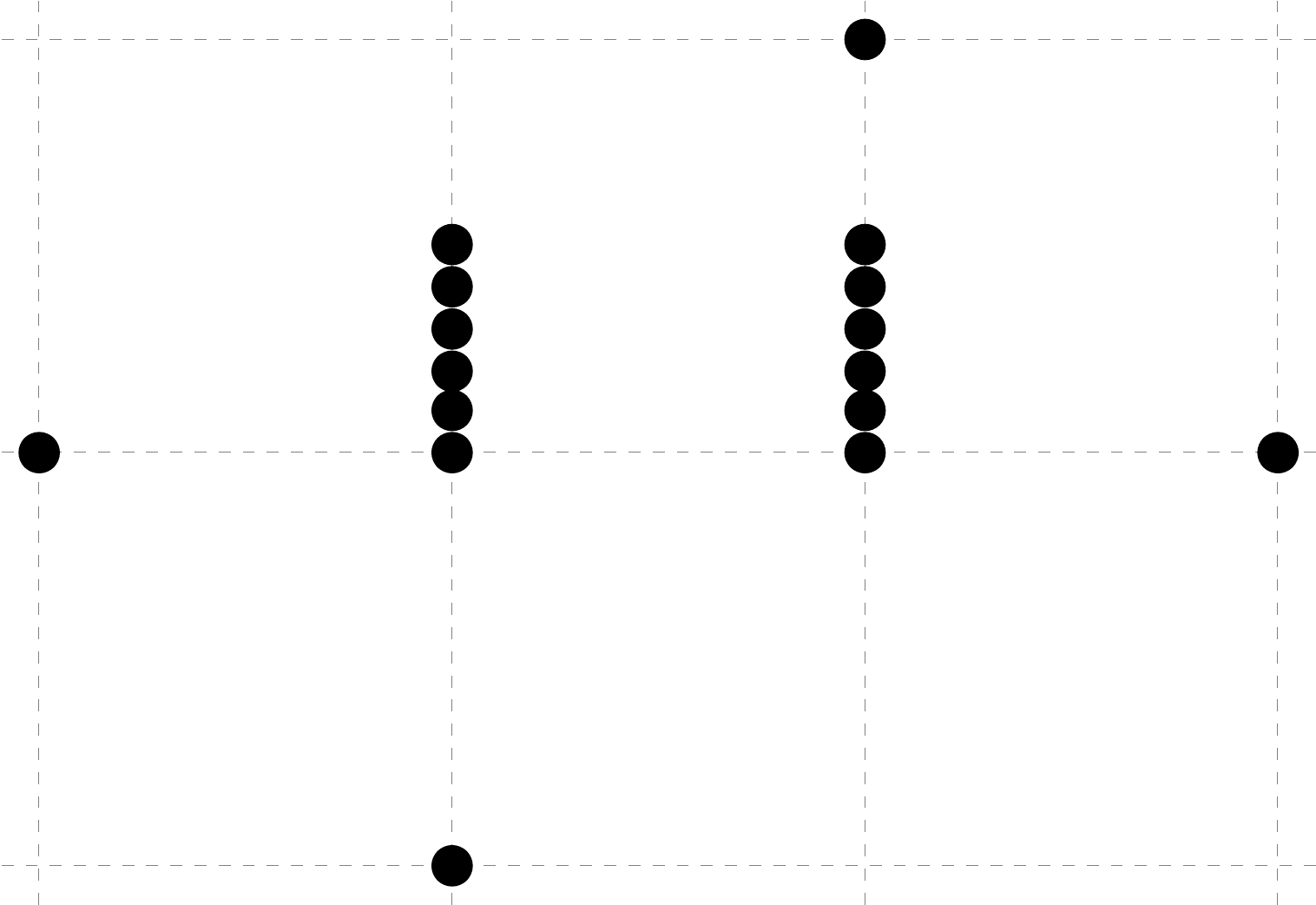} \\
$ Y^{3,0} $ (I)
\end{tabular}
 &
\begin{tabular}[b]{c} 
$X_{12}^1.X_{23}^{}.X_{34}^1.X_{41}^{}$ \\
$-X_{12}^1.X_{25}^{}.X_{56}^2.X_{61}^{}$ \\
$-X_{12}^2.X_{23}^{}.X_{34}^2.X_{41}^{}$ \\
$+X_{12}^2.X_{25}^{}.X_{56}^1.X_{61}^{}$ \\
$-X_{34}^1.X_{45}^{}.X_{56}^1.X_{63}^{}$ \\
$+X_{34}^2.X_{45}^{}.X_{56}^2.X_{63}^{}$
\end{tabular}
\\ \hline

(3.28) &
\includegraphics[width=3.0cm]{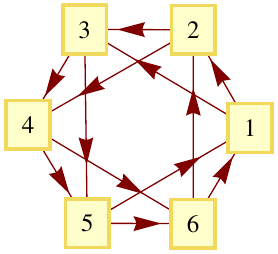} &
\includegraphics*[height=3.5cm]{N6-G6-9-tiling.pdf} &
\begin{tabular}[b]{c}
\includegraphics[height=2.4cm]{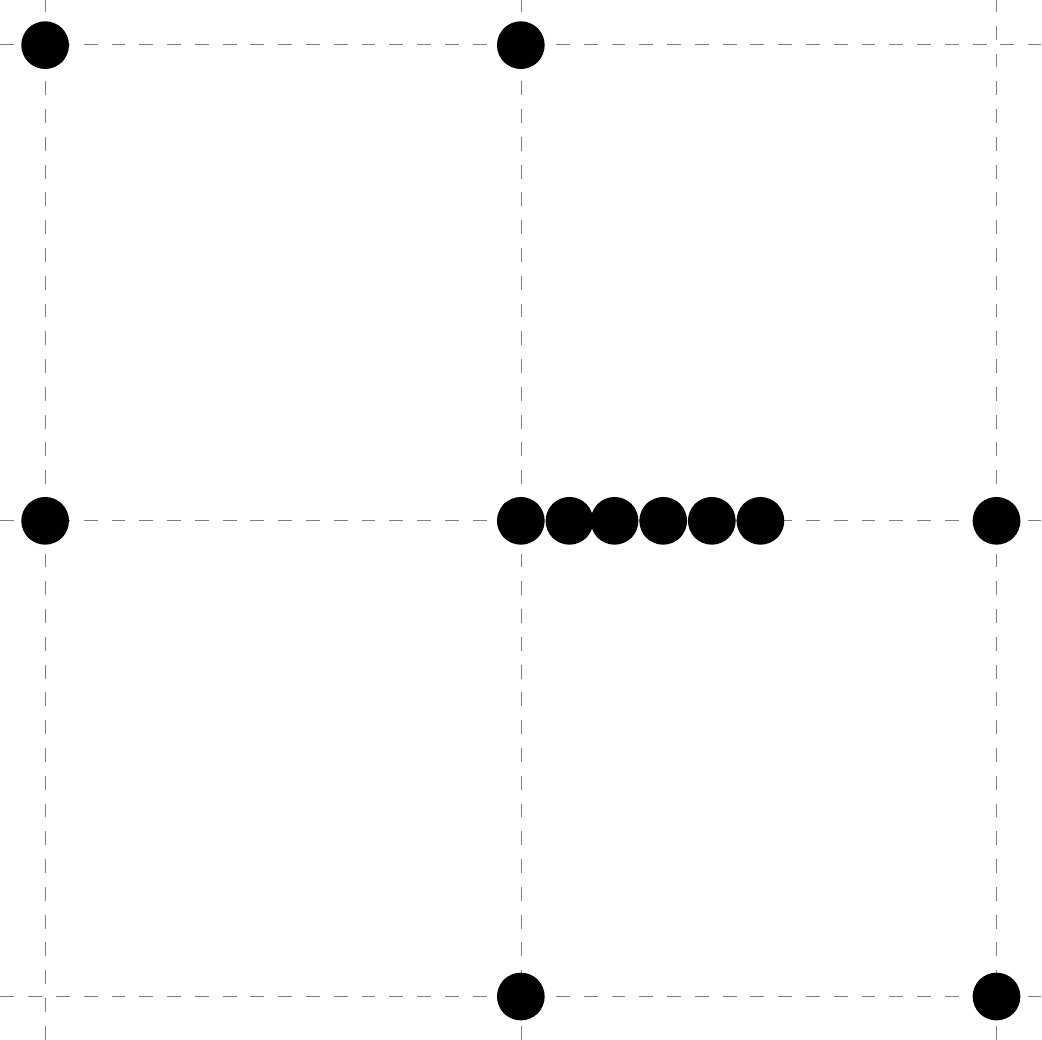}  \\
$ dP_3 $ (I)
\end{tabular}
&
\begin{tabular}[b]{c} 
$-X_{13}^{}.X_{35}^{}.X_{51}^{}$ \\
$-X_{24}^{}.X_{46}^{}.X_{62}^{}$ \\
$+X_{12}^{}.X_{24}^{}.X_{45}^{}.X_{51}^{}$ \\
$+X_{13}^{}.X_{34}^{}.X_{46}^{}.X_{61}^{}$ \\
$+X_{23}^{}.X_{35}^{}.X_{56}^{}.X_{62}^{}$ \\
$-X_{12}^{}.X_{23}^{}.X_{34}^{}.X_{45}^{}.X_{56}^{}.X_{61}^{}$
\end{tabular}
\\ \hline

(3.30) &
\includegraphics[width=3.0cm]{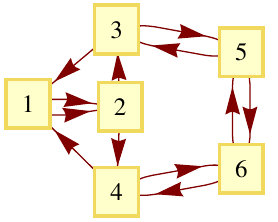} &
\includegraphics*[height=3.5cm]{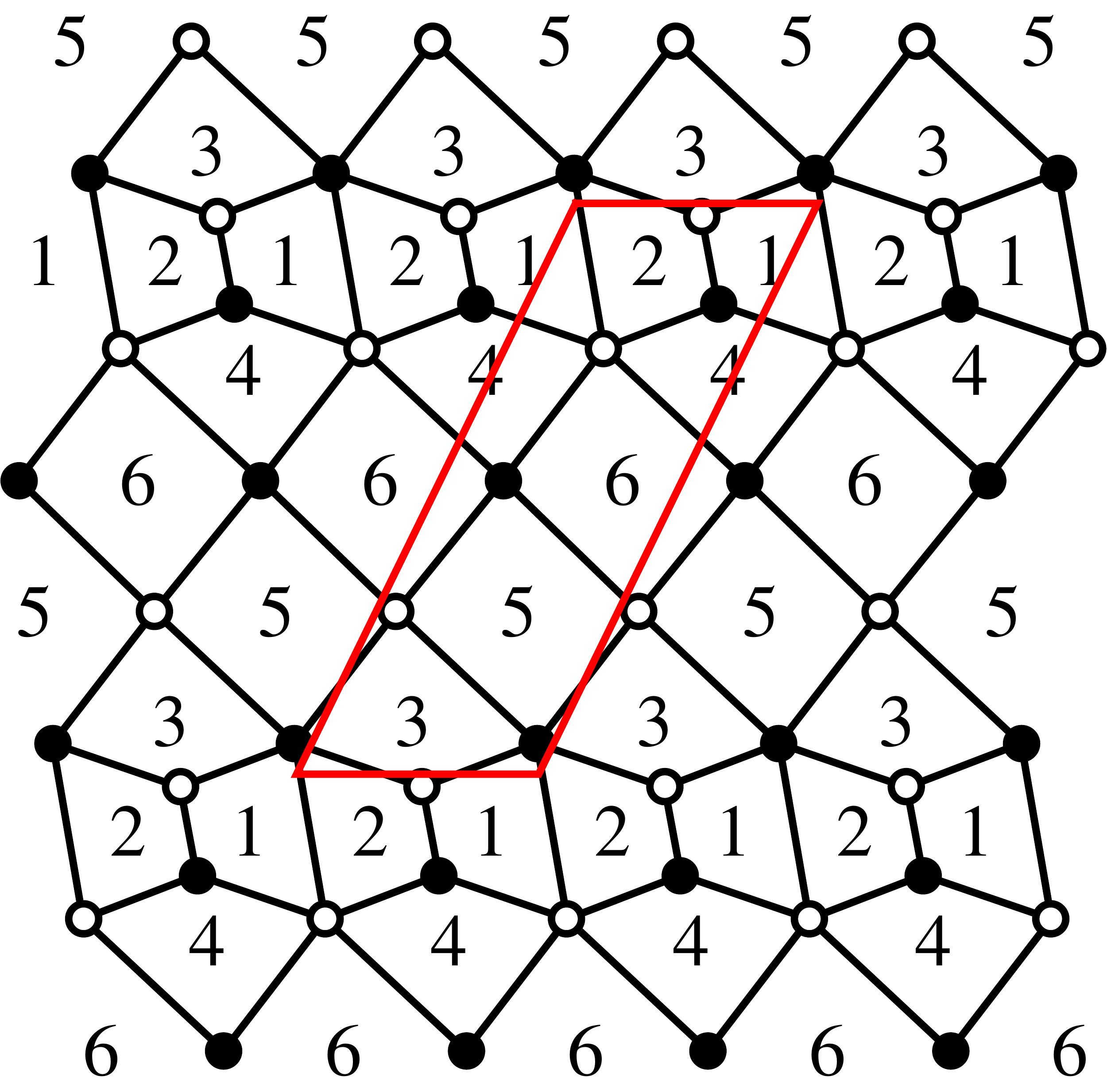} &
\begin{tabular}[b]{c}
\includegraphics[height=2.4cm,angle=90]{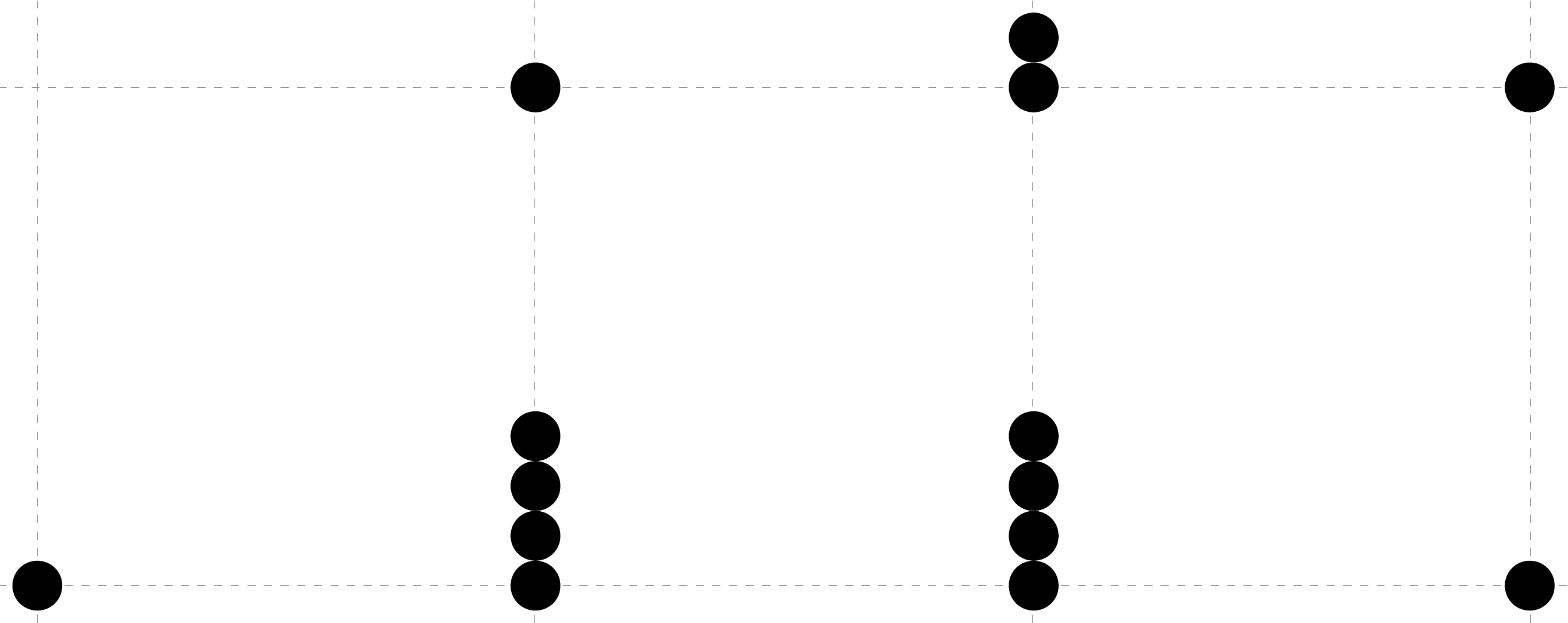} \\
$ L^{232} $ (inc.)
\end{tabular}
 &
\begin{tabular}[b]{c} 
$X_{12}^1.X_{23}^{}.X_{31}^{}$ \\
$-X_{12}^1.X_{24}^{}.X_{41}^{}$ \\
$+X_{35}^{}.X_{56}^{}.X_{65}^{}.X_{53}^{}$ \\
$-X_{46}^{}.X_{65}^{}.X_{56}^{}.X_{64}^{}$ \\
$-X_{12}^2.X_{23}^{}.X_{35}^{}.X_{53}^{}.X_{31}^{}$ \\
$+X_{12}^2.X_{24}^{}.X_{46}^{}.X_{64}^{}.X_{41}^{}$
\end{tabular}
\\ \hline

(3.29) &
\includegraphics[width=3.0cm]{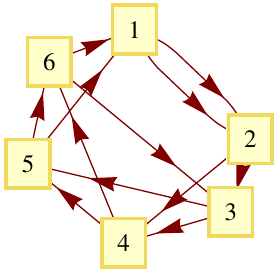} &
\includegraphics*[height=3.5cm]{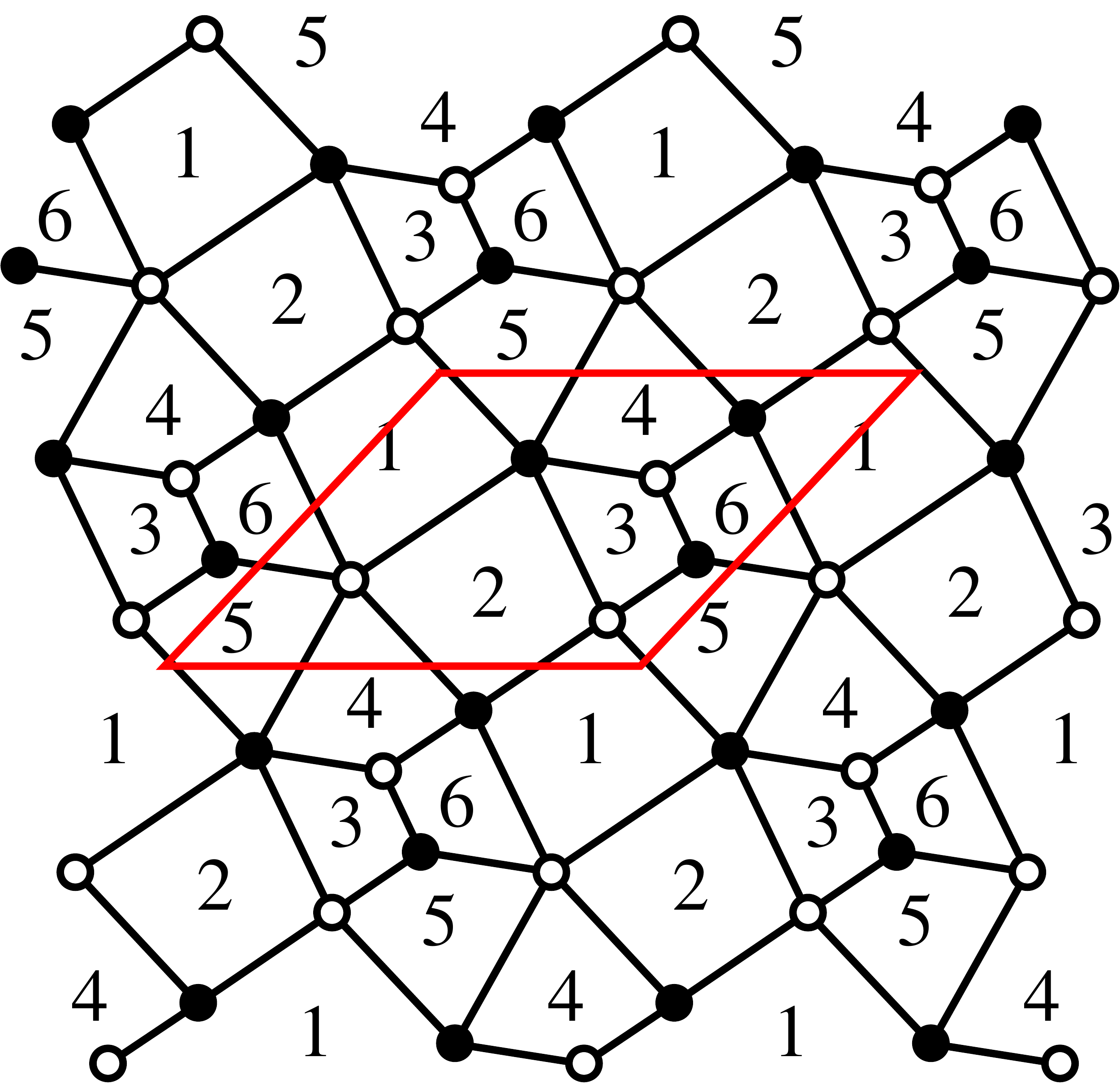} &
\begin{tabular}[b]{c}
\includegraphics[height=2.4cm]{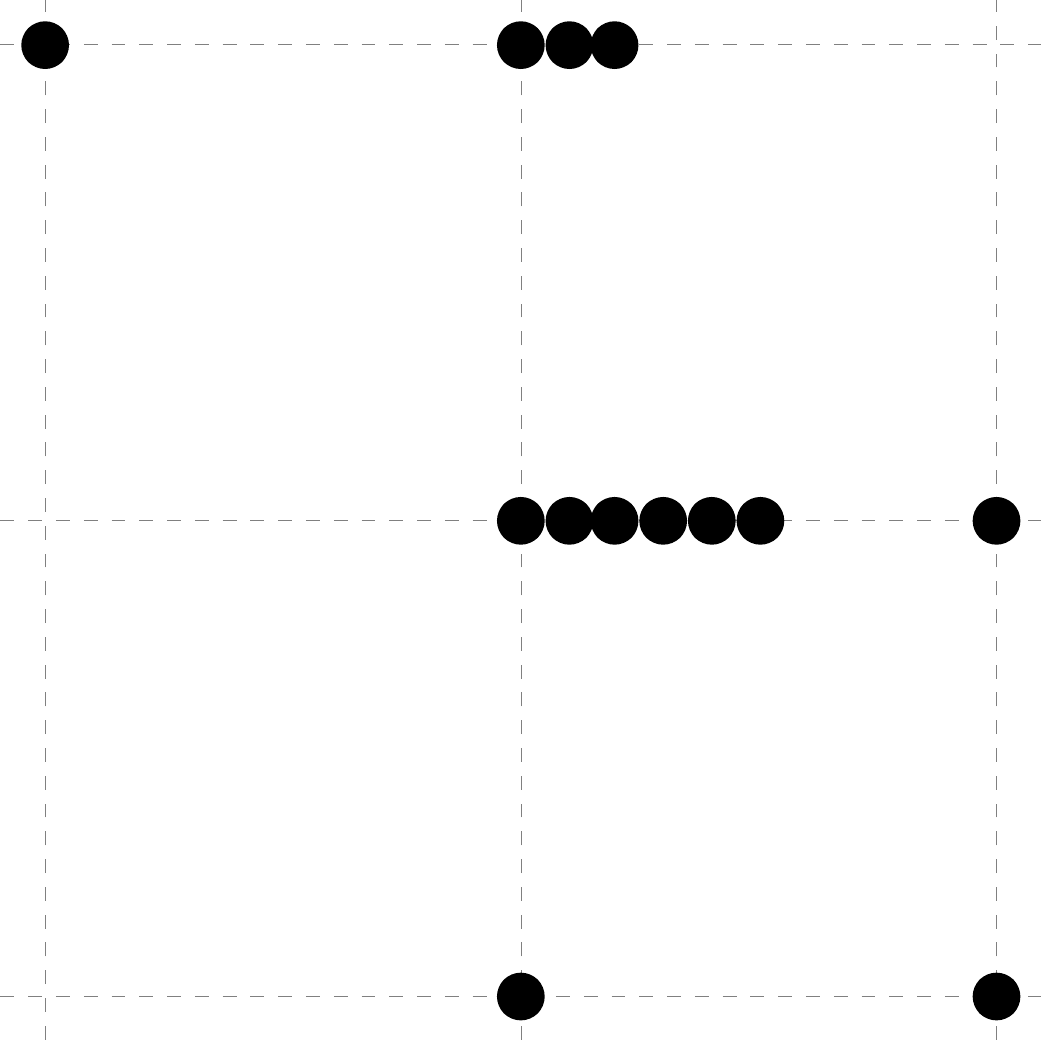} \\
$ dP_2 $ (inc.)
\end{tabular}
 &
\begin{tabular}[b]{c} 
$X_{34}^{}.X_{46}^{}.X_{63}^{}$ \\
$-X_{35}^{}.X_{56}^{}.X_{63}^{}$ \\
$+X_{12}^1.X_{23}^{}.X_{35}^{}.X_{51}^{}$ \\
$-X_{12}^1.X_{24}^{}.X_{46}^{}.X_{61}^{}$ \\
$-X_{12}^2.X_{23}^{}.X_{34}^{}.X_{45}^{}.X_{51}^{}$ \\
$+X_{12}^2.X_{24}^{}.X_{45}^{}.X_{56}^{}.X_{61}^{}$
\end{tabular}

\end{tabular}
\end{center}

\caption{Tilings with 6 superpotential terms and 6 gauge groups \bf{(page 1/3)}}
\label{t:tilings6-6a}
\end{table}

\begin{table}[h]

\begin{center}
\begin{tabular}{c|c|c|c|c}
\# & Quiver & Tiling & Toric Diagram & Superpotential\\
\hline \hline

(3.31) &
\includegraphics[width=3.0cm]{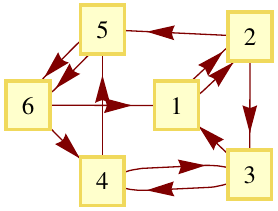} &
\includegraphics*[height=3.5cm]{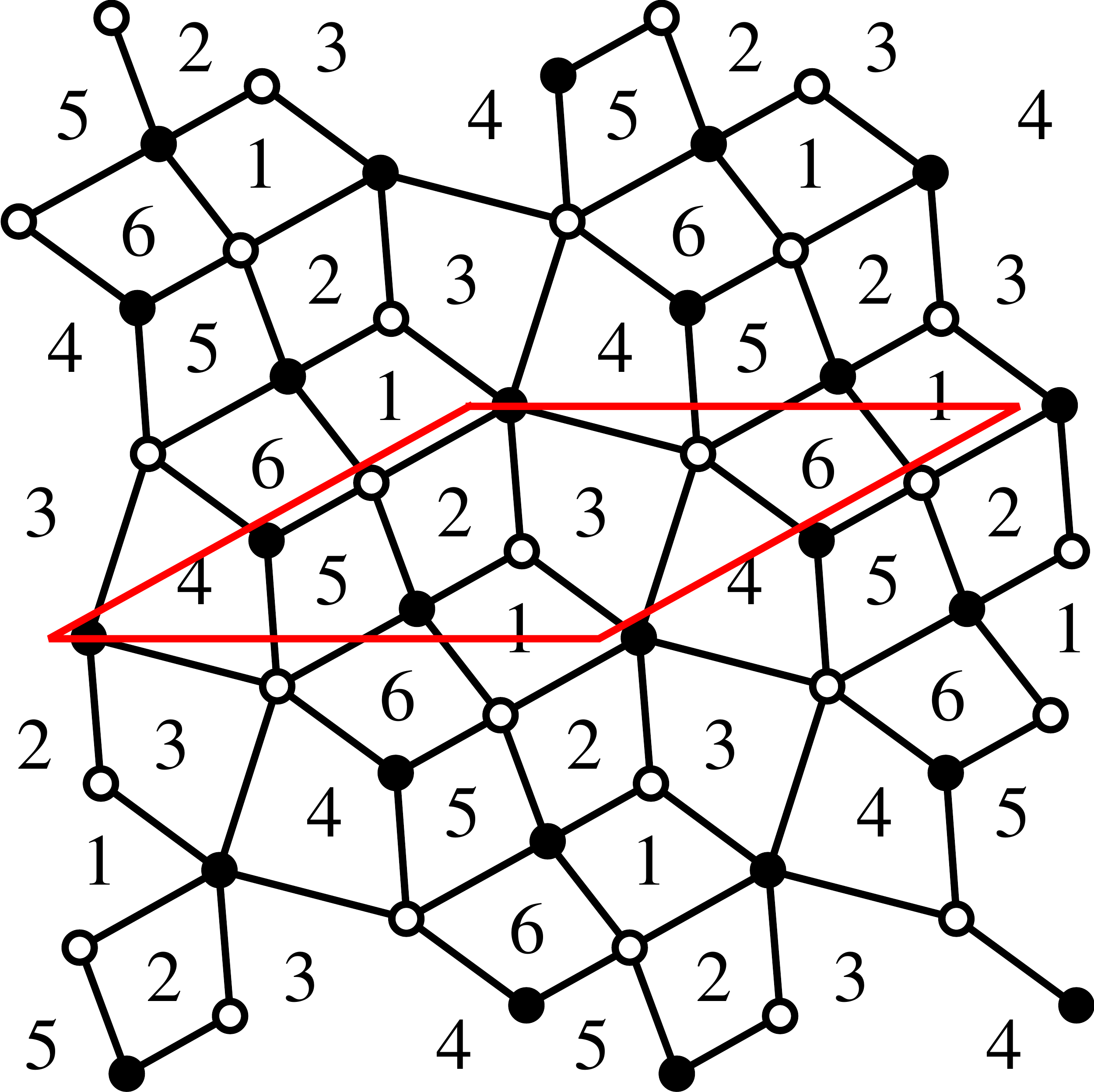} &
\begin{tabular}[b]{c}
\includegraphics[height=2.4cm,angle=90]{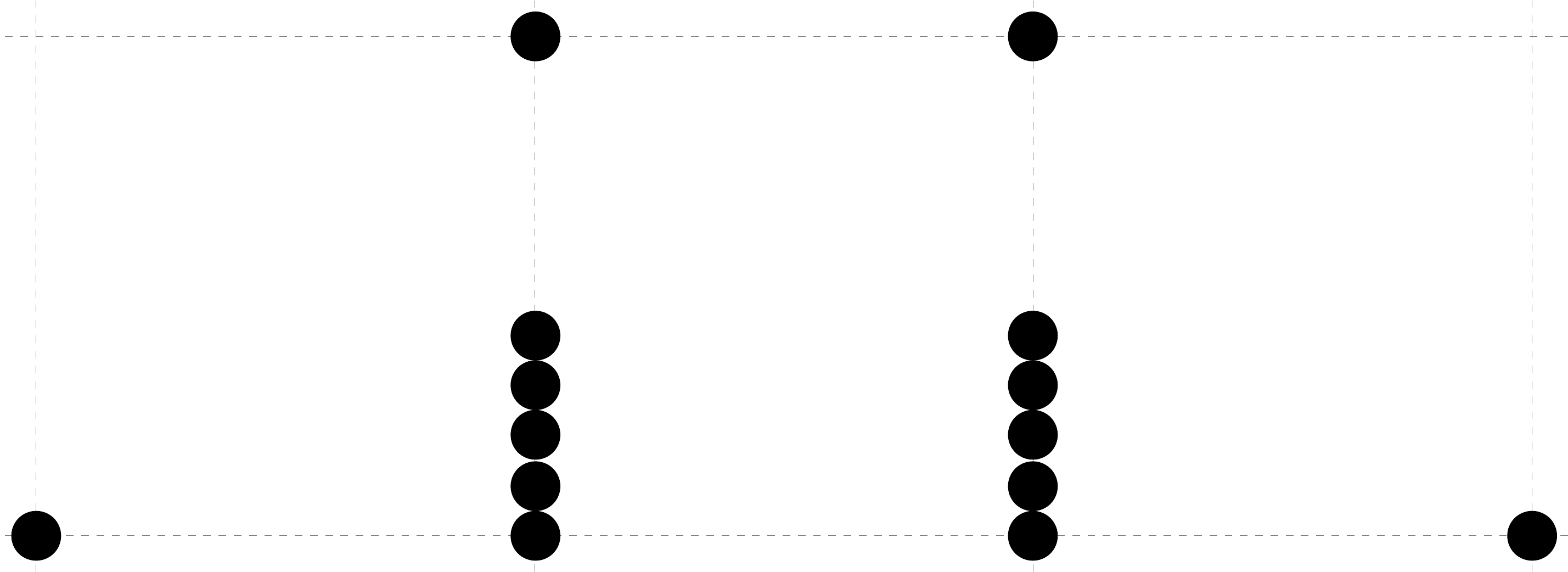} \\
$ L^{131} $ (inc.)
\end{tabular}
 &
\begin{tabular}[b]{c} 
$X_{12}^1.X_{23}^{}.X_{31}^{}$ \\
$-X_{45}^{}.X_{56}^1.X_{64}^{}$ \\
$-X_{12}^1.X_{25}^{}.X_{56}^2.X_{61}^{}$ \\
$+X_{12}^2.X_{25}^{}.X_{56}^1.X_{61}^{}$ \\
$-X_{12}^2.X_{23}^{}.X_{34}^{}.X_{43}^{}.X_{31}^{}$ \\
$+X_{34}^{}.X_{45}^{}.X_{56}^2.X_{64}^{}.X_{43}^{}$
\end{tabular}
\\ \hline

(3.32) &
\includegraphics[width=3.0cm]{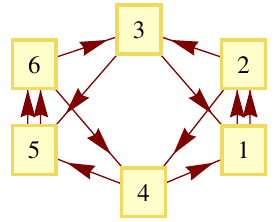} &
\includegraphics*[height=3.5cm]{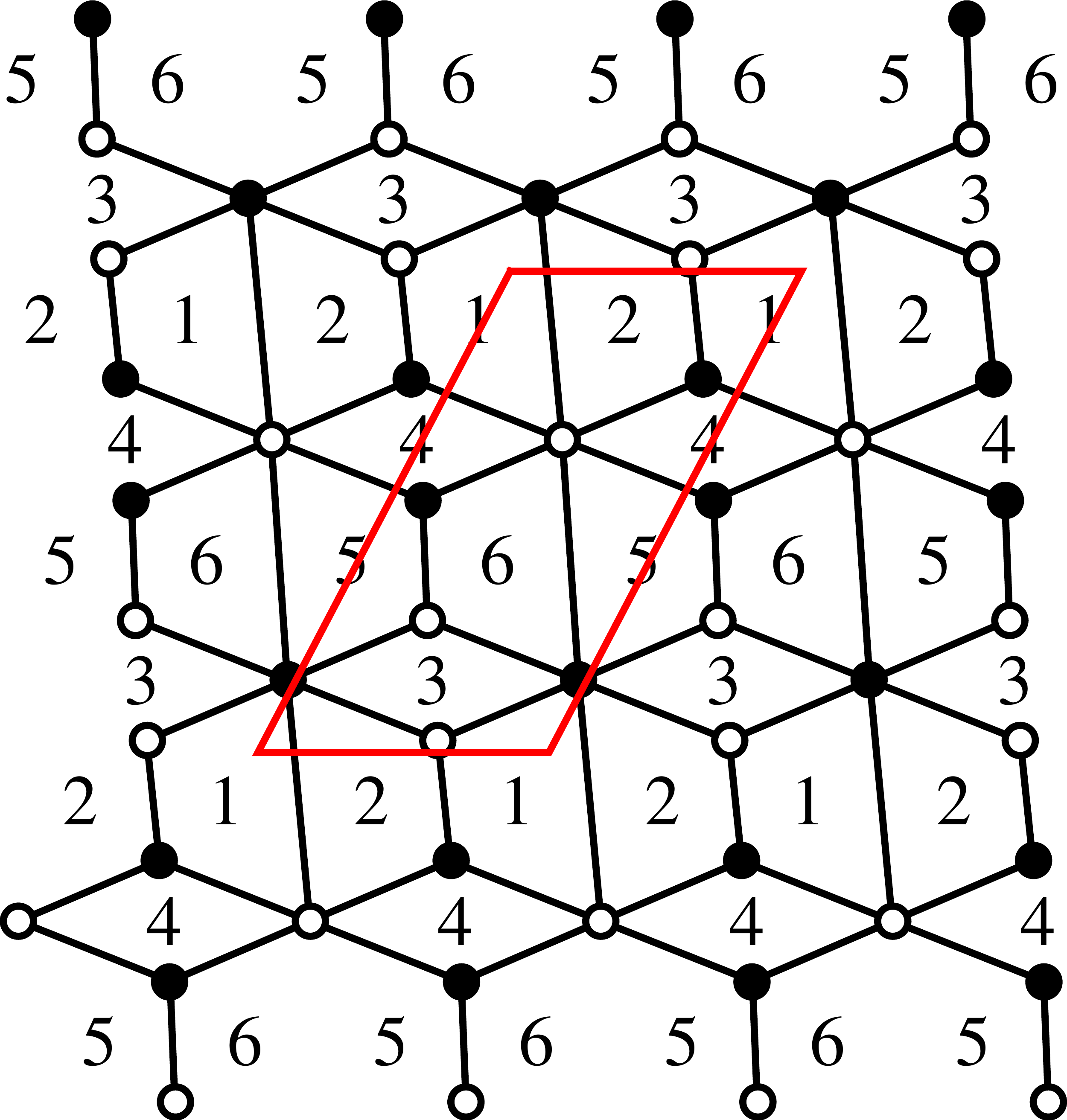} &
\begin{tabular}[b]{c}
\includegraphics[height=2.4cm,angle=90]{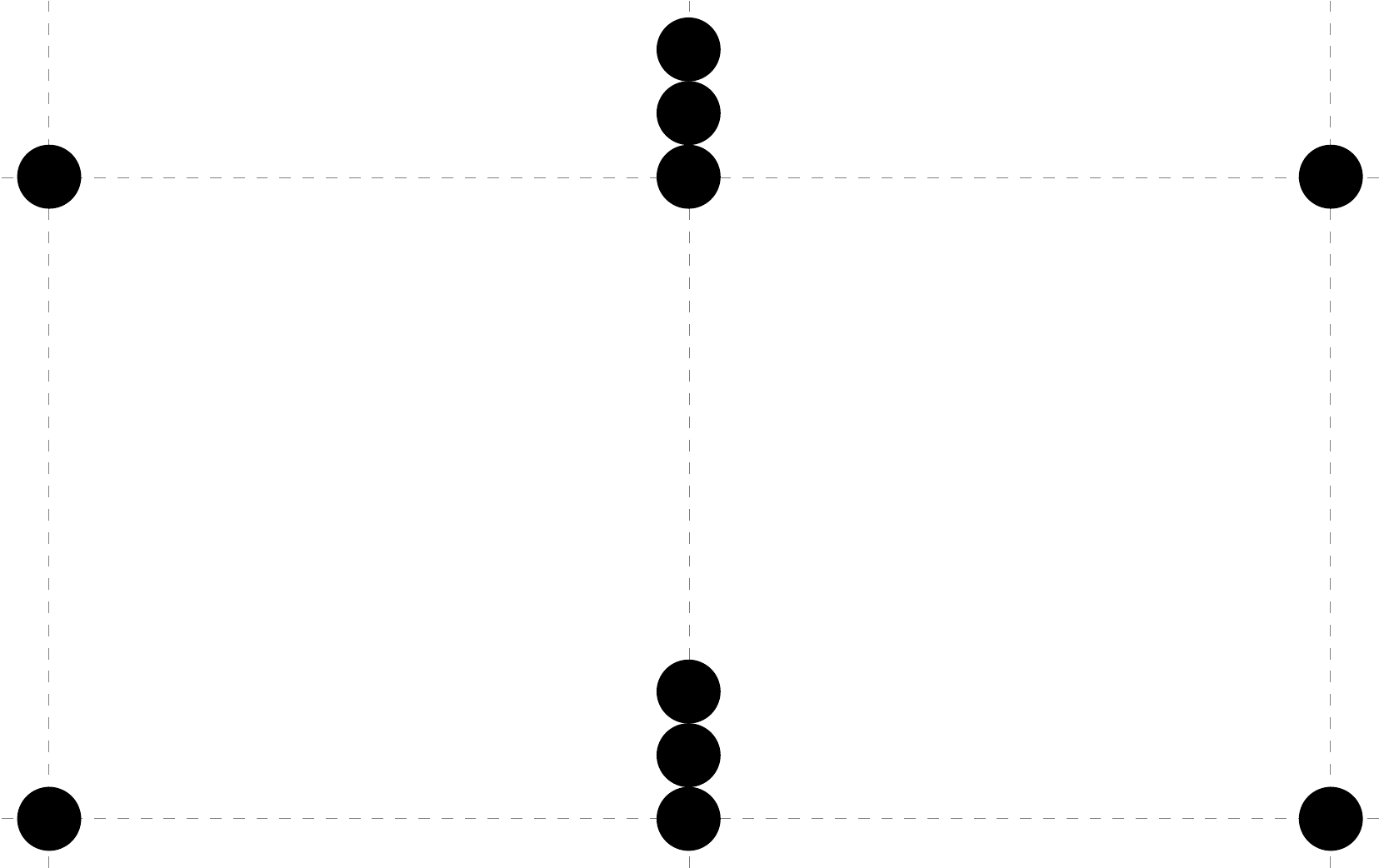} \\
$ L^{222} $ (inc.)
\end{tabular}
 &
\begin{tabular}[b]{c} 
$X_{12}^1.X_{23}^{}.X_{31}^{}$ \\
$-X_{12}^1.X_{24}^{}.X_{41}^{}$ \\
$+X_{35}^{}.X_{56}^2.X_{63}^{}$ \\
$-X_{45}^{}.X_{56}^2.X_{64}^{}$ \\
$-X_{12}^2.X_{23}^{}.X_{35}^{}.X_{56}^1.X_{63}^{}.X_{31}^{}$ \\
$+X_{12}^2.X_{24}^{}.X_{45}^{}.X_{56}^1.X_{64}^{}.X_{41}^{}$
\end{tabular}
\\ \hline

(3.33) &
\includegraphics[width=3.0cm]{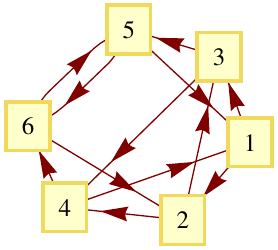} &
\includegraphics*[height=3.5cm]{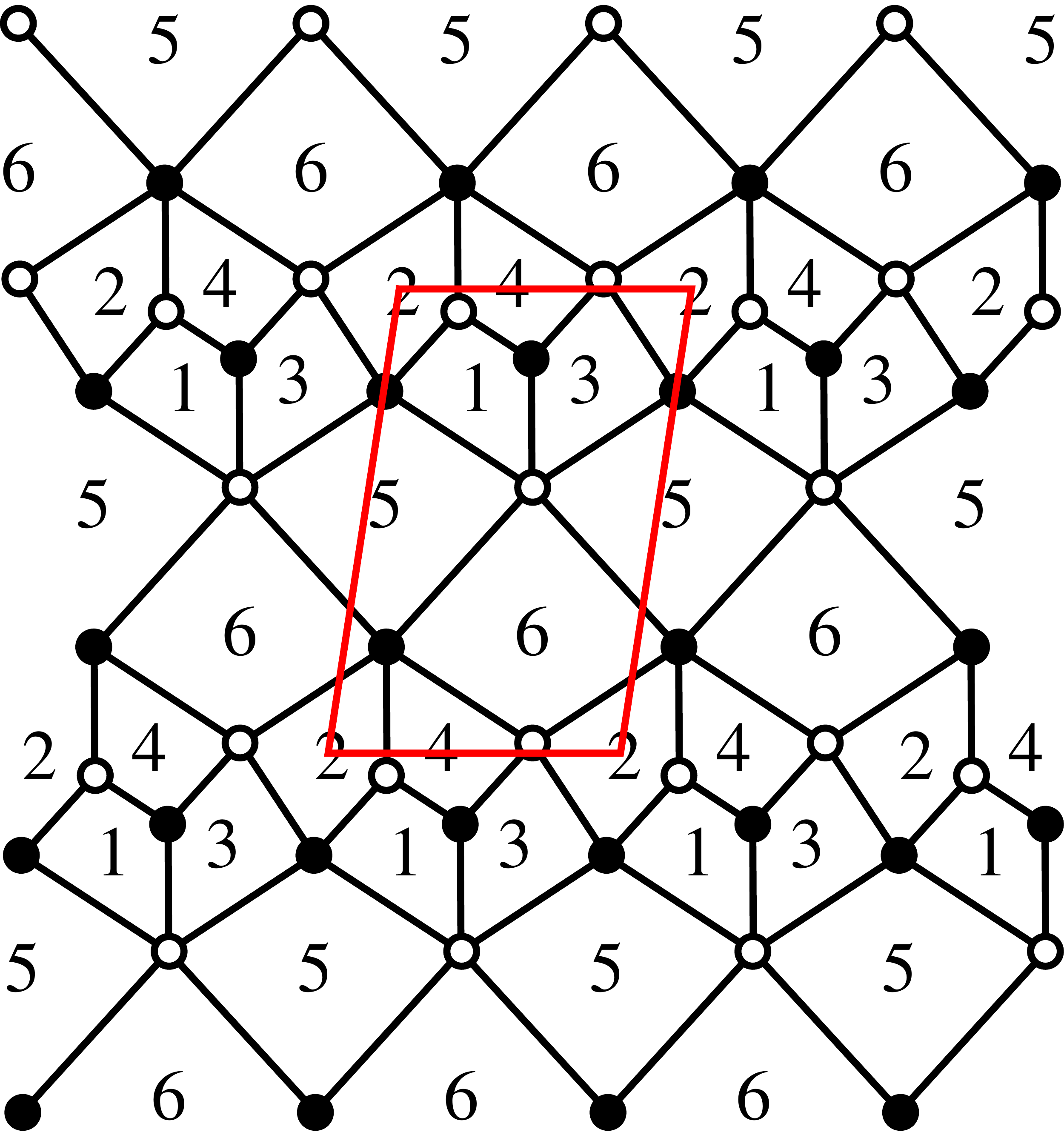} &
\begin{tabular}[b]{c}
\includegraphics[height=2.4cm,angle=90]{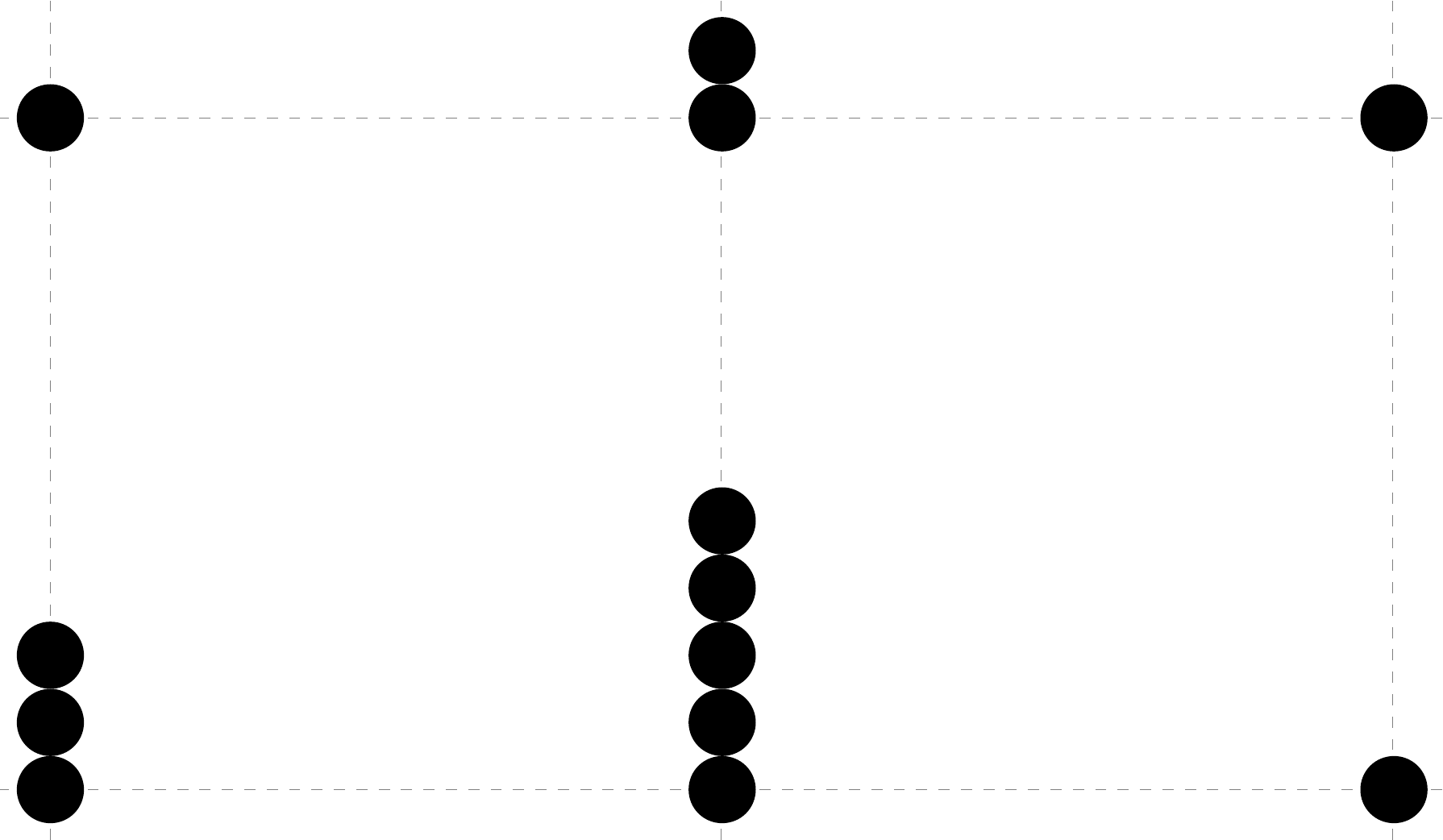}  \\
$ L^{222} $ (inc.)
\end{tabular}
&
\begin{tabular}[b]{c} 
$X_{12}^{}.X_{24}^{}.X_{41}^{}$ \\
$-X_{13}^{}.X_{34}^{}.X_{41}^{}$ \\
$-X_{12}^{}.X_{23}^{}.X_{35}^{}.X_{51}^{}$ \\
$+X_{23}^{}.X_{34}^{}.X_{46}^{}.X_{62}^{}$ \\
$+X_{13}^{}.X_{35}^{}.X_{56}^{}.X_{65}^{}.X_{51}^{}$ \\
$-X_{24}^{}.X_{46}^{}.X_{65}^{}.X_{56}^{}.X_{62}^{}$
\end{tabular}
\\ \hline

(3.34) &
\includegraphics[width=3.0cm]{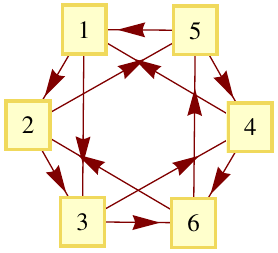} &
\includegraphics*[height=3.5cm]{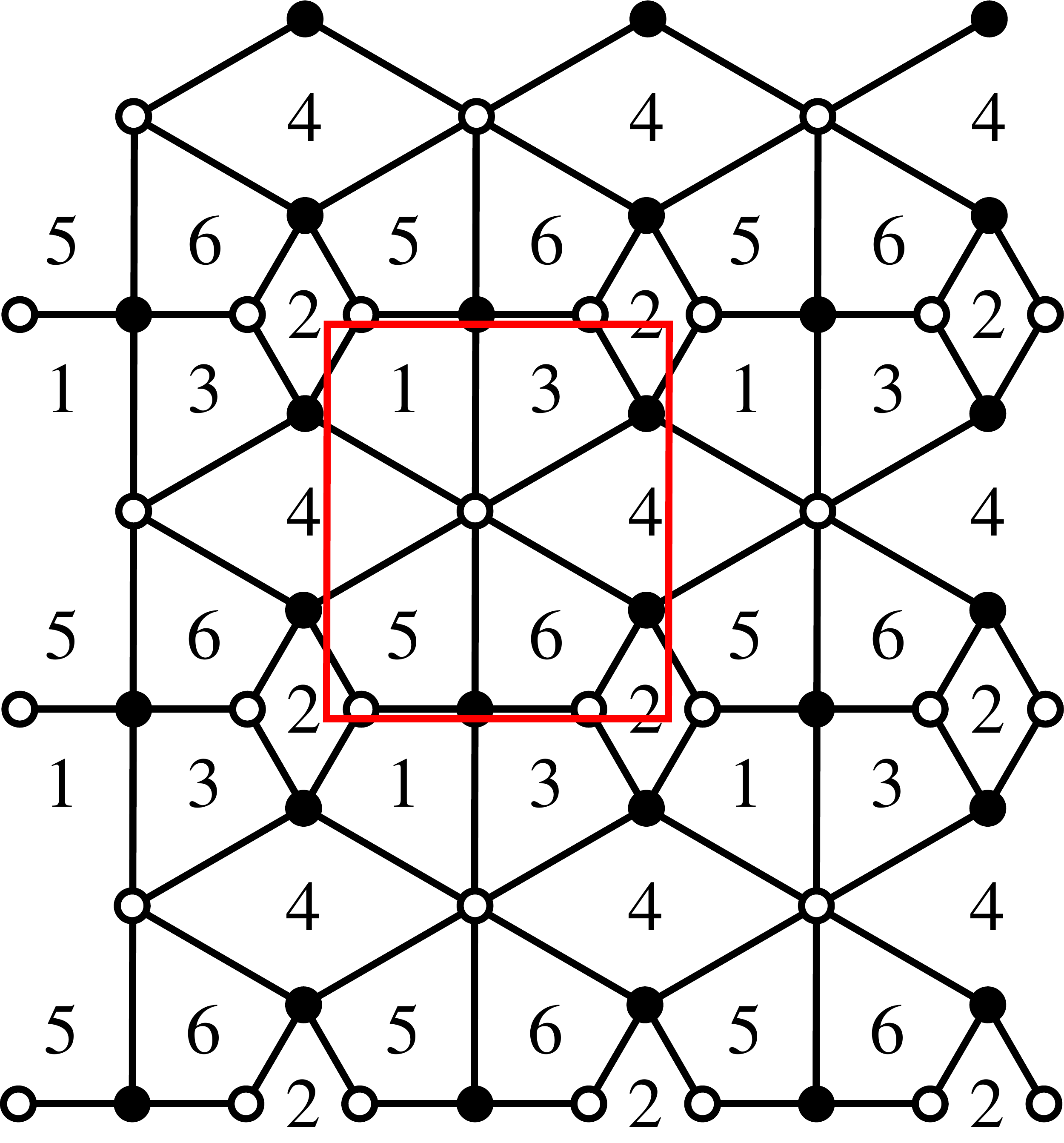} &
\begin{tabular}[b]{c}
\includegraphics[height=2.4cm,angle=90]{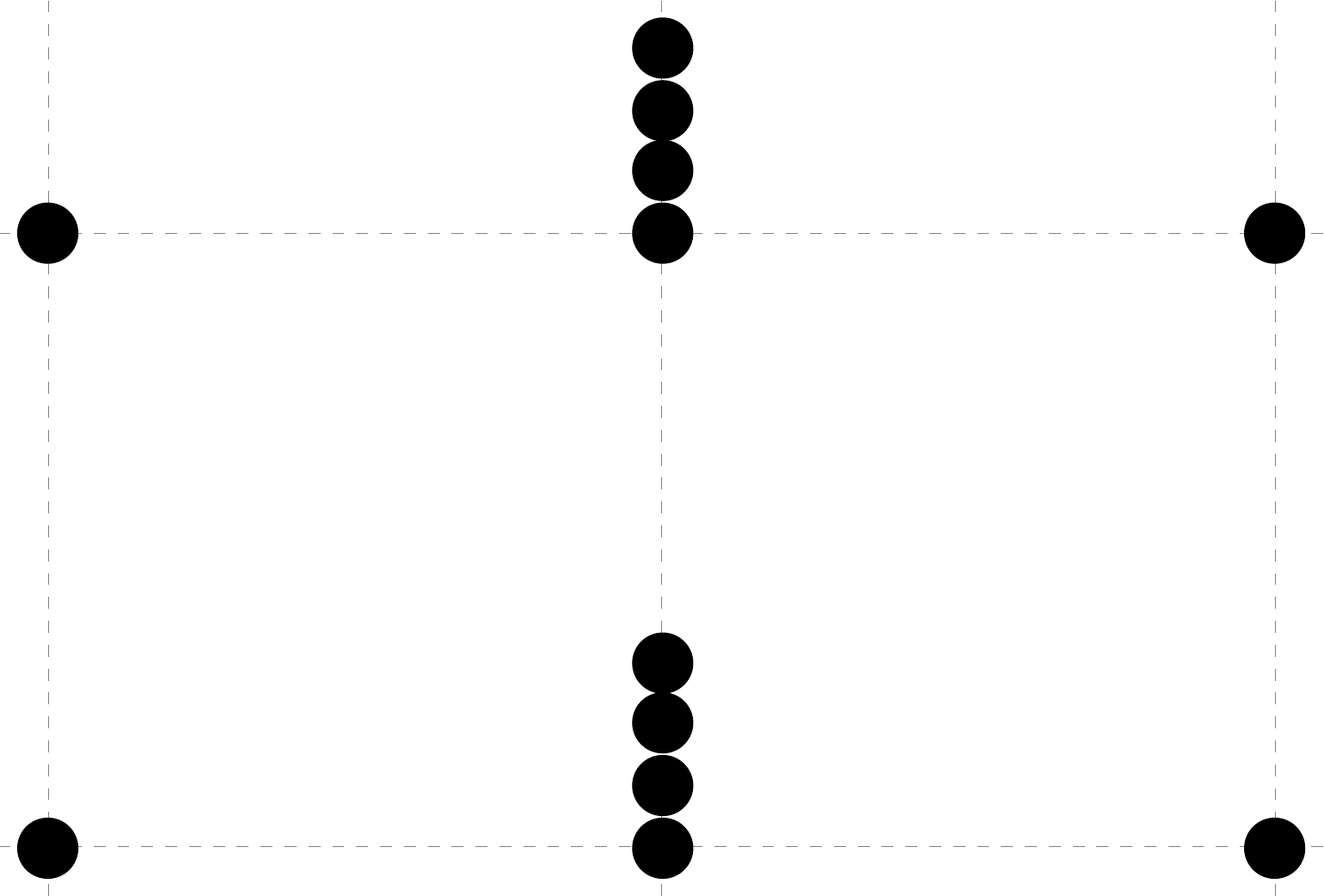} \\
$ L^{222} $ (inc.)
\end{tabular}
 &
\begin{tabular}[b]{c} 
$X_{12}^{}.X_{25}^{}.X_{51}^{}$ \\
$+X_{23}^{}.X_{36}^{}.X_{62}^{}$ \\
$-X_{12}^{}.X_{23}^{}.X_{34}^{}.X_{41}^{}$ \\
$-X_{13}^{}.X_{36}^{}.X_{65}^{}.X_{51}^{}$ \\
$-X_{25}^{}.X_{54}^{}.X_{46}^{}.X_{62}^{}$ \\
$+X_{13}^{}.X_{34}^{}.X_{46}^{}.X_{65}^{}.X_{54}^{}.X_{41}^{}$
\end{tabular}
\\ \hline

(3.35) &
\includegraphics[width=3.0cm]{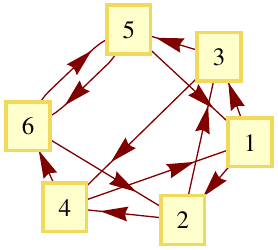} &
\includegraphics*[height=3.5cm]{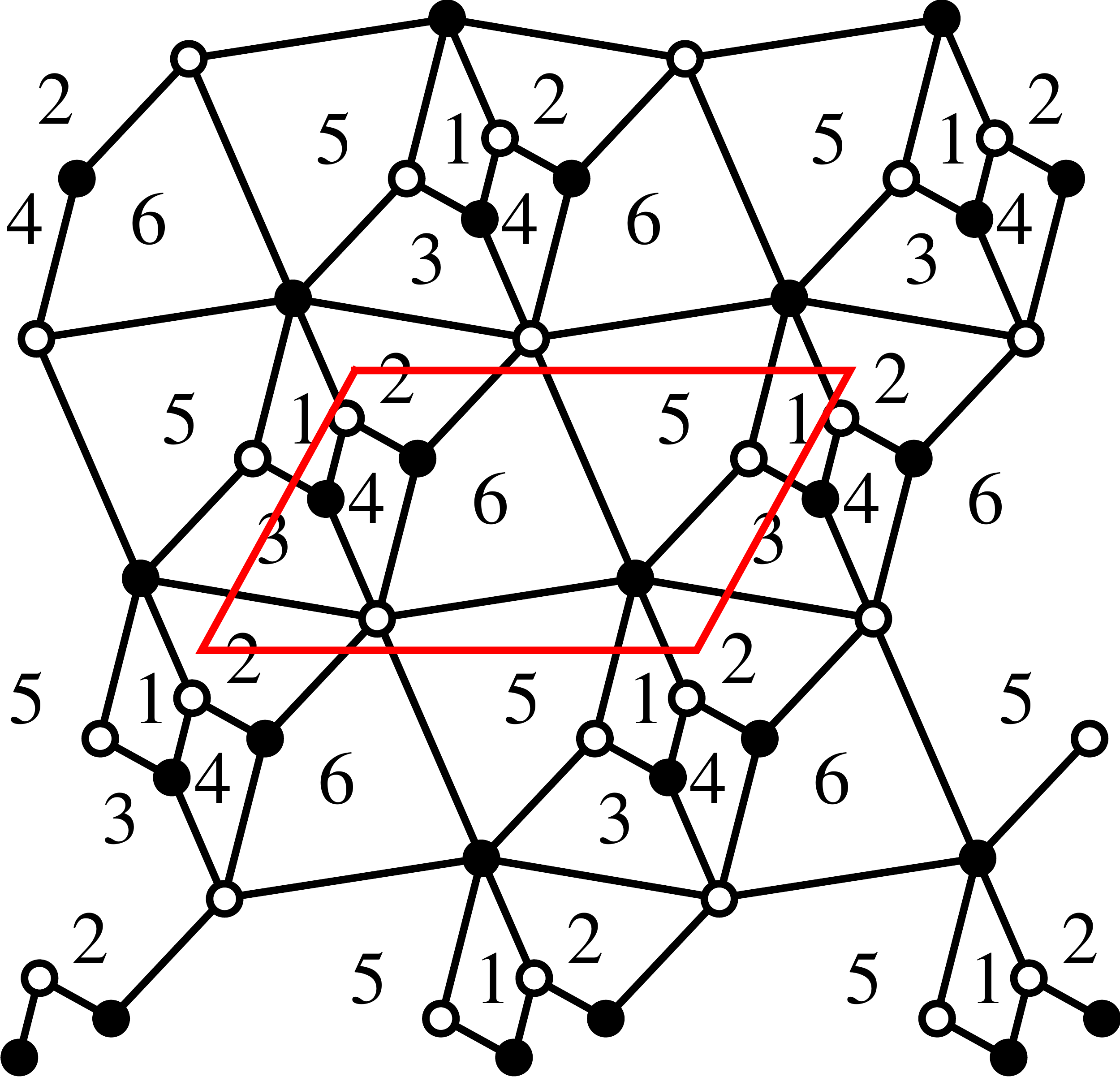} &
\begin{tabular}[b]{c}
\includegraphics[height=2.4cm,angle=90]{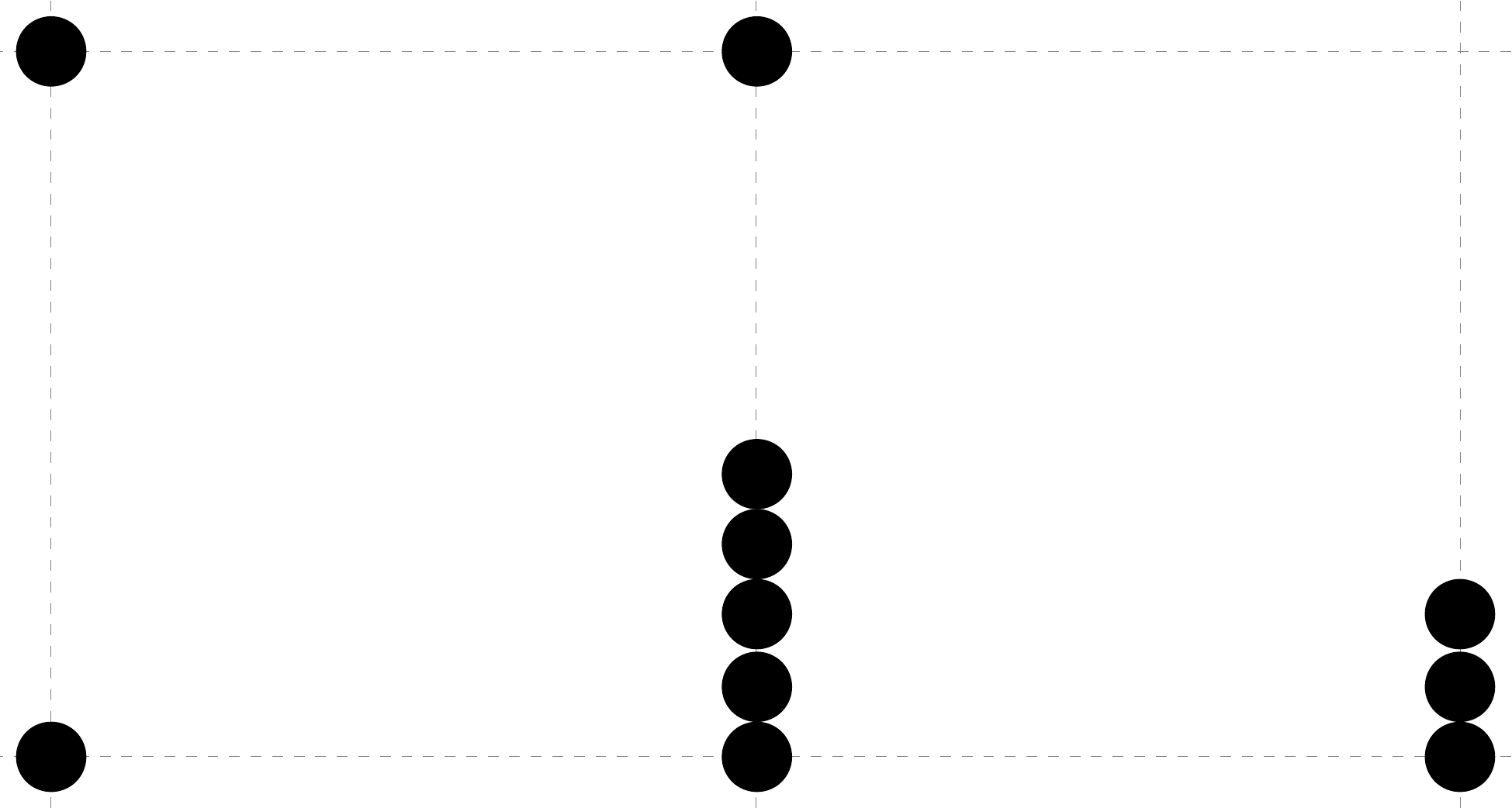} \\
$ SPP $ (inc.)
\end{tabular}
 &
\begin{tabular}[b]{c}
$X_{12}^{}.X_{24}^{}.X_{41}^{}$ \\
$-X_{13}^{}.X_{34}^{}.X_{41}^{}$ \\
$+X_{13}^{}.X_{35}^{}.X_{51}^{}$ \\
$-X_{24}^{}.X_{46}^{}.X_{62}^{}$ \\
$-X_{12}^{}.X_{23}^{}.X_{35}^{}.X_{56}^{}.X_{65}^{}.X_{51}^{}$ \\
$+X_{23}^{}.X_{34}^{}.X_{46}^{}.X_{65}^{}.X_{56}^{}.X_{62}^{}$
\end{tabular}

\end{tabular}
\end{center}

\caption{Tilings with 6 superpotential terms and 6 gauge groups \bf{(page 2/3)}}
\label{t:tilings6-6b}
\end{table}

\begin{table}[h]

\begin{center}
\begin{tabular}{c|c|c|c|c}
\# & Quiver & Tiling & Toric Diagram & Superpotential\\
\hline \hline

(3.36) &
\includegraphics[width=3.0cm]{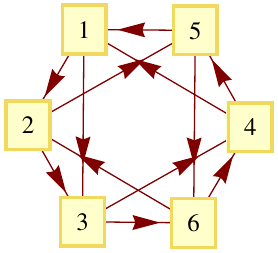} &
\includegraphics*[height=3.5cm]{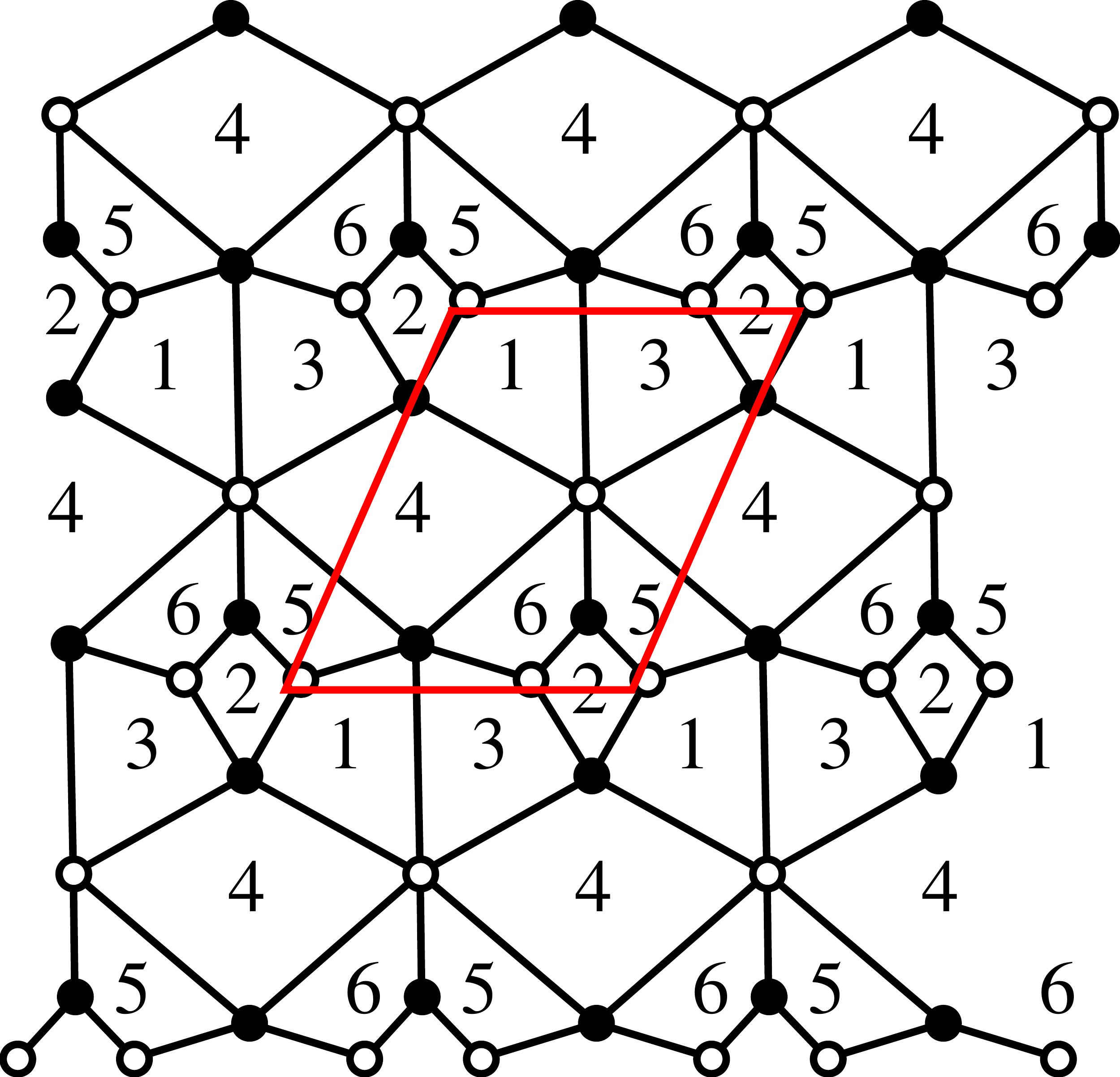} &
\begin{tabular}[b]{c}
\includegraphics[height=2.4cm,angle=90]{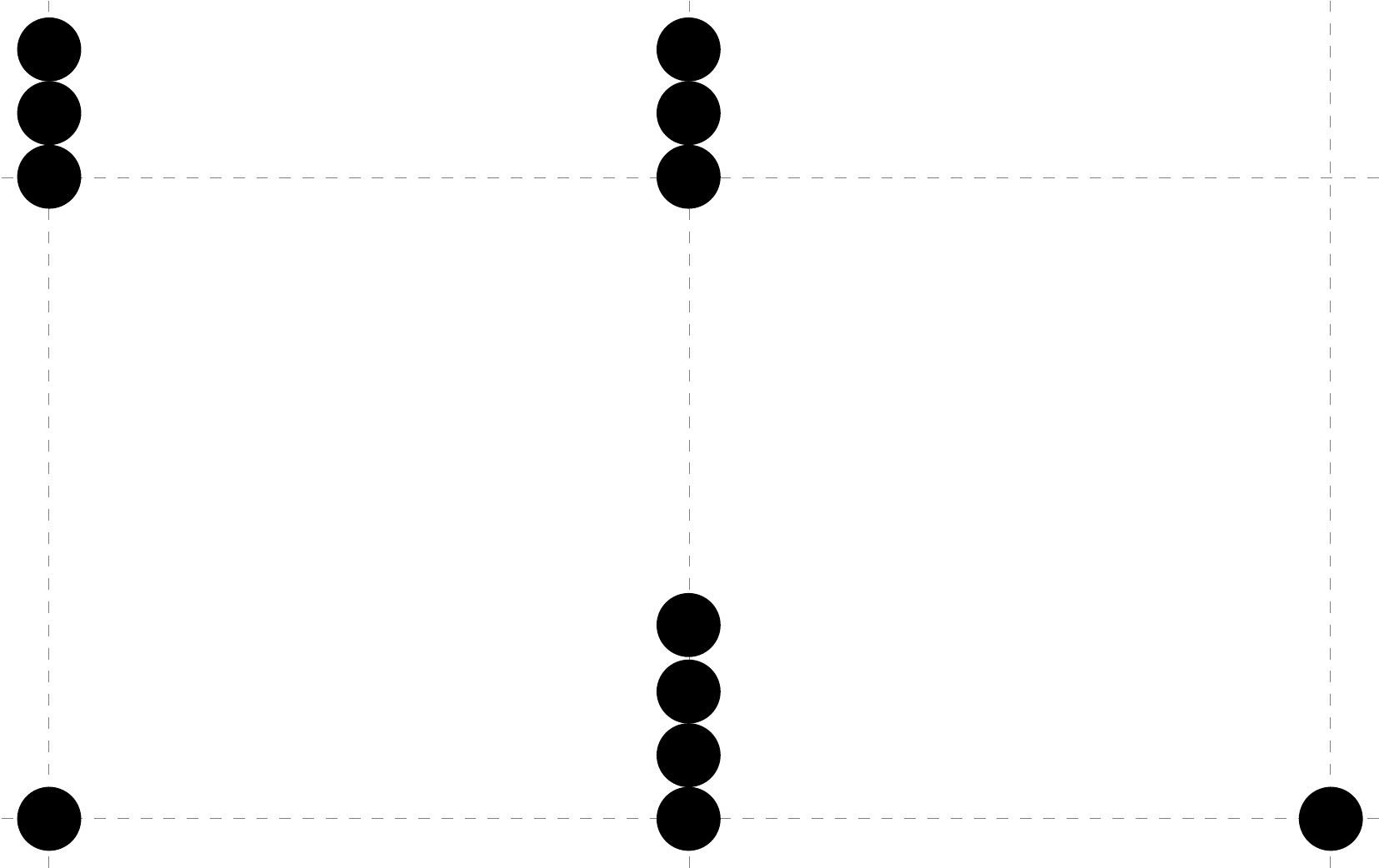} \\
$ SPP $ (inc.)
\end{tabular}
 &
\begin{tabular}[b]{c} 
$X_{12}^{}.X_{25}^{}.X_{51}^{}$ \\
$+X_{23}^{}.X_{36}^{}.X_{62}^{}$ \\
$-X_{25}^{}.X_{56}^{}.X_{62}^{}$ \\
$-X_{12}^{}.X_{23}^{}.X_{34}^{}.X_{41}^{}$ \\
$-X_{13}^{}.X_{36}^{}.X_{64}^{}.X_{45}^{}.X_{51}^{}$ \\
$+X_{13}^{}.X_{34}^{}.X_{45}^{}.X_{56}^{}.X_{64}^{}.X_{41}^{}$
\end{tabular}
\\ \hline

(3.37) &
\includegraphics[width=3.0cm]{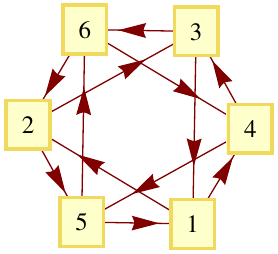} &
\includegraphics*[height=3.5cm]{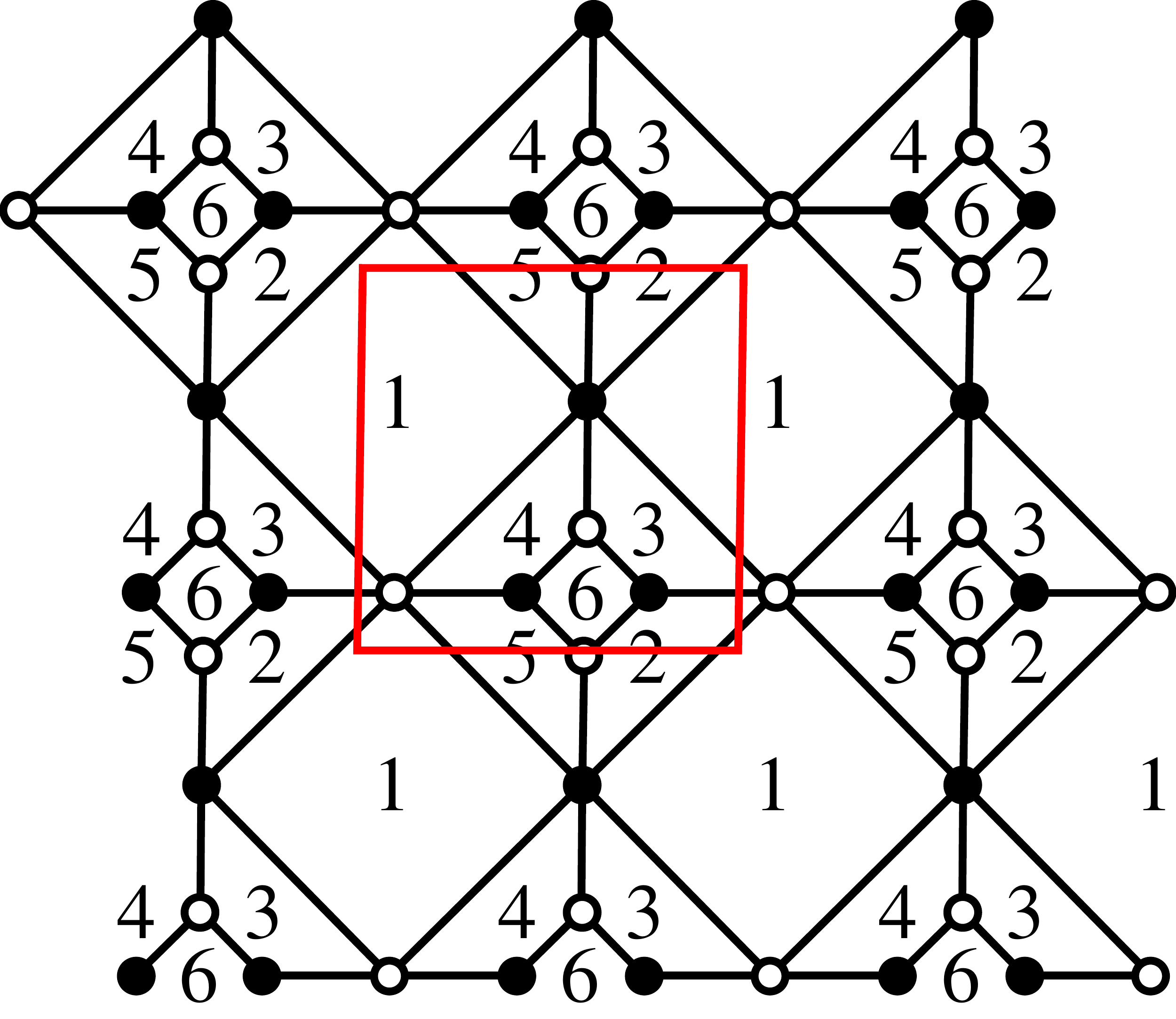} &
\begin{tabular}[b]{c}
\includegraphics[height=2.4cm]{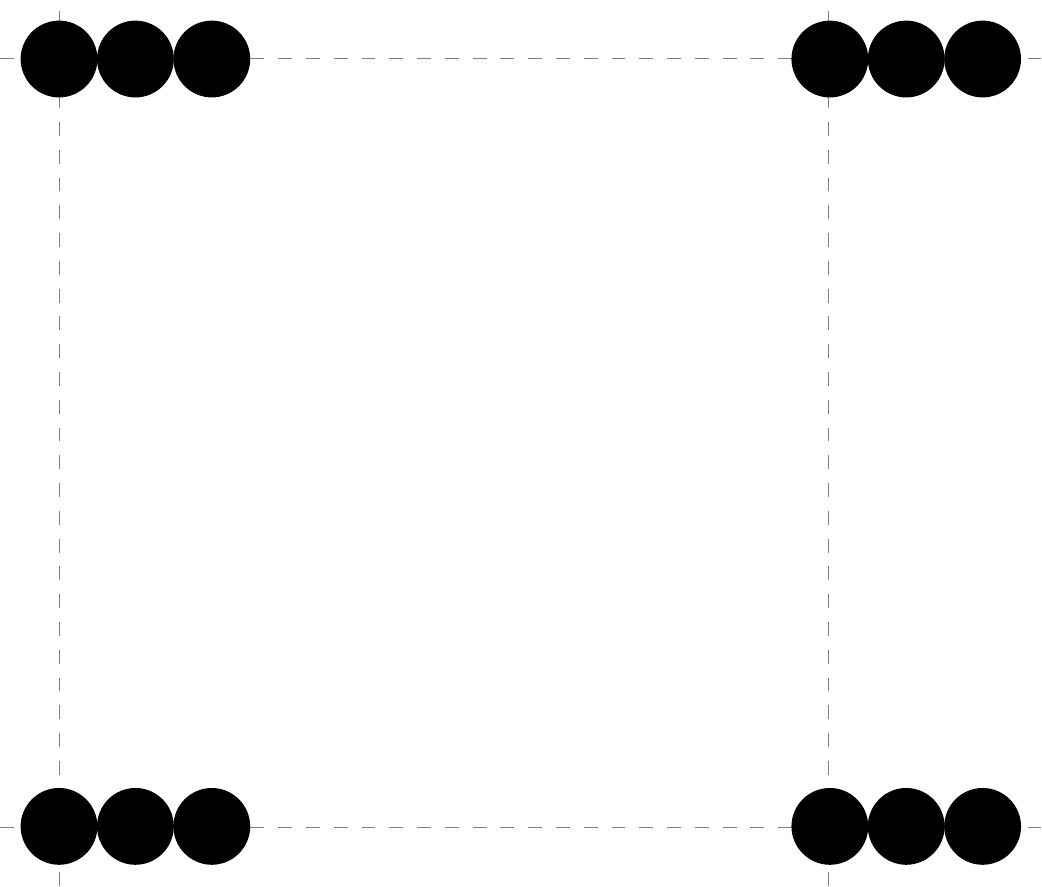} \\
$ \CC $ (inc.)
\end{tabular}
 &
\begin{tabular}[b]{c} 
$-X_{23}^{}.X_{36}^{}.X_{62}^{}$ \\
$+X_{25}^{}.X_{56}^{}.X_{62}^{}$ \\
$+X_{36}^{}.X_{64}^{}.X_{43}^{}$ \\
$-X_{45}^{}.X_{56}^{}.X_{64}^{}$ \\
$+X_{12}^{}.X_{23}^{}.X_{31}^{}.X_{14}^{}.X_{45}^{}.X_{51}^{}$ \\
$-X_{12}^{}.X_{25}^{}.X_{51}^{}.X_{14}^{}.X_{43}^{}.X_{31}^{}$
\end{tabular}

\end{tabular}
\end{center}

\caption{Tilings with 6 superpotential terms and 6 gauge groups \bf{(page 3/3)}}
\label{t:tilings6-6}
\end{table}

\ \clearpage


\end{document}